\documentclass{aa}  

\usepackage{graphicx}
\usepackage{txfonts}
\usepackage{xcolor}
\usepackage{lipsum}
\usepackage{subcaption}         
\usepackage{lscape}             
\usepackage{placeins}           
                                
\begin{document}

   \title{Self-interacting Dark Matter signatures from stars extremely magnified by galaxy clusters}


%
%
%

    \author{Sung Kei Li\inst{1, 2, 3}\corrauth{skli@ifca.es}\fnmsep\thanks{Croucher fellow}  \and
    Yuxuan He \inst{4}  \and
    Yi-Ming Zhong \inst{4} \and
    Jose M. Diego \inst{1} \and
    Jeremy Lim \inst{2, 3} \and
    Amruth Alfred \inst{2, 3} \and
    Thomas Broadhurst\inst{5, 6, 7} \and 
    Sangjun Cha \inst{8}
    \and
    Birendra Dhanasingham \inst{9} \and
    Leo W.H. Fung \inst{10} \and
    Patrick Kelly \inst{9} \and 
    Richard Massey \inst{10} \and
    Paloma Morilla \inst{5} \and 
    Jose M. Palencia \inst{11} \and
    Derek Perera \inst{9} \and
    Liliya L.R. Williams \inst{9}
        }

   \institute{IFCA, Instituto de F\'isica de Cantabria (UC-CSIC), Av. de Los Castros s/n, 39005 Santander, Spain
   \and
   Department of Physics, The University of Hong Kong, Pokfulam Road, Hong Kong  
   \and The Hong Kong Institute for Astronomy and Astrophysics, The University of Hong Kong, Pokfulam Road, Hong Kong, P. R. China
   \and Department of Physics, City University of Hong Kong, Kowloon, Hong Kong SAR, China
   \and Department of Physics, University of Basque Country UPV/EHU, Bilbao, Spain 
   \and Donostia International Physics Center, Paseo Manuel de Lardizabal, 4, San Sebasti\'an, 20018, Spain 
   \and Ikerbasque, Basque Foundation for Science, Bilbao, Spain
   \and Academia Sinica Institute of Astronomy and Astrophysics (ASIAA), No. 1, Section 4, Roosevelt Road, Taipei 106319, Taiwan  
   \and Minnesota Institute for Astrophysics, University of Minnesota, 116 Church St. SE, Minneapolis, MN 55455, USA
   \and Institute for Computational Cosmology, Department of Physics, Durham University, South Road, Durham DH1 3LE, UK
   \and David A. Dunlap Department of Astronomy and Astrophysics, University of Toronto, 50 St. George Street, Toronto, Ontario M5S 3H4, Canada
   }

   \date{Received \today}

 
  \abstract{The spatial distribution of cluster-lensed star transients with respect to the macroscopic critical curve (CC) is sensitive to underlying substructures predicted by different classes of dark matter (DM) models. \citet{Broadhurst_2025} have shown that wave-like DM composed of ultra-light bosons better explains both the width and negative skewness (towards the interior of the CC) of the lensed star spatial distribution discovered in the ``Dragon'' lensed galaxy ($z = 0.73$) in the Abell 370 (A370, $z = 0.375$) cluster field, compared with the case of regular cold dark matter (CDM) subhaloes. In this work, we consider the case of self-interacting DM (SIDM) and investigate how the lensed star spatial distribution depends on the velocity-dependent self-interaction cross-section of SIDM particles, and whether constraints can be placed on the cross-section strength at relevant velocities. Establishing a cross-section strength-dependent (sub)halo gravothermal core-collapse probability, we found that the SIDM appears to be the only known class of models thus far in which the sign of skewness is set by its own particle parameters. The observed lensed star distribution in the Dragon galaxy can be well explained, in terms of both skewness and width, when a significant fraction of subhaloes have undergone core collapse and exhibit a steep power-law density slope. Given the cluster redshift, this qualitatively favours an (effective) cross-section strength of $500 -1000\,\textrm{cm}^{2}/\textrm{g}$ for halos with mass $10^{6} - 10^{8}\,M_{\odot}$ (velocity dispersion of $\sim 1-10\,\rm{km}/\rm{s}$), when the role of environmental effects on the core-collapse (de)acceleration is not considered, and limits the three-dimensional inner density slope of the collapsed halo to $\Gamma \gtrsim 2.5$ (Bayes factor of $\sim 10^{2}$). }

   \keywords{Cosmology: dark matter --- Gravitational lensing: strong --- Gravitational lensing: micro --- Galaxies: clusters: general
               }

   \maketitle
\nolinenumbers

\section{Introduction}
\label{sec: intro}

The regular detection of lensed stars through deep, repeated {\it Hubble} Space Telescope (HST) and {\it James-Webb} Space Telescope (JWST) observations of galaxy clusters \citep[e.g.,][]{Kelly_2018, Kelly_2022_Flashlights, Yan_2023, Fudamoto_2025, Palencia_2026} has stimulated studies of Dark Matter (DM) substructures based on lensed star spatial distribution \citep{Williams_2024, Broadhurst_2025} with respect to the cluster critical curve (CC), where the lensing magnification is at its maximum. 
\citet{Broadhurst_2025} showed that both the broadness and the negative-parity skewness (towards the cluster core, interior of CC) of the spatial distribution of lensed stars in the ``Dragon'' lensed galaxy \citep[$z = 0.73$,][]{Fudamoto_2025} in cluster Abell 370 (A370, $z = 0.375$) aligns with the prediction made adopting the wave-like interference assuming ultra-light axions, whereas standard cold and collisionless DM subhaloes predict a narrower spatial distribution, and more importantly, an opposite, positive-parity (away from the cluster core, exterior of CC) skewness of lensed star distribution \citep[see also][]{Williams_2024}. Recent observations and calculations pinned down the CC position to a higher confidence level with different methodologies \citep[e.g.,][]{Zhou_2025, Diego_2025} and revealed more lensed stars \citep{Palencia_2026} that reinforce both the broadness and the negative lensed star skewness stressed in \citet{Broadhurst_2025}. 

In this paper, we explore for the first time the effect of another class of DM model known as self-interacting DM (SIDM) in the context of lensed stars, especially to see if SIDM can qualitatively explain the observed lensed star spatial distribution -- and if so, what can we say about the nature of SIDM particles. SIDM was first proposed to solve the small-scale problems of the cold and collisionless DM paradigm~\citep{Spergel:1999mh}. It provides solutions to the cusp-core problem, the too-big-to-fail problem, and the diversity problem \citep[see, e.g.,][for a review]{Tulin:2017ara}. Self-interactions also arise naturally when DM belongs to a dark sector, i.e., a collection of hidden particles that interact feebly with the Standard Model. New hidden particles, such as dark photons, can mediate non-gravitational interactions among DM particles. A heavy mediator gives rise to a constant and isotropic self-interaction cross-section, while a light mediator can give rise to a velocity-dependent cross-section \footnote{For simplicity, we always imply the velocity-dependent cross-section whenever the cross-section is mentioned}, as in Yukawa SIDM, resonant SIDM, and many others~\citep{Feng:2009mn,Buckley:2009in, Boddy:2016bbu, Huo:2019yhk, Gilman_2023}. Therefore, astrophysical probes of SIDM, which rely only on the gravitational effects of DM, offer a new way to study the nature of DM and dark sectors. For example, recent lensing observations through image reconstruction point towards overly concentrated subhaloes in both galaxy- and cluster-scale lenses, favouring the SIDM scenario \citep[e.g.,][]{enzi2024overconcentrateddarkhalostrong, Dutra_2024, Tajalli_2025, Vegetti_2026, Natarajan_2026} after which subhaloes have undergone core collapse. 

This paper is structured as follows. In Sec.~\ref{sec: mag_map}, we describe the methodology that computes the magnification distribution when a fraction of the lensing cluster is constituted of SIDM subhaloes under different conditions. In Sec.~\ref{sec: star_rate}, we outline the way in which the lensed star spatial distribution is computed based on the result in Sec.~\ref{sec: mag_map}. We then discuss the possible application of our result in constraining the interaction cross-section of SIDM particles and properties of SIDM subhaloes, followed by a brief estimation of how many events in total are required to distinguish between DM models with lensed star spatial distribution in Sec.~\ref{sec: discuss}. We then discuss briefly how different choices of hyperparameters and model uncertainties affect our results in Sec.~\ref{sec: uncertainties}.
Throughout this work, we adopt the AB magnitude system \citep{Oke_1983}, along with Planck18 cosmological parameters: $\Omega_{m} = 0.31$, $\Omega_{\Lambda} = 0.69$, and $H_{0} = 67.4\,\textrm{km s}^{-1}\, \textrm{Mpc}^{-1}$\citep{Planck18}.

\section{Magnification maps}
\label{sec: mag_map}

To estimate how the lensed star spatial distribution is affected by substructures, we have to first compute the magnification maps by considering a macroscopic DM halo (representing, for example, a galaxy cluster) being perturbed by DM substructures, and then compute the relative lensed star detection rate for different magnifications.


\subsection{Macroscopic Model}

For simplicity, we consider a galaxy cluster with $2\times 10^{15}\,M_{\odot}$ following a Navarro-Frenk-White \citep[NFW,][]{Navarro_1997} profile:

\begin{equation}
    \rho_{\rm NFW}(r) = \frac{\rho_{s, \textrm{NFW}}}{\frac{r}{r_{s, \textrm{NFW}}}\left(1+\frac{r}{r_{s, \textrm{NFW}}}\right)^{2}}
\label{eqn: NFW}
\end{equation}

\noindent as the macroscopic model. Such mass is common for massive clusters, including A370 \citep[e.g.,][]{Umetsu_2022}, which we shall compare against our simulations later on. For a given halo, its mass, $M_{200}$ (the enclosed mass within $r_{200}$ such that its enclosed density is 200 times the critical density),
and its concentration, $c_{200}\equiv r_{200}/r_{s, \rm{NFW}}$, follows the concentration-mass-redshift relation \citep[e.g.,][]{Diemer_2015} with a scatter of $\sigma = 0.16\,$dex in log space, such that the NFW scale radius and density can be written as:
\begin{equation}
    r_{s, \text{NFW}} = \frac{1}{c_{200}}\left(\frac{3M_{200}}{800\pi\rho_{\textrm{crit}}(z)}\right)^{\frac{1}{3}},
    \label{eq:rs}
\end{equation}
given the critical density at redshift $z$, $\rho_\text{crit}(z)$, and

\begin{equation}
    \rho_{s, \text{NFW}} = \frac{200}{3}\frac{c_{200}^{3}}{\ln(1+c_{200}) - c_{200}/(1+c_{200})}\rho_\text{crit}(z).
    \label{eq:rhos}
\end{equation}

\noindent For our cluster halo, $c_{200} \approx 4$, which is typical for clusters as found from lensing observations \citep[e.g.,][]{Mandelbaum_2008, Merten_2015}. Throughout this work, we adopt a lens redshift of $0.375$ and source redshift of $0.7251$.

\subsection{Subhaloes}

The gravothermal evolution of the SIDM (sub)haloes is based on an initial NFW profile (Eqn.~\ref{eqn: NFW}) where their time evolution is normalized by the collapse time scale:

\begin{equation}
\label{eqn: t_C}
    t_{C} \simeq \frac{200}{\rho_{s, \text{NFW}}r_{s, \text{NFW}}(\sigma/m)}\frac{1}{\sqrt{4\pi G \rho_{s, \rm{NFW}}}},
\end{equation}
 with $\sigma/m$ the effective cross-section strength of SIDM particles \citep{Essig_2019, Yang_2024} and the long-mean-free-path conduction factor is calibrated according to \cite{Koda:2011yb} (i.e., $0.75$). Note that Eq. \eqref{eqn: t_C} is only valid when the halo stays at the long-mean-free-path regime at its formation~\citep{Pollack:2014rja, Essig_2019, vandenBosch:2026iyn}, or more concretely, $\rho_{s\,\text{NFW}} r_{s,\text{NFW}} (\sigma/m) \lesssim \mathcal{O}(1)$. For $\rho_{s\,\text{NFW}} r_{s,\text{NFW}} (\sigma/m)$ to exceed the limit, the assumption of an initial NFW profile is not valid.  We estimate the (sub)halo evolution time $t_{\mathrm{evo}}$, i.e., the age of halo, as $t_{\mathrm{evo}}=t(z_l) -t(z_{\mathrm{form}})$, where $t(z_l)$ and $t(z_{\mathrm{form}})$ are cosmic times at the halo observation redshift $z_l$ and its formation redshift, $z_{\rm form}$, respectively. If $t_\text{evo} > t_C$, we consider that the halo has completely core-collapsed. Otherwise, the halo is in the core expansion stage or begins to enter the core collapse phase. 
 
 Using $\Lambda$CDM cosmology and Eqs. \eqref{eq:rs} and \eqref{eq:rhos}, we can rewrite Eq.~\eqref{eqn: t_C} of a halo as a function of $M_{200}$, $z_\text{form}$, and $c_{200}$ at its formation redshift. The concentration can be further fixed by using a concentration-mass-redshift relation. For a halo with mass $M_{200}$ at $z_l$, its core-collapsing probability, with formation concentration $c_{200,\mathrm{form}}$ and $z_{\mathrm{form}}$, is given by 
 \begin{align}
 \label{eq:prob_single}
 p_{0,\mathrm{coll}}(M_{200},z_l,z_\mathrm{form},c_{200,\mathrm{form}}) & = \Theta [t(z_l)-t(z_\mathrm{form})\nonumber\\
 &- t_C(M_{200},z_l,z_{\mathrm{form}}, c_{200, \mathrm{form}})],
 \end{align}
 where $\Theta$ is the Heaviside step function.
 
\begin{figure}
    \centering
    \includegraphics[width=\linewidth]{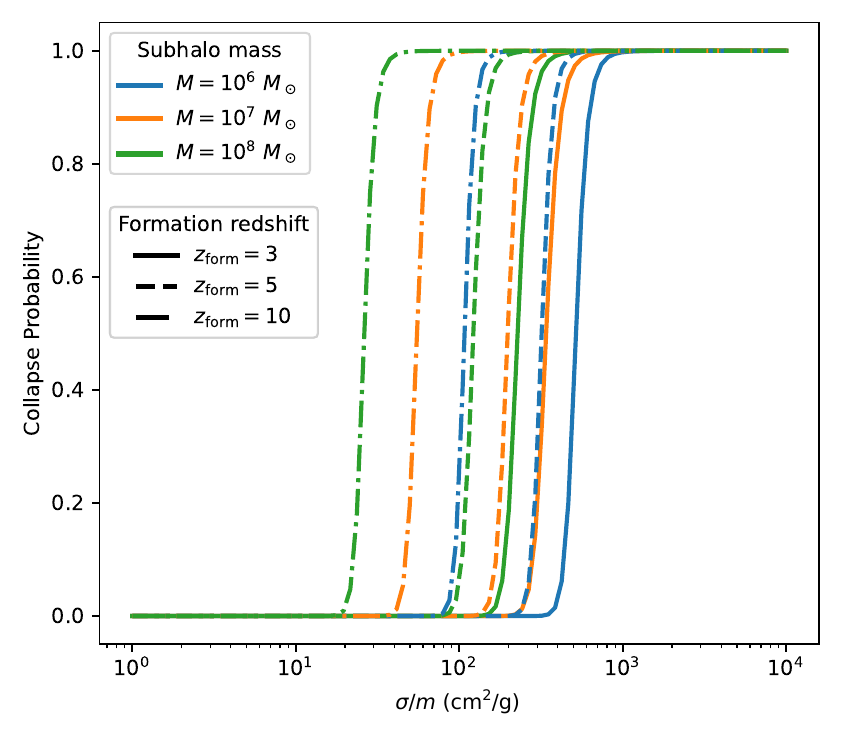}
    \caption{Core collapse probability as a function of SIDM interacting cross-section, for different subhalo masses thus velocities (colour), and formation redshifts (linestyle). }
    \label{fig: CC_P}
\end{figure}

To apply the probability to a population of similar halos, we replace $z_\mathrm{form}$ by their average formation redshift $\bar z_\text{form}$ and update Eq.~\eqref{eq:prob_single} to
\begin{equation}
\label{eq:prob}
p_{\mathrm{coll}} = \frac{1}{2}\left\{1+\tanh\left[\frac{t(z_l)-t(\bar z_\mathrm{form})-t_C(M_{200},z_l,\bar z_{\mathrm{form}},c_{200,\mathrm{form}})}{s(M_{200},z_l)}\right]\right\},  
\end{equation}
where $\bar z_{\mathrm{form}} (M_{200}, z_l)$ and the scatter $s(M_{200},z_l)$ are functions of halo mass and lens redshift, determined by fitting Eqn. \eqref{eq:prob} with a set of isolated halo samples generated by SatGen \citep[][]{Jiang_2021} merging tree, where we trace back the half-mass redshift $z_{1/2}$ and identify it as the formation redshift. The $c_{200,\mathrm {form}}$ is also take value at $\bar z_{\mathrm{form}} (M_{200}, z_l)$.
 One can see from Fig.~\ref{fig: CC_P} that, for the subhalo masses of interest between $10^{6} - 10^{8}\,M_{\odot}$, the core collapse probability is either $0$ or $1$ except in the range of $\sim 100 - 1000\,\textrm{cm}^{2}/\textrm{g}$ for a formation redshift of $\sim 3$ (solid curves), which is expected from high-resolution cluster simulation \citep[e.g.,][]{Xie_2015}. In other words, under such a setup, this method is expected to have sensitivity only to the SIDM interaction cross-section within this window, as any cross-sections below or higher than this range would be identical to the boundary cases. 

Note that environmental effects such as tidal stripping~\citep{Nishikawa:2019lsc} and the central baryonic component could have led to accelerated core-collapse~\citep{Zhong:2023yzk}, and the self-interactions could have velocity dependence such that small haloes, with larger cross-sections, will go through core-collapse quicker, still leading to over-concentrated subhaloes, as motivated by the claims of overly-dense subhaloes found through galaxy-scale strong lensing analysis mentioned earlier in Sec~\ref{sec: intro} \citep{enzi2024overconcentrateddarkhalostrong, Vegetti_2026}. We shall discuss the possible role of tidal stripping in further detail later in Sec.~\ref{sec: constraint}.

To model the density profiles of core-collapsed subhaloes, we adopt a cored power-law profile from \cite{Gilman_2023, he2026probingcollapseddarkmatter}:
\begin{equation}
    \rho_{\rm SIDM}(r) = \frac{\rho_{0}}{(1+r^{2}/r_{c}^{2})^{\Gamma/2}},
\label{eqn: cusp_SIDM_NFW}
\end{equation}
 which has a concentrated, cuspy profile outside the core radius $r_c$ with a power-law slope characterized by a constant, $2 \lesssim \Gamma \lesssim 3$~\citep{Turner:2020vlf, Fischer:2025rky}. Similarly steep three-dimensional inner slopes, \(\Gamma\gtrsim2.5\), have also been inferred from cluster strong-lensing analyses \citep{Natarajan_2026}, and we shall discuss this later in Sec.~\ref{sec: Gamma}. 
 
 The core radius, $r_{c}$, changes as the halo evolves~\citep{Zhong:2023yzk}. It first increases to $\mathcal O(1) r_{s, \textrm{NFW}}$ during the core expansion stage and then decreases during the core-collapse stage in a self-similar manner, during which the shrinkage accelerates near the onset of gravothermal catastrophe. The end state of the core-collapse leads to a secondary core with radius $r_c \sim 10^{-2} r_{s, \textrm{NFW}}$, which departs from the single-power-law profile ($\Gamma$) of its outskirts. To be concrete, we fix $r_c = 0.5\,r_{s,\textrm{NFW}}$ for subhaloes that have not yet core-collapsed \citep[e.g.,][]{Outmezguine_2023, Yang_2024}, and $r_{c} = 0.01\,r_{s, \textrm{NFW}}$ for those that have core-collapsed. This is because the core-collapse probability is nearly a sigmoid function for most interaction cross-sections, and only at very specific cross-sections would one have a thorough mix of subhaloes at different degrees of core-collapse. Moreover, as we shall show in Sec.~\ref{sec: r_c_effect}, the exact choice of $r_{c}$ in either case is not the main driver of our results, as the primary driver of lensed star spatial distribution is $\Gamma$, as well as $\sigma/m$.

$\rho_{0}$ of~Eq.~\eqref{eqn: cusp_SIDM_NFW} is computed by matching the total integrated mass of Eqn.~\eqref{eqn: cusp_SIDM_NFW} to the total integrated mass Eqn.~\eqref{eqn: NFW} to within $r\lesssim 2r_{s, \textrm{NFW}}$~\citep{Jiang:2022aqw, Zhong:2023yzk}. Since the projected mass of the NFW and power-law profiles diverges at large $r$, we have to truncate $r$ at some characteristic radius. For the NFW profile, the truncation would naturally be at $r_{200} = c_{200}\, r_{s, \text{NFW}}$. Effects such as tidal stripping could have truncated the subhaloes to a radius smaller than $r_{200}$, where we shall discuss this further in Sec.~\ref{sec: discuss}. Nevertheless, to conserve the number density of core-collapsed SIDM subhaloes with respect to the number density of NFW subhaloes (as SIDM core-collapsed subhaloes "evolve" from the initial NFW subhaloes), we truncate the power-law NFW profiles at some radius, $r_{\rm trun}$, such that the total projected mass within $r_{\rm trun}$ is equal to that of the NFW profile within $r_{200}$. 

\begin{figure}
    \centering
    \includegraphics[width=\linewidth]{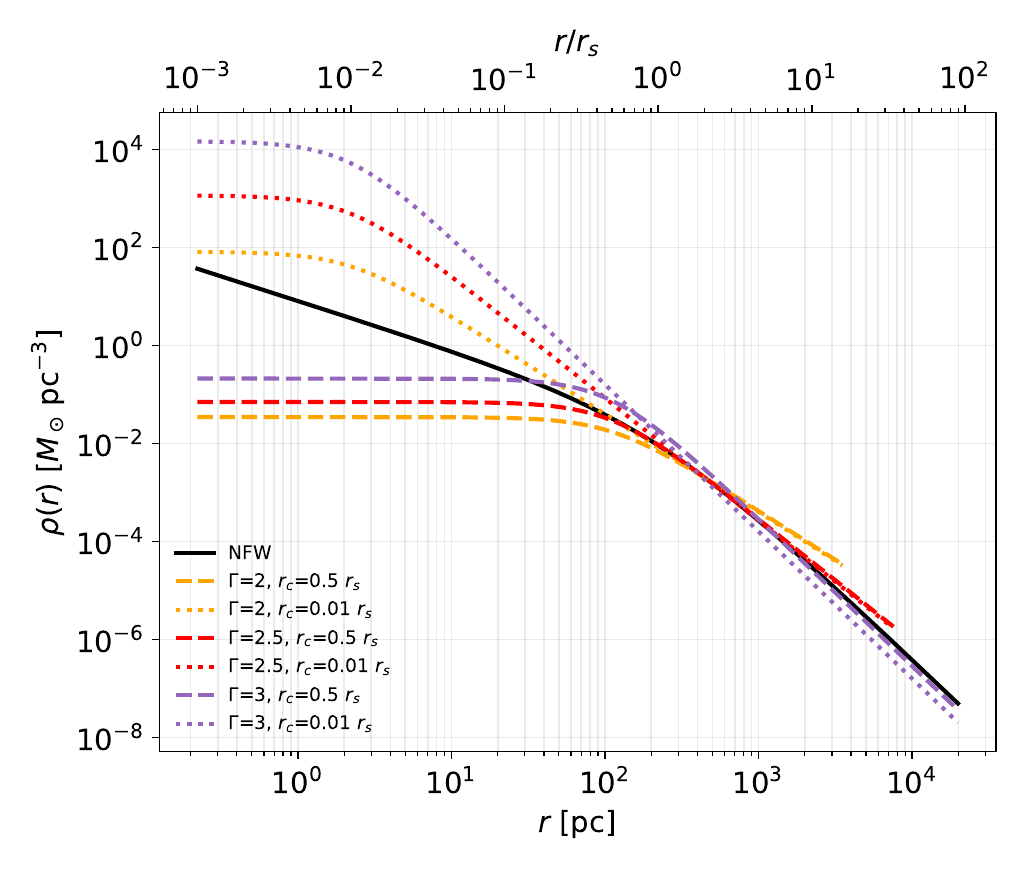}
    \caption{An example of profiles adopted in this work, for $\Gamma = 2$ (orange), $\Gamma = 2.5$ (red) and $\Gamma = 3$ (purple). Core-collapsed subhalos with $r_{c} = 0.01\,r_{s}$ are shown in dotted line style, and cored subhalos with $r_{c} = 0.5\,r_{s}$ are in dashed line. The solid black curve refers to an NFW subhalo for reference. All these haloes have $M_{200} = 10^{7}\,M_{\odot}$, and a therefore concentration of $\sim 17$ provided the \citet{Diemer_2015} mass-concentration relation. }
    \label{fig: profiles}
\end{figure}

A special scenario is $\Gamma \approx 3$. A prior matching between the initial NFW and SIDM subhalo mass within $2\,r_{s}$ will lead to a failure in finding a matching $r_\text{trun}$. This is because integrating the SIDM halo's cored power-law profile with $\Gamma = 3$ at $r_\text{trun} = \infty$ will only yield at most $\sim 70\%$ of the mass in the initial NFW subhalo. To preserve the mass and number density of subhaloes, instead of matching to an initial NFW profile, we match the cored power-law profile with $\Gamma \approx 3$ to a \citet{King_1962} profile:
\begin{equation}
    \rho(r) \propto (1+r^{2}/r_{c}^{2})^{-3/2},
\end{equation}
truncated at the $r_{200}$ calculated for an NFW subhalo with the same initial mass and concentration. We show a suite of the adopted profiles in Fig.~\ref{fig: profiles} alongside an NFW profile for reference. 

Given that self-interactions sufficiently randomize the velocity distributions of the scattered DM particles, it is reasonable to assume the subhaloes are perfectly spherically symmetric. We project the three-dimensional mass profile along the line-of-sight into surface mass density, $\Sigma(r)$. Within the thin lens approximation, the reduced deflection field can then be expressed as:

\begin{equation}
    \alpha(r) = \frac{8\pi G}{c^2 \, R} \int_{0}^{R} r \, \Sigma(r) \, dr.
\end{equation}

If subhaloes are to take up a fraction of the DM halo,  we can replace some fraction of the mass of the macroscopic NFW halo with subhaloes. We consider a case where these subhaloes take up $1\%$ of the DM mass \citep[same as previous lensed star spatial distribution studies, e.g.,][]{Williams_2024, Broadhurst_2025}, following a typical subhalo mass function \citep{Dai_2018, Dai_2020}:

\begin{equation}
    \frac{d\,N}{d\,\log\, M_{200}} \propto M_{200}^{-0.9}
\end{equation}

\noindent Instead of simply summing up their $M_{200}$ mass, we consider their total mass integrated within the field-of-view (FOV) of interest (i.e., near the cluster CC, as it is computationally impossible to do a fine-scale simulation for the whole cluster) to ensure the conservation of projected surface mass density. We adopt an image plane resolution of $\sim 3.3\,\textrm{pc}$ ($\equiv 6.3\times10^{-4}\,$arcsec) to maximize the resolution of microscopic CCs formed from subhaloes while retaining a reasonable computational cost. Sampling all subhaloes within the FOV, we can make use of the linearity of the deflection field to co-add the deflection field contributed by the cluster NFW profile, as well as those from subhaloes, to compute the lensing magnification:

\begin{equation}
    \mu = \textrm{det}\left(\delta_{ij} - \frac{\partial\alpha_{i}}{\partial \theta_{j}}\right)^{-1},
\label{eqn: magnification}
\end{equation}

\noindent with $\theta$ being the image plane position.

\subsection{Magnification maps}

\begin{figure}
    \centering
    \includegraphics[width=\linewidth]{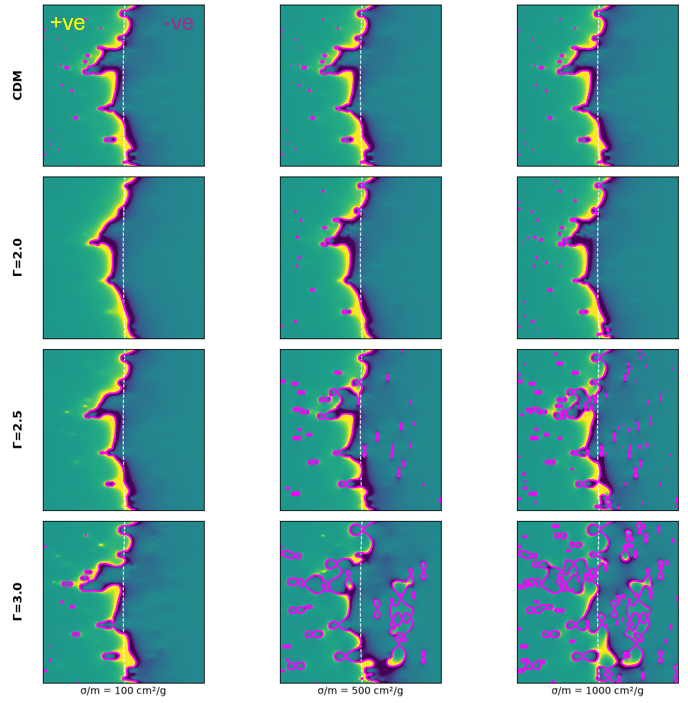}
    \caption{Macro magnification maps of CDM subhaloes (top row), SIDM core-collapsed subhaloes with $\Gamma = 2$ (second row), $\Gamma = 2.5$ (third row), and $\Gamma = 3$ (bottom row) when $\sigma/m = 100\,\textrm{cm}^{2}/\textrm{g}$ (left column), $500\,\textrm{cm}^{2}/\textrm{g}$ (middle column), and $1000\,\textrm{cm}^{2}/\textrm{g}$ (right column) respectively. Subhaloes take up $1\%$ of the DM mass in all these maps. The white, almost vertical dashed line shows the smooth CC when there is only the cluster lensing effect without substructures. The region towards the right of the smooth CC is known as the negative parity, whereas the region towards the left is the positive parity. }
    \label{fig: mag_maps}
\end{figure}

We show a suite of magnification maps in Fig.~\ref{fig: mag_maps}, where the top row features the scenario where one adopts regular NFW subhaloes following the standard CDM convention. Other than that, each row shows simulations adopting SIDM core-collapsed subhaloes for different $\Gamma$, whereas each column is a case with different characteristic interaction cross-sections.\footnote{The condition for Eq.~\eqref{eqn: t_C}, $\sigma/m \lesssim \mathcal{O}(1)/(\rho_{s\,\text{NFW}} r_{s,\text{NFW}})$ are satisfied for all benchmarks, with $\sigma/m=1000\,\text{cm}^2/\text{g}$ is close to saturate the limit.} For consistency's sake and to best compare the effect of the choice of $\Gamma$ and $\sigma/m$, we use the same seed to generate these magnification maps such that they have the same subhalo mass function sampled, the same concentration for each subhalo despite the scatter in the mass-concentration relation, as well as the same position within the FOV. Notice that the magnification map of the case of CDM subhaloes follows closely the distribution in \citet{Williams_2024, Broadhurst_2025} where the originally smooth cluster CC (white dashed line) is extended towards the positive parity (left) in most regions, with extra milliCCs forming exclusively at the positive parity side due to subhaloes. 

For shallow SIDM profiles with $\Gamma=2$, and at a smaller cross-section strength of $\sim 100\,\textrm{cm}^{2}/\textrm{g}$ (left column), none of the subhaloes have undergone core collapse (as indicated in Fig.~\ref{fig: CC_P}), and have retained a large core of $0.5\,r_{s}$. With increasing $\Gamma$, these subhaloes create stronger perturbations towards the macroscopic CC, bringing the CC further out towards the positive parity, like the case of NFW subhaloes. At an intermediate cross-section strength of $\sim 500\,\textrm{cm}^{2}/\textrm{g}$ (middle column), some of the subhaloes have already gone through core collapse, as shown in Fig.~\ref{fig: CC_P}. While the case of $\Gamma = 2$ remains vastly similar to the case of NFW subhaloes, smaller CCs begin to form in the negative parity when $\Gamma \gtrsim 2.5$. Since CCs form when the projected surface mass density transitions between sub- and supercritical, this explains why milliCCs are only found where $\Gamma\gtrsim2.5$, as these steeper profiles can generate supercriticality in the negative parity (see Sec~\ref{sec: analytical} for an analytical understanding). The steeper the density profile, the more dramatic the CCs are. Lastly, at a larger cross-section strength of $\sim 1000\,\textrm{cm}^{2}/\textrm{g}$ (right column), where all subhaloes have their core collapsed, as shown in Fig.~\ref{fig: CC_P}, the CCs are most dramatically perturbed. For the steepest $\Gamma = 3$ case, one can see that not only are more milliCCs formed in the negative parity, but the macroscopic CC is also ``deformed'', similar to what is expected for wave-like DM \citep{Amruth_2023, Broadhurst_2025} when they perturb macroscopic CC through the under- and over-density created by wave-like interference.

\begin{figure}
    \centering
    \includegraphics[width=1\linewidth]{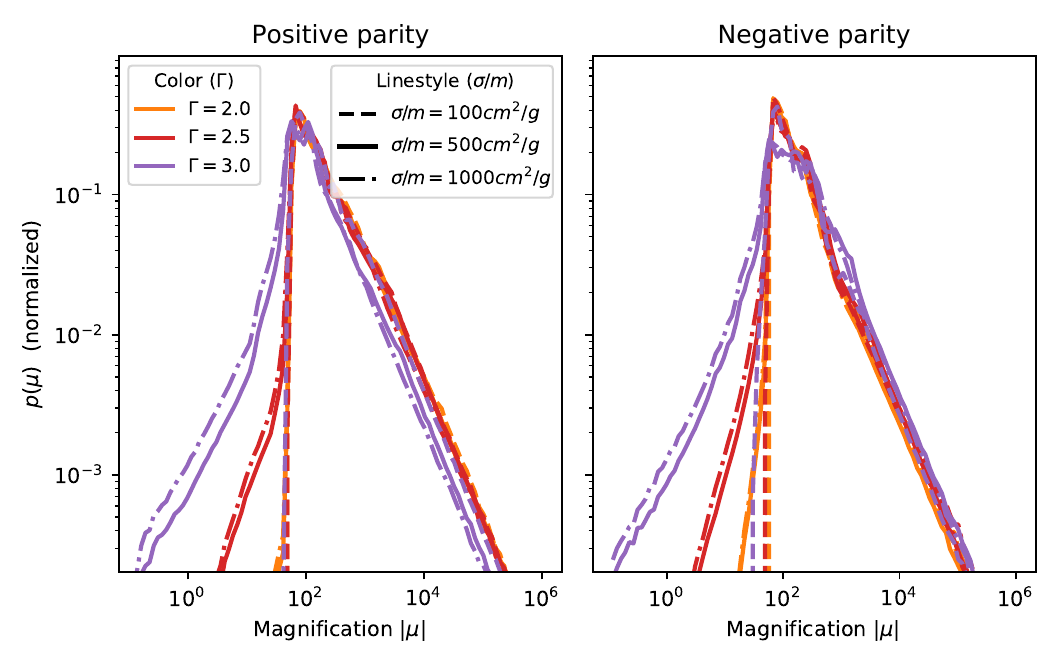}
    \caption{Probability density function of macroscopic lensing magnification near the CC (from the magnification maps in Fig.~\ref{fig: mag_maps}), for different SIDM interaction cross-section strengths (linestyles) and different power-law index, $\Gamma$ (colour). We show the case for the cluster positive parity and the negative parity in the left and right panels, respectively. }
    \label{fig: mag_pdfs}
\end{figure}

In Fig.~\ref{fig: mag_pdfs}, we also show the macroscopic magnification probability density function in each of the maps in Fig.~\ref{fig: mag_maps}, which helps interpret the result. In the positive parity, for a low interaction cross-section, where no subhaloes have undergone core collapse, all three $\Gamma$ values give rise to very similar PDFs. When we increase the interaction cross-section, the collapsed core makes the profile much steeper depending on $\Gamma$, such that the steeper the power-law slope, the more demagnification. For the higher magnification tail, although all the cases demonstrate the $\mu^{-2}$ slope, the $\Gamma = 3$ case, when subhaloes have collapsed, has a lower probability of producing higher magnification than the cases of $\Gamma = 2$ and $2.5$. In the negative parity, a similar trend is observed, except that now $\Gamma = 3$ cases have a higher probability of gaining higher magnification. 

Significant demagnification with subhaloes (or point mass lenses) only occurs in the negative parity\footnote{A unique case is for wave-like DM, as the density modulation owing to wave-like interference can create demagnification easily in either parities in another mechanism} \citep[e.g.,][]{Diego_2019, Palencia_2024} -- the reason why demagnification is also observed in the positive parity, as shown in Fig.~\ref{fig: mag_pdfs}, is that the ``main'' CC has been dragged towards the positive parity by massive subhaloes (see Fig.~\ref{fig: mag_maps}). In this case, part of the original positive parity in the macroscopic model is now actually negative parity, where demagnification occurs.

\section{Lensed star detection and spatial distribution}
\label{sec: star_rate}

To estimate the lensed star detection rate under the effect of microlensing on top of strong lensing, we have to first assume a stellar luminosity function, $N(m)$, of the background galaxy and estimate the lensed star detection probability given some magnification. Instead of computing $N(m)$ from star formation history and stellar isochrones, we assume a power-law stellar luminosity function where the relative abundance of stars with luminosity is:

\begin{equation}
    dN/dL \propto L^{-\beta},
\end{equation}

\noindent which in apparent magnitude, $m$, space, is:

\begin{equation}
    dN/dm \propto 10^{0.4(\beta-1)m},
\label{eqn: sLF}
\end{equation}

\noindent truncated at an absolute magnitude of $\sim -12$, roughly corresponding to the brightest stars known in the local universe and the prediction made by isochrones \citep[e.g.,][]{Choi_2016}. Recent lensed star observations deduced $\beta \approx 2-2.5$ \citep{Diego_2024_3M, Broadhurst_2025, Palencia_2026}, and we adopt $\beta = 2$ as a fiducial value. We shall show later in Sec.~\ref{sec: beta} that $\beta = 1.5$ or $\beta = 2.5$ \citep[$\sim2\sigma$ lower and upper limits in][]{Palencia_2026} only changes the statistical significance of our result but not the qualitative conclusion. Based on the underlying macroscopic magnification\footnote{For simplicity, we assumed that the macroscopic magnification is dominated by the tangential magnification such that the radial magnification equals unity. This is common for tangential arcs, but is not explicitly true for all lensed galaxies \citep[e.g., ][]{Palencia_2025_microlensing} despite the fact that this would not affect our qualitative discussion.}, $\mu_{m}$, (perturbed by substructures or not) and the abundance of microlenses (in terms of surface mass density, $\Sigma_{\star}$), background stars have some probability of attaining some total magnification, $p(\mu; \mu_{m}, \Sigma_{\star})$, under the influence of microlensing as approximated by \citet{Palencia_2024}. Here we adopt a fiducial $\Sigma_{\star}$ of $\sim 10\,M_{\odot}/\textrm{pc}^{2}$, which is typical for cluster-lensed galaxies with lensed star detections, including the Dragon Arc, against which we will compare our simulations \citep[e.g.,][]{Diego_2024_buffaloflashlights, Li_2024, Meena_2025_IMF, Li_2025_IMF}. Such a value is deduced from analysis of intra-cluster light \citep[e.g.,][]{Morishita_2017}, and assuming there are little to no dark compact objects such as primordial black holes at the relevant stellar mass range of $\sim \mathcal{O}(0.1-10\,M_{\odot})$, which is constrained by different kinds of observations like galactic microlensing and gravitational waves \citep[e.g.,][]{Green_2021}. We shall inspect what would happen to our result if the inferred surface mass density of microlenses differs significantly in Sec.~\ref{sec: uncertainties}.

If we assume that the surface brightness of the lensed galaxy is more or less constant (which is true for most lensed galaxies, as the brightness variation is of order $\sim 10\%$), then we can follow the methodology in \citet{Palencia_2025_microlensing, Li_2025_IMF, Li_2025_Horseshoe} and estimate the detection rate $\mathcal{R}$ as:

\begin{equation}
\label{eqn: detection_rate}
\mathcal{R}(\mu_{m}) \propto \int_{-\infty}^{m_{5\sigma}} \int^{\infty}_{\mu_{min}} p(\mu; \mu_{m}, \Sigma_{\star}) \times  \frac{N(m-2.5 \textrm{log}_{10}\mu)}{|\mu_{m}|}\, d\mu \, dm',
\end{equation}

\noindent with $m' = m-2.5\textrm{log}_{10}(\mu)$ describing the action of lensing magnifying the apparent magnitude of background stars. We adopt a fiducial yet typical {\it JWST} observation depth of $29\,\textrm{mag}_{AB}$ throughout our calculation, whereas $\mu_{min}$ is the minimum magnification required for a background star to be detected at this detection limit. By applying $\mathcal{R}(\mu_{m})$ to the macroscopic magnification map \citep{Broadhurst_2025} generated following the methodology stated in Sec.~\ref{sec: mag_map}, we can simulate the expected spatial distribution of lensed stars with respect to the macroscopic cluster CC (i.e., when there is only the main NFW halo). 

\begin{figure*}
    \centering
    \includegraphics[width=\linewidth]{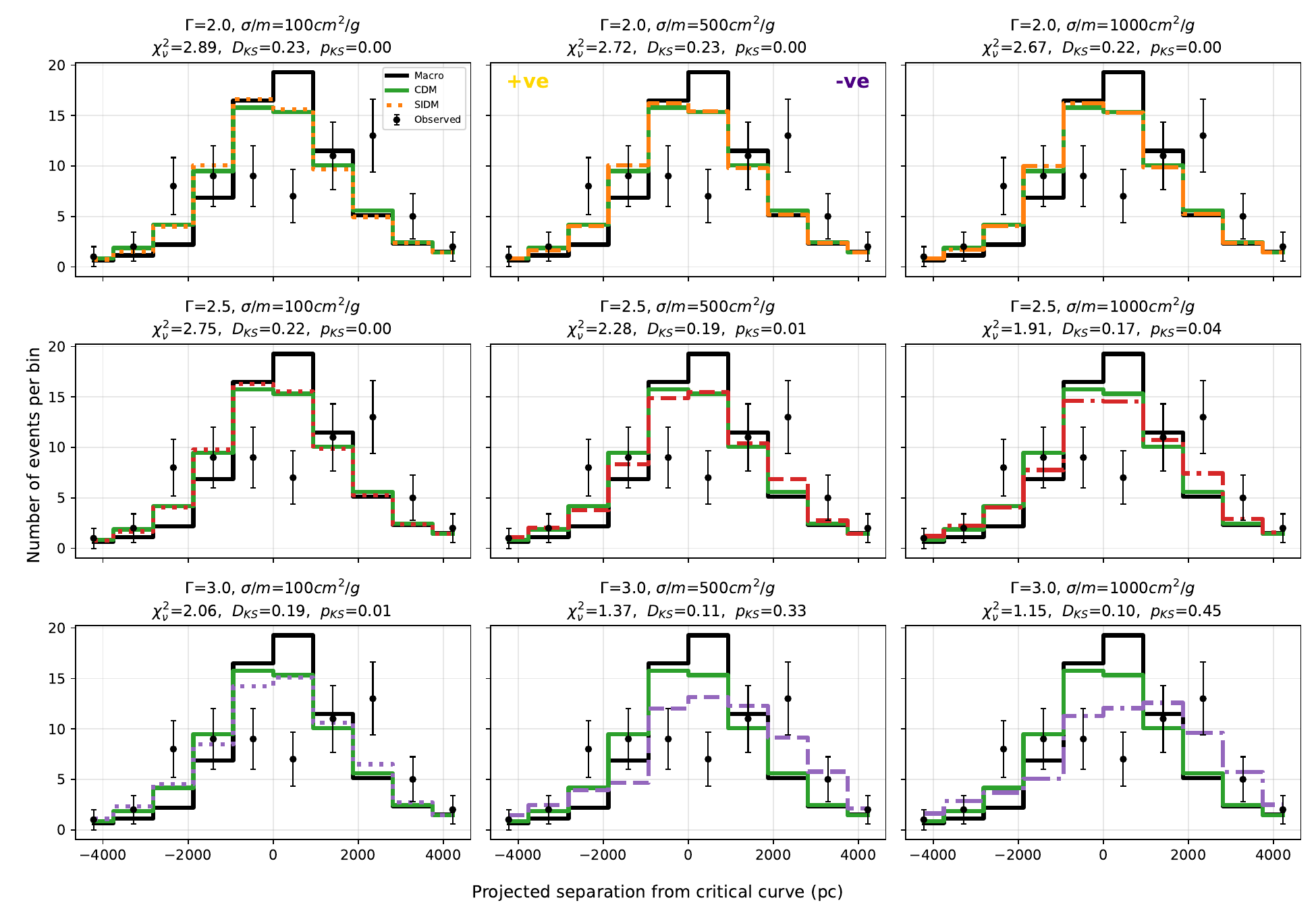}
    \caption{Expected strong+micro lensed star spatial distribution of smooth (black), CDM subhalo (green), SIDM core-collapsed subhalo with $\Gamma = 2$ (top row, orange), $2.5$ (middle row, red), and $3$ (bottom row, purple) respectively, with respect to the cluster CC, given $\sigma/m = 100\,\textrm{cm}^{2}/\textrm{g}$ (left column, dotted line), $500\,\textrm{cm}^{2}/\textrm{g}$ (middle column, dashed line), and $1000\,\textrm{cm}^{2}/\textrm{g}$ (right column, dash-dot line). Positive projected distance refers to the negative parity, whereas negative projected distance refers to the positive parity. Each bin has a width of 0.18 arcsec, symmetric around the smooth CC (x = 0). We also show the observed spatial distribution of lensed stars extracted from \citet{Kelly_2022_Flashlights, Fudamoto_2025, Palencia_2026} anchored to the \citet{Diego_2025} lens model at each corresponding bin as black data points, with the error bars showing the Poisson uncertainties at the corresponding bins. The reduced $\chi^{2}$ ($\chi^{2}_{\nu}$), KS-test statistics, $D_{KS}$, and p-value for each simulation run are shown at the top of each panel. }
    \label{fig: spatial_distribution}
\end{figure*}

We simulate the expected spatial distribution of lensed stars under microlensing for all the magnification maps shown in Fig.~\ref{fig: mag_maps} as well as for the case of a smooth CC with no perturbation \citep[which roughly resembles the prediction made with warm dark matter,][]{Lovell_2014} in Fig.~\ref{fig: spatial_distribution}. Owing to the fact that microlensing can demagnify images in the negative parity \citep{Diego_2019, Palencia_2024}, which is compensated by a higher probability of getting higher magnification, detection of a lensed star is more likely in the negative macroscopic magnification \citep{Broadhurst_2025, Li_2025_IMF, Palencia_2025_microlensing}. Based on this effect, a smooth model where the magnification follows $\propto 1/d$ with $d$ the separation from the CC \citep{Schneider_1992, Diego_2019} should show a slight negative skewness of the lensed star under the effect of microlensing, as shown in \citet{Broadhurst_2025} and the black histogram in Fig.~\ref{fig: spatial_distribution}. We found a positive skewness of the lensed star spatial distribution in the case of CDM subhaloes (green histogram), aligning with the earlier findings in \citet{Williams_2024, Broadhurst_2025}. 

In the case of core-collapsed SIDM subhaloes, the skewness of the lensed star spatial distribution depends on both $\sigma/m$ and $\Gamma$: for $\sigma/m \sim 100 \,\textrm{cm}^{2}/\textrm{g}$, all models demonstrate a positive skewness, as the CDM model does, as few to no subhaloes are core-collapsed such that they do not have sufficiently steep profiles to induce super-criticality at the negative parity. For $\Gamma = 2$, their profiles are not steep enough such that even when all the haloes are core-collapsed, the lensed star spatial distribution is still towards the positive parity. In fact, they resemble a similar spatial distribution to that of CDM subhaloes (green) for all cases of $\sigma/m$. In the case of $\Gamma = 2.5$, low $\sigma/m$ also resembles the case of CDM subhaloes and $\Gamma = 2$. However, when $\sigma/m$ grows to $\sim 1000\,\rm{cm}^{2}/\rm{g}$, all subhaloes have undergone core collapse, and a slight negative skewness is observed. At $\Gamma = 3$, if no subhaloes have undergone core collapse ($\sigma/m = 100\,\rm{cm}^{2}/\rm{g}$), the large core still prohibits milliCCs from forming in the negative parity despite the steep profile, thus resulting in a positive skewness. When some subhaloes have core-collapsed ($\sigma/m \gtrsim 500\,\rm{cm}^{2}/\rm{g}$), milliCCs begin to form at the negative parity (and are stronger than in the case of $\Gamma = 2.5$, see Sec.~\ref{sec: analytical}) and begin to create the negative skewness.




\section{Discussion}
\label{sec: discuss}


\subsection{Constraints on SIDM cross-section strength}
\label{sec: constraint}

Here, we compare our simulation results with observations to determine whether we can place constraints on the SIDM cross-section to the first order. Three lensed star catalogs published so far have detections in the ``Dragon'' galaxy \citep{Kelly_2022_Flashlights, Fudamoto_2025, Palencia_2026}, and we adopt the \citet{Diego_2025} lens model as the baseline to compute the separation of lensed stars from the cluster CC. Since the position of the CC is highly uncertain at the location of the cluster members near the head of the Dragon galaxy, we only adopt lensed stars detected away from that region, the same as was done in \citet{Broadhurst_2025}, yielding a total of 67 lensed stars in our sample. For all our simulations, we compare the prediction against the observed lensed star spatial distribution and compute the reduced $\chi^{2}$ \footnote{The individual data points are still in the Poissonian regime, so this primarily serves as a reference rather than a stringent statistical test} ($\chi^{2}_{\nu}$). Since our prediction is not in absolute value, we normalize all our predictions to have a total of 67 events.

Fig.~\ref{fig: spatial_distribution} shows that simulations with a larger interacting cross-section and larger $\Gamma$ are preferred in general. The best-performing models are the ones with $\Gamma = 3$, $\sigma/m = 1000\,\textrm{cm}^{2}/\textrm{g}$ and $\sigma/m = 500\,\textrm{cm}^{2}/\textrm{g}$, as they give a reduced $\chi^{2}$ of 1.15 and 1.37, respectively. The third best-performing model would be the one with $\Gamma = 2.5$ and $\sigma/m = 1000\,\textrm{cm}^{2}/\textrm{g}$, which has a reduced $\chi^{2}$ of 1.91.
These models all predict a skewness towards the negative parity, which is visualized in Fig.~\ref{fig: mag_maps}, where an abundant amount of milli-CCs is generated in the cluster negative parity when the subhaloes have undergone core-collapse with a steep power-law density profile. As a reference, $\chi^{2}_{\nu}$ is $\sim 4$ and $\sim 2.5$ for the case of the smooth model (black histogram) and regular NFW subhaloes in the case of standard CDM (green histogram). We can evaluate the significance of statistical preference between the models with the Asimov estimate \citep{Cowan_2011}:

\begin{equation}
    q = 2\sum_{i}[(x_{A, i} - x_{B, i}) - n_{i}\,\ln(x_{A,i}/x_{B,i})].
\label{eqn: asimov}
\end{equation}

\noindent Here, $x$ is the predicted count of models $A$ and $B$ to be compared upon at the $i$-th data point (spatial bin) with $n$ observed count in the corresponding data bin. The best-fitting SIDM model has $|q| \approx 24$ compared with the smooth model, meaning that it is $\sim 5\sigma$ better than the smooth model. Compared with the CDM model, $|q| \approx 10$, so the SIDM model is preferred over the CDM model by $\sim 3\sigma$. The qualitative difference between these models will eventually allow the distinction when sufficient observations have been accumulated. We shall address this aspect in Sec.~\ref{sec: number}.

As a binning-independent check, we adopt the one-sample Kolmogorov-Smirnov (KS) test on the 67 event positions. The KS statistic, $D_{KS}$, is given as the maximum difference between the observed and model cumulative distribution of the 67 independent observed positions. Since no parameters are fitted, we also compute the p-values from $10^{5}$ simulated samples for each model, with 67 events sampled for each realization. The KS test confirms the above $\chi^{2}$ test -- the model rankings are the same, rejecting both the smooth and the CDM model at a similar confidence level ($D_{KS} \approx 0.2$ for both models, with $p = 0.008$ and $0.006$, respectively). Both the $\Gamma = 3$ models with $\sigma/m = 500\,\rm{cm}^{2}/\rm{g}$ and $1000\,\rm{cm}^{2}/\rm{g}$ are fully consistent with the data ($D_{KS} \approx 0.1$ for both models, with $p = 0.33$ and $0.45$, respectively), where $\sqrt{N}\times D_{KS} \approx 0.8-0.9$ is typical for a correct model. A paired KS test with the best-fitting SIDM model and the smooth (CDM) model gives $3\sigma$ ($2.7\sigma$) preference, lower than the confidence level inferred from the Asimov estimate. This is attributed to the KS test only using the maximum cumulative deviation, whereas the Asimov estimate combines the difference between the models in every bin. Nevertheless, this demonstrates that the quantitative preference among models is broadly consistent between the two statistical tests.

\begin{figure}
    \centering
    \includegraphics[width=\linewidth]{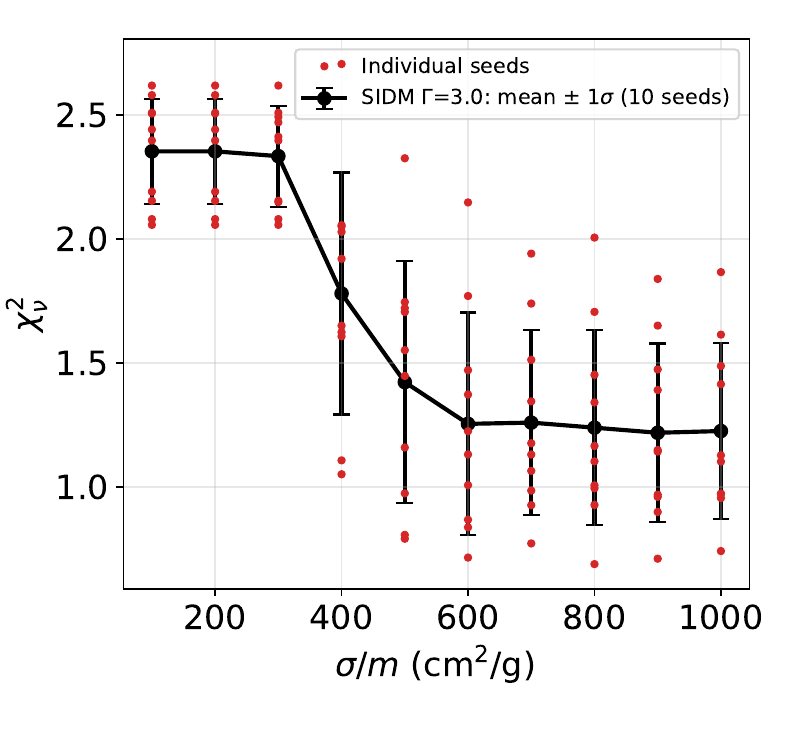}
    \caption{Reduced $\chi^{2}$ of simulations adopting $\Gamma = 3$ with different $\sigma/m$. Black data points refer to the mean value and the errorbar $1\sigma$ standard deviation among ten different realizations for each cases. The reduced $\chi^{2}$ of the individual realizations are shown in red data points.} 
    \label{fig: gamma3_reducedchisq}
\end{figure}

Since $\Gamma = 3$ best reproduces the observations among all our models, we rerun our simulation at a finer grid of $\sigma/m$ for $\Gamma = 3$ and show the reduced $\chi^{2}$ as a function of $\sigma/m$ within the range of $100 - 1000 \,\textrm{cm}^{2}/g$ in Fig.~\ref{fig: gamma3_reducedchisq}. Instead of fixing to one seed in generating ten solutions with different $\sigma/m$, we run the suite of simulations ten times with ten different seeds to sample the uncertainty in the randomly sampled subhalo mass function and mass-concentration relation that affect the result. One can see that there are two main sets of reduced $\chi^{2}$ -- one group with $\sigma/m  \lesssim300\,\textrm{cm}^{2}/\textrm{g}$ ($\chi^{2}_{\nu}\approx2.3$) and another with $\sigma/m \gtrsim 600\,\textrm{cm}^{2}/\textrm{g}$ ($\chi^{2}_{\nu}\approx1.2$). This trend is apparent when one refers to Fig.~\ref{fig: CC_P}, as $\sim 300-400\,\textrm{cm}^{2}/\textrm{g}$ is the interaction cross-section where the core collapse probability of $\sim 10^{6}\,M_{\odot}$ subhalo quickly transitions from zero to unity. Since these haloes are most abundant, this transition in core collapse probability propagates into the magnification distribution; thus, into the spatial distribution of lensed stars and the reduced $\chi^{2}$. Estimated from the scatter in Fig.~\ref{fig: gamma3_reducedchisq}, the first group has a seed noise of $\chi^{2}_{\nu} \approx 0.3$, and the second group has a seed noise of $\chi^{2}_{\nu} \approx 0.4$. Propagating the seed noise would then yield $\sim 0.5\sigma$ uncertainty in relevant statistics.

While the current scattering is large due to the Poissonian statistics, at face value, this shows that the lensed star spatial distribution provides a pathway to constrain the interacting cross-section of SIDM particles for subhaloes in cluster environments. However, the constraining power from the current calculation is incomplete, as the effect of $\sigma/m$ on $\Gamma$ is not explored, and we have simply taken $\Gamma$ as a hyperparameter that describes the SIDM density profile. Furthermore, we have neglected the fact that tidal forces could accelerate (or decelerate) the core collapse \citep{Nishikawa_2020, Zeng:2023fnj, mace2026sensitivitysubstructurelensingsidm} and make the truncation radii smaller, meaning that we could be underestimating (overestimating) the core-collapse probability as calculated in Eqn.~\ref{eq:prob}, thus overestimating (underestimating) the cross-section strength that is required to match the observed lensed star spatial distribution. 

In principle, the environmental effects can be considered as a nuisance parameter that affects our result through Eqn.~\ref{eq:prob}. To date, there are no tailored simulations that estimate the effect of tidal stripping upon subhalo core-collapse on \emph{cluster scales}. One can expect that tidal disruption is more severe for slow encounters, so the high relative velocities in clusters make individual encounters between subhaloes inefficient in tidal stripping \citep[e.g.,][]{Gnedin_1999}. Instead, it would be the cumulative effect of many such fast encounters, known as harassment \citep{Moore_1996}, on top of the repeated passage through the cluster potential that dominates the stripping. Another environmental effect comes from ram pressure. The self-interacting nature of SIDM makes ram-pressure stripping possible, especially when the interaction cross-section is larger \citep{Shirasaki_2022}. In such a case, the mass loss from ram-pressure stripping will make the (sub)haloes more susceptible to tidal stripping \citep{Nadler:2020ulu}, as well as making the truncation radius much smaller \citep[see discussion in][]{Chiang_2026} and yielding even more compact subhaloes. We shall discuss briefly the potential effect later in Sec.~\ref{sec: trunc_radius}

Because of the complication in accessing the environmental effect in cluster subhaloes, even though \citet{mace2026sensitivitysubstructurelensingsidm} infer that a stronger tidal effect does not appear to speed up the core-collapse process significantly in smaller-scale host haloes at a larger interaction cross-section (what is required to explain the observed lensed star spatial distribution according to our calculation), we hereby do not simply extrapolate their result to infer the constraining power of the lensed star spatial distribution upon the interacting cross-section of SIDM particles. 

Although lensing observations have constrained the cluster-scale interaction cross-section down to $\lesssim 1\textrm{cm}^{2}/\rm{g}$ level \citep[e.g.,][]{Randall_2008, Gomez_2012, Harvey_2014, Robertson_2016, Kim_2017, Cha_2025, diego2026largecoresdarkmatter}, there are few constraints on the interaction cross-section of subhaloes embedded in cluster-scale halos. The lower velocity of these subhaloes, with masses of $10^{6}-10^{8}\,M_{\odot}$, could allow for a larger interaction cross-section for velocity-dependent SIDM. One of the most relevant constraints comes from \citet{gilman2026hintsselfinteractingdarkmatter}, where the flux ratios from 29 quadruply-lensed quasars \citep{gilman2026jwstlensedquasardark} constrain the core-collapse timescale of low-mass subhaloes in galaxy lenses that reflect the interaction cross-section of SIDM particles. Their result disfavours $\sigma/m\lesssim10\,\rm{cm^{2}}/\rm{g}$ for velocities of $\sim 10\,\rm{km}/\rm{s}$ ($\sim 10^{8}\,M_{\odot}$, the upper mass end of our subhalo mass function), where two sets of more conservative and aggressive assumptions upon structure of collapsed-haloes and tidal effects leads to peak of likelihood at $\sim 1000\,\rm{cm^{2}}/\rm{g}$ and $\sim 200\,\rm{cm^{2}}/\rm{g}$, respectively. This aligns with our inference at face value, especially since they have adopted $\Gamma \approx 2.6$ and $2.9$, also roughly the $\Gamma$ that best reproduces the observed lensed star distribution. Since we include subhaloes with even lower mass (thus velocity) than their benchmark velocity of $\sim 10\,\rm{km}/\rm{s}$, meaning that our preferred interaction cross-section could have been larger than what they have inferred. Another relevant constraint comes from \citet{Sanchez_Almeida_2025}, where they used the stellar cores of ultra-faint dwarfs with masses of $\sim 10^{7}-10^{8}\,M_{\odot}$ to infer that $\sigma/m$ lies between $0.35-199\,\rm{cm}^{2}/\rm{g}$, assuming that the cores are the end result of DM self-interactions. This is at the more massive end of our subhalo mass function, where one would expect the less massive end of the mass function would have also significantly contributed to our inference. Neither inference is for the cluster environment, as we have done here; the different environmental effects mean that such a comparison only serves as a first-order sanity check of our result, and our methodology could be a complementary method to provide insights into the SIDM interaction cross-section at the relevant mass range in different environments.


\subsection{Preference in $\Gamma$}
\label{sec: Gamma}

Following the calculation in the previous section, we can also deduce that a steeper profile with $\Gamma \gtrsim 2.5$ better explains the observations, regardless of the interaction cross-section. We can marginalize over $\sigma/m$ and compute the preference of $\Gamma$ through the Poissonian deviance for all given sets of $\Gamma$ and $\sigma/m$ with Eqn.~\ref{eqn: asimov}, with the likelihood defined as:

\begin{equation}
    \mathcal{L}\propto e^{-q/2},
\end{equation}

\noindent if we set one of the models to be the observed value for Eqn.~\ref{eqn: asimov}. Marginalizing over $k$ choices of $\sigma/m$ with a flat prior:
\begin{equation}
P(\Gamma\,|\,n)\;\propto\;\sum_k\mathcal{L}(\Gamma,\sigma_k/m),
\end{equation}

\noindent would then yield posterior weights of $\sim 1\%$, $\sim9\%$ and $\sim 90\%$ for $\Gamma = 2$, $2.5$ and $3$, respectively. $\Gamma = 3$ is moderately preferred over $\Gamma = 2.5$ with a Bayes factor of $\sim 10$, whereas the hypothesis of $\Gamma\gtrsim 2.5$ is definitively preferred over $\Gamma = 2$ with a Bayes factor of $\sim 100$.


Recent findings claim that the observational cross-section of galaxy-galaxy strong lensing is significantly larger than the prediction of the standard CDM model, an issue that can be alleviated when (cluster) subhaloes adopt a SIDM-consistent core with $\Gamma \gtrsim 2.5$ \citep{Dutra_2024, Natarajan_2026}. Although aligning with our inference, such subhaloes are in the range of $\sim 10^{10-12}\,M_{\odot}$, significantly larger than the subhaloes considered in our work. For our mass range, the most relevant constraints come from galaxy strong lens modelling, which prefer $\Gamma\approx 2$ for a $\sim 10^{6}\,M_{\odot}$ subhalo \citep{Powell_2025}. That is, however, the case for subhaloes inside a field galaxy instead of a cluster environment. Given the different ambient environment, we cannot directly compare our inference with such a constraint. 
The different preference in $\Gamma$ at the two regimes could reflect different degrees of, for example, tidal disruption in SIDM subhaloes, and detailed simulation is required to better understand the root cause behind the different preferred $\Gamma$s.

In either weighting scheme, they prefer a density profile of $\Gamma \gtrsim 2.5$

\subsection{Comparison with wave DM prediction}

\citet{Broadhurst_2025} showed that wave DM with a characteristic wavelength of $\sim 15\,$pc, which corresponds to a typical cluster halo mass of $\sim 10^{15}\,M_{\odot}$ and an axion mass of $\sim 10^{-22}\,$eV, best fits the observed lensed star spatial distribution in the Dragon arc with the \citet{Kelly_2022_Flashlights} and \citet{Fudamoto_2025} data. Since we have additional events from \citet{Palencia_2026}, we also predict and compare the expected lensed star spatial distribution with the wave DM simulations in \citet{Broadhurst_2025} as both a sanity check of our method and a comparison to the SIDM predictions. 

As in \citet{Broadhurst_2025}, we run the simulation for wave DM with the characteristic de Broglie wavelengths of $\lambda_{db} = 15$ pc, 20 pc, and 40 pc and compare with the observed lensed star spatial distribution. We found the reduced $\chi^{2}$ (KS-test p-value) of the three models to be 0.84 (0.5), 0.97 (0.6), and 2.40 (0.4), respectively. The quality of fit for the first two models is comparable with SIDM simulations adopting $\Gamma \approx 3$. Our best-fitting model, $\Gamma = 3$ and $\sigma/m = 1000\,\textrm{cm}^{2}/\textrm{g}$, performs similarly to the 15 and 20 pc wave DM simulations. This specific SIDM case corresponds to a de Broglie wavelength of $\sim 17\,$pc if we make use of the relation in \citet{Broadhurst_2025} to fit for the lensed stars distribution. The current difference in $\chi^{2}$ between the best-fitting SIDM ($\Gamma = 3$ and $\sigma/m = 1000\,\textrm{cm}^{2}/\textrm{g}$) and waveDM ($\lambda_{db} = 15\,$pc) models is only $\sim 2.8$ with ten bins, which is dominated by Poissonian statistical fluctuations, as the statistical uncertainty is larger than the difference itself. Future observations increasing the statistical confidence of the lensed star spatial distribution is required to distinguish between the two classes of models in the context of lensed stars.

\subsection{Observations required to distinguish between models}
\label{sec: number}

At this stage, an interesting question would be to ask how many lensed stars (and epochs of observations) are required to distinguish between different DM models. We begin by comparing the best-fitting SIDM ($\Gamma = 3$, $\sigma/m = 1000\,\textrm{cm}^{2}/\rm{g}$) and wave DM (de Broglie wavelength of $15\,$pc) models and see how many observations are required to distinguish between the two fixed models to $3\sigma$ confidence again through the Asimov estimate. The current observation yields $|q| \approx 4$, meaning that the two models are not significantly discriminated, with only $\sqrt{4} \approx 2\,\sigma$. We can take this one step further and assume one model is true, such that its prediction would then serve as the typical data -- comparing the other model against it will then inform us of the goodness-of-fit. The current two models give $|q| \approx 3$. $|q|$ grows linearly with the increasing number of events, $N$, such that the significance scales as $\sqrt{N}$. To achieve $3\sigma$, we would need 3 times the current $N = 67$, meaning that a total of $\sim 200$ events are required to distinguish between the two classes of DM models based on the lensed star spatial distribution. \citet{Fudamoto_2025} and \citet{Palencia_2026} together gave $\sim 60$ events (since a fraction falls on regions where the position of the CC is ambiguous) over three epochs of observation. In other words, about $\sim 10 $ total observations are required to distinguish the two models at the current rate of lensed star discovery. Right now, the JWST cycle 4-6 ``Dragon'' survey (GO-7345, PI: Fudamoto) is being executed and is expected to return five extra epochs of data ($\sim 100$ extra events). Expanding the calculation to other lensed galaxies (e.g., those observed by Vast Exploration for Nascent, Unexplored Sources, ``VENUS'', GO-6882, PI: Fujimoto) with proper marginalization would boost the amount of data available and help with the comparison. 

We can also consider the case for the same best-fitting SIDM model, but instead, compare with the case of the smooth cluster model ($|q| \approx 25$) and the CDM model ($|q| \approx 10$). Since the smooth model is already at $\sim 5\sigma$ significance from the best-fitting SIDM model, and the CDM model is currently at $\sim 3\sigma$ with this Asimov estimate, instead of asking for the number of events required to distinguish between CDM and our best-fitting SIDM at $3\sigma$, we compute the number of events required for $5\sigma$ significance. Repeating the same calculation but putting as $5\sigma$, we require a total of $ \sim 165$ events to distinguish between SIDM model of $\Gamma = 3$, $\sigma/m = 1000\,\rm{cm}^{2}/\rm{g}$, and the CDM model, which is expected to be achievable with the upcoming data from the Dragon survey. 

As an important note, none of these calculations have taken into account intrinsic scatter in the fitting quality arising from the stochastic sampling of the subhalo mass function and the placement of subhaloes, as shown in Fig.~\ref{fig: mag_vs_d}. Also, the actual number of observations that is required to separate between models is very likely to be larger than what we estimated here, which is the median expectation. Thus, this exercise serves no more than a rough estimation to show, at least, that the lensed star spatial distribution is a viable way of distinguishing between DM models.

\section{Effect of choice of model parameters}
\label{sec: uncertainties}

\subsection{Lens model dependency}

One of the uncertainties is the dependency on the lensed model, from which we infer the spatial distribution. While we have already avoided using the region with the highest lens model uncertainty near the head of the Dragon, models behave slightly differently at other parts of the Dragon owing to different modelling methods and different numbers of constraints used. Apart from \citet{Diego_2025}, we test if adopting two other recent models \citep[after 2025,][]{Eid_2025, cha2026testingoffsetsclusterscalehalos} will affect our inference. \citet{Eid_2025} is a tailored parametric model for the Dragon using the full pixel brightness and allows for extra degrees of freedom for the masses and truncation radii of the cluster members nearby; we adopt an updated version of \citet{cha2026testingoffsetsclusterscalehalos} for the Dragon, constructed by the hybrid code MrMartian \citep{Cha_2026} that includes ten extra knots as constraints compared with the published version, which only used one image of the Dragon as a constraint. Image reconstruction of these models resembles the morphology of the Dragon galaxy satisfactorily and thus is expected to be reliable for our purpose. Also, at the region with most lensed stars ($\alpha \approx 2^{h}39^{m}52.9^{s}, \delta = -1^{\circ}35'06''$), all models align within $2\sigma$ of the inference through velocity dispersion mapping in \citet{Zhou_2025}. 

\begin{figure}
    \centering
    \includegraphics[width=1\linewidth]{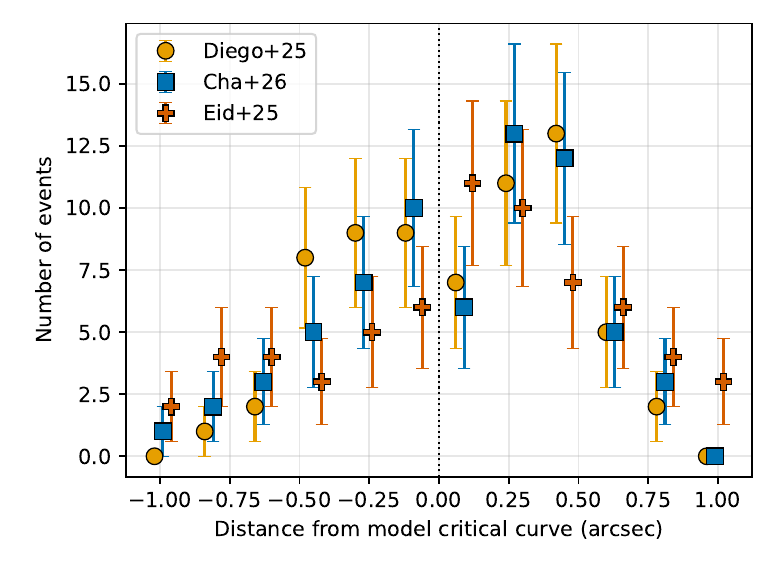}
    \caption{Dependency of lensed star spatial distribution upon choice of recent lens models \citep{Diego_2025, cha2026testingoffsetsclusterscalehalos, Eid_2025}. Again, negative parity is to the right (positive x-axis), and positive parity to the left (negative x-axis). }
    \label{fig: models}
\end{figure}

We show in Fig.~\ref{fig: models} the different inferred lensed star spatial distributions with respect to the model critical curves. All three models exhibit a similar skewness towards the negative parity, and agree to within $\sim 1\sigma$. We recompute the $\chi^{2}_{\nu}$ for all our nine models compared with the \citet{Eid_2025} and \citet{cha2026testingoffsetsclusterscalehalos} models, finding that the statistics behave similarly among the models. The $\Gamma = 3$, $\sigma/m = 1000\,\rm{cm}^{2}/\rm{g}$ model remains the best-fitting under the three models, with $\chi^{2}_{\nu} = 1.15$, $0.87$, and $1.01$ for \citet{Diego_2025}, \citet{cha2026testingoffsetsclusterscalehalos}, and \citet{Eid_2025}, respectively. The KS test gives the exact same conclusion, with $p = 0.45$, $0.69$, and $0.82$ for the three models. Such variation is within the Poissonian uncertainty; thus, the choice of lens model does not appear to be critical to our results.

\subsection{Substructure mass fraction }


\begin{figure}
    \centering
    \includegraphics[width=\linewidth]{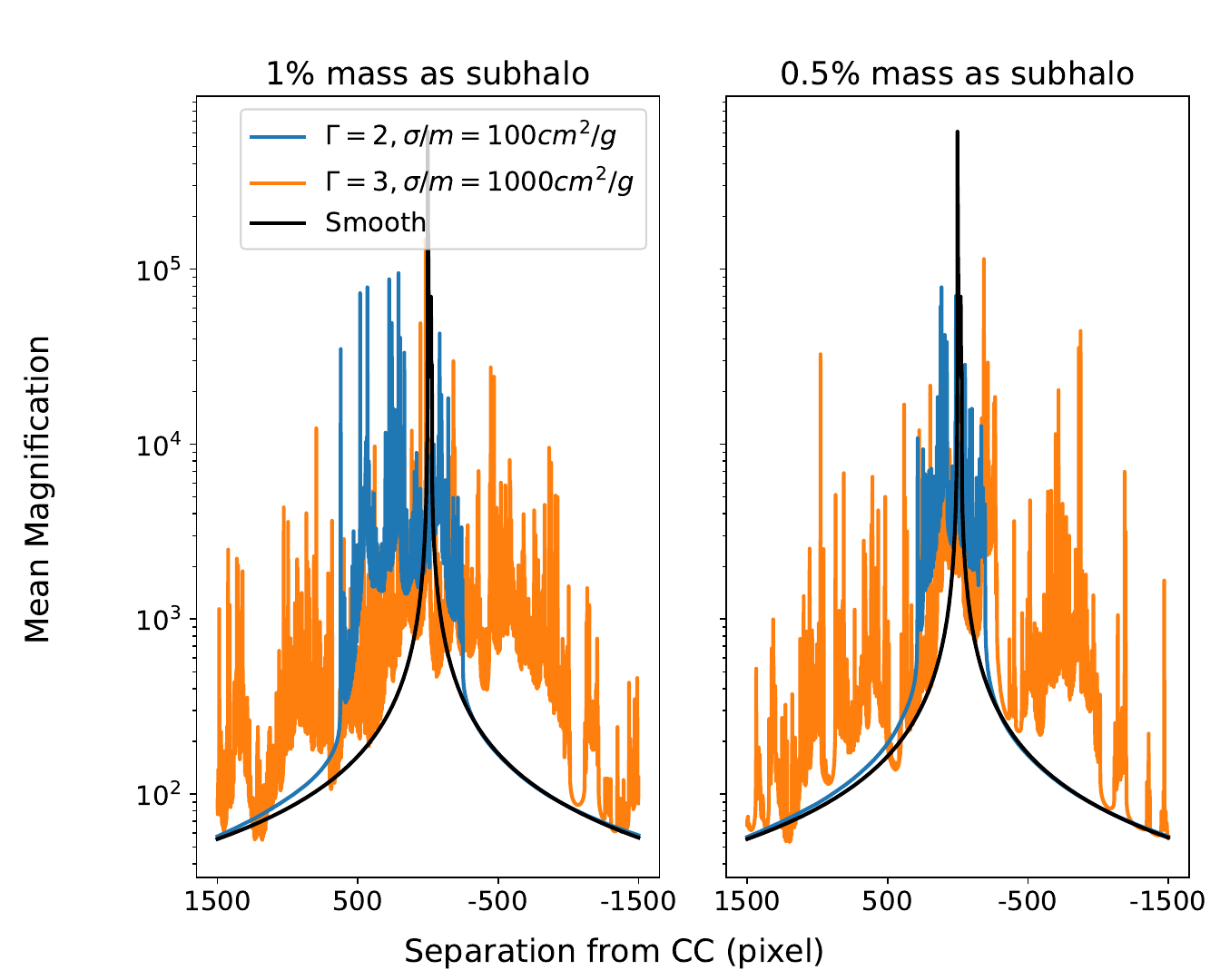}
    \caption{Mean absolute macroscopic magnification across the CC when integrating along the CC. The smooth model is shown in black as a reference curve, where one can see that the magnification drops inversely with the distance from the peak. Blue curves show the case for $\Gamma = 2$ and $\sigma/m = 100\,\textrm{cm}^{2}/\textrm{g}$ (case of minimum CC perturbation), whereas orange curves show $\Gamma = 3$ and $\sigma/m = 1000\,\textrm{cm}^{2}/\textrm{g}$ (case of maximum CC perturbation). The left and right panels are the cases when subhaloes take up $1\%$ and $0.5\%$ of the total mass of the cluster, respectively. }
    \label{fig: mag_vs_d}
\end{figure}

Another unknown is the fraction of cluster mass that is taken up by subhaloes. Recent inference from asymmetries in cluster gravitationally-lensed images places a $1\sigma$ constraint on the subhalo mass fraction of $0.18^{+0.44}_{-0.14}\%$ \citep{Perera_2026} and $\sim1\%$ \citep{Perera_2026b} under the same subhalo mass function. Our adopted $1\%$ is slightly above the former with $\sim 1.9\sigma$, but lies within $1\sigma$ of the latter, as well as the inference from galaxy-lens inferences \citep[e.g.,][]{Vegetti_2014, Hsueh_2020}. Since increasing the subhalo fraction in our simulation will deviate further from the previous constraints and significantly increase the computational cost, here we alternatively consider qualitatively what would happen if subhaloes take up more or less fraction of the cluster mass by looking at the magnification distribution across the CC given different subhalo mass fractions. We rerun our simulations with $0.5\%$ cluster mass (closer to the inference in \citet{Perera_2026}) as subhalo instead of 1\%, then show the magnification distribution across the CC (by averaging along the CC) for two representative cases that represent the least perturbation ($\Gamma = 2$, $\sigma/m = 100\,\textrm{cm}^{2}/\textrm{g}$, blue curve) and the most perturbation ($\Gamma = 3$, $\sigma/m = 1000\,\textrm{cm}^{2}/\textrm{g}$, orange curve) in Fig.~\ref{fig: mag_vs_d}.

One of the outstanding trends is that the fluctuations in the CC (with respect to the smooth CC) vanish at the same distance of $\sim 250$ pixels from the CC for the negative parity in the case of $\Gamma = 2$, regardless of the mass fraction. For the opposite parity, the fluctuation extends further to $\sim 500$ pixels when subhaloes take up more of the cluster mass. This trend is not observed in the other case ($\Gamma = 3$, $\sigma/m = 1000\,\textrm{cm}^{2}/\textrm{g}$), as milliCCs are forming everywhere in both parities. This could be understood through our derivation in Sec.~\ref{sec: analytical} -- shallow subhaloes can never form milli-CCs far beyond the macroscopic CC in the negative parity, so increasing the fraction of subhaloes does not help at all; on the other hand, adding more subhaloes will nucleate more milli-CCs that are further away from the macroscopic CC in the positive parity by both increasing the probability of having more massive subhaloes and the number density of subhaloes. This aligns with the fact that breaking the criticality in (milli)CC formation depends strongly on the slope of the substructure profile, especially when the substructure is lower in mass. Under this circumstance, the less mass fraction is taken up by subhaloes, the narrower the lensed star spatial distribution would be when $\Gamma = 2$ and $\sigma/m = 100\,\rm{cm}^{2}/\rm{g}$, as the magnification distribution resembles the smooth model. Such models would be less preferred by the observations -- as one could observe from Fig.~\ref{fig: spatial_distribution}, a narrow, strongly peaked lensed star spatial distribution like the smooth case (black histogram) will produce an excess of events close to the macroscopic CC instead of a plateaued distribution. In other words, the lensed star spatial distribution would also simultaneously place a limit on the subhalo mass fraction.

The amplitude, on the other hand, depends on the fraction of cluster mass in DM subhaloes: in both cases, $\Gamma = 2$, $\sigma/m = 100\,\textrm{cm}^{2}/\textrm{g}$ and $\Gamma = 3$, $\sigma/m = 1000\,\textrm{cm}^{2}/\textrm{g}$, the fluctuation in magnification is more dramatic with more subhaloes, while bringing down the peak magnification. This could be easily understood as a more abundant number of subhaloes increasing the perturbation towards the CC, redistributing the peak magnification to regions further away from the CC, as shown also in Fig.~\ref{fig: mag_maps}.

Qualitatively, the subhalo mass fraction should not shift the skewness of the lensed star spatial distribution: milliCCs -- the prerequisite of lensed stars' negative skewness -- can barely form in the negative parity as long as the subhaloes are of lower mass as considered in this work and have a shallow profile, regardless of the mass fraction. In fact, increasing the amount of shallow subhaloes will only form more milliCCs in the positive parity, leading to even more lensed stars towards the positive parity. This is also true for the case of steep subhaloes -- increasing the mass fraction does not significantly alter the magnification distribution at the two parities, and therefore should not affect the skewness of lensed stars.

\subsection{Core size}
\label{sec: r_c_effect}

The size of the core for core-collapsed subhaloes is also unknown in a cluster environment. It was shown in \citet{Williams_2024} that a smaller core (for a different profile) will lead to stronger CC perturbation in the negative parity. From Fig.~\ref{fig: mag_maps}, one can see that the core size is certainly affecting our case -- magnification distributions are drastically different when uncollapsed subhaloes (with large cores, $\sigma/m = 100\,\rm{cm}^{2}/\rm{g}$, left column) and collapsed subhaloes (with small cores, $\sigma/m = 1000\,\rm{cm}^{2}/\rm{g}$, right column) are respectively dominating the subhalo population, regardless of $\Gamma$. Here, we would like to test the sensitivity of our method to the collapsed core size by rerunning our simulations while replacing the core size of the core-collapsed subhaloes with different values. For the sake of computational power, we only compare the magnification maps for $\sigma/m = 1000\,\textrm{cm}^{2}/\textrm{g}$, where the perturbation of CC is most dramatic across all models, as all subhaloes are core-collapsed.

\begin{figure}
    \centering
    \includegraphics[width=\linewidth]{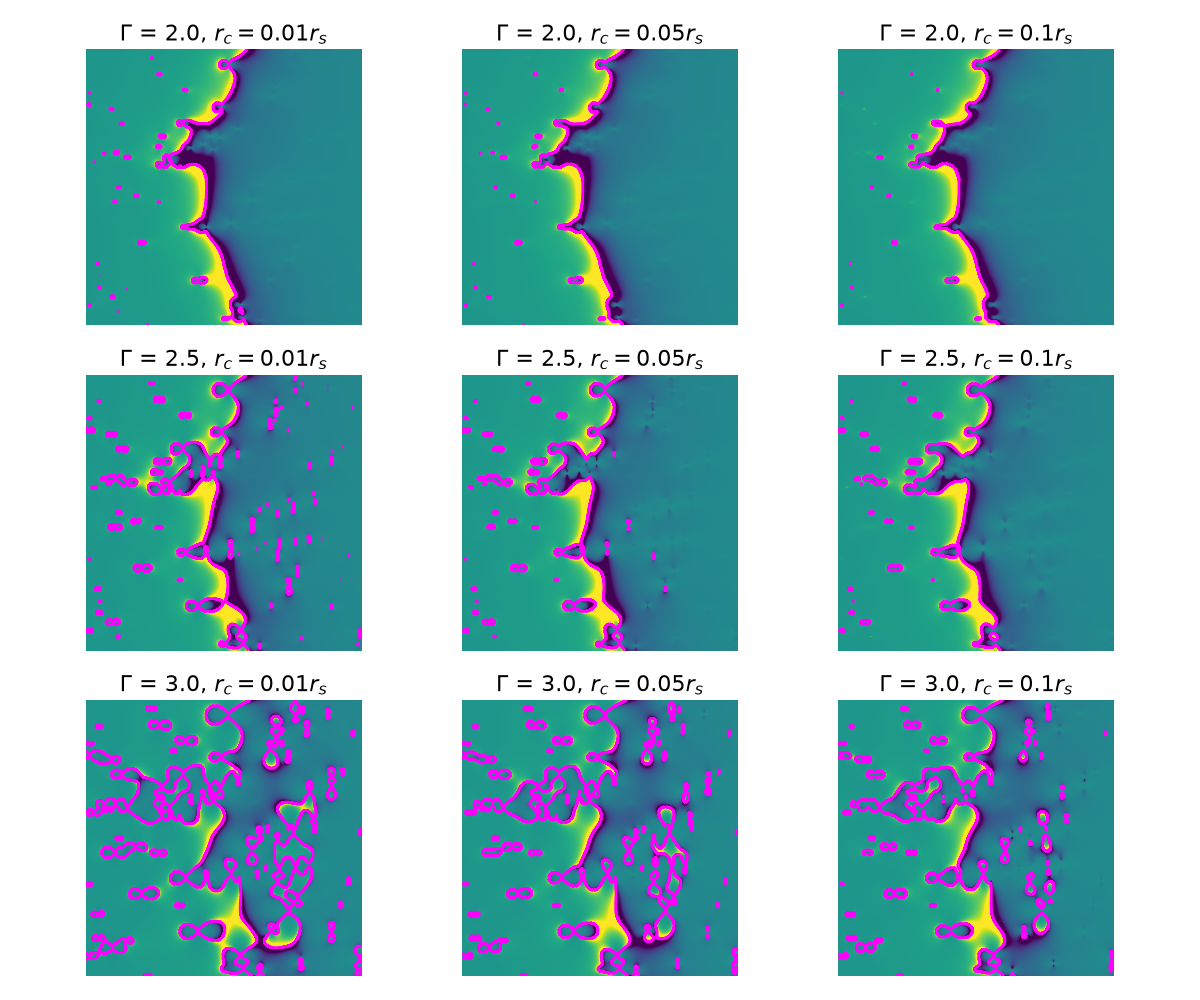}
    \caption{Same as Fig.~\ref{fig: mag_maps}, except we adopt $\sigma/m = 1000\,\textrm{cm}^{2}/\textrm{g}$ for all the simulations, and each column now instead represents a different core radius, $r_{c}$, as indicated in the title of each panel. }
    \label{fig: magmaps_compare_rc}
\end{figure}

We show the magnification maps in Fig.~\ref{fig: magmaps_compare_rc} for $r_{c} = 0.01\,r_{s}$ (left), $0.05\,r_{s}$ (middle) and $0.1\,r_{s}$ (right). For a shallower slope of $\Gamma = 2$, the choice of $r_{c}$ is insignificant. While the formation of milliCCs at the positive parity decreases with increasing $r_{c}$, no milliCCs are formed at the negative parity at any $r_{c}$. This is because the profile is too shallow and could never have achieved supercriticality in the negative region, even with a small core, let alone with a larger core now. Qualitatively, $r_{c}$ does not affect the lensed star spatial distribution when $\Gamma$ is small enough. At $\Gamma = 2.5$, a small fraction of milliCCs are formed at negative parity when $r_{c} = 0.01\,r_{s}$. Increasing $r_{c}$ makes it more and more difficult to achieve supercriticality in the negative parity region, generating fewer and fewer milliCCs. At $r_{c} = 0.1\,r_{s}$, no milliCC could be formed at the negative parity. A similar effect is observed in the $\Gamma = 3$ -- however, since $\Gamma = 3$ is extremely steep, even if we increase $r_{c}$ to $0.1\,r_{s}$, milliCCs are still formed at the negative parity, and the macro-CC is still being perturbed in both parities following the location of subhaloes. As shown earlier in Fig.~\ref{fig: mag_maps}, the dramatic perturbation for $\Gamma = 3$ is only absent when $\sigma/m = 100\,\textrm{cm}^{2}/\textrm{g}$, where none of the subhaloes have undergone core collapse and thus remain at $r_{c} = 0.5\,r_{s}$. Since the formation of milliCC is the main driver behind the negative skewness of lensed stars' spatial distribution, the choice of core size of core-collapsed SIDM subhaloes when $\Gamma \approx 3$ is less significant for the magnification map and therefore the expected lensed star spatial distribution, unless $r_{c}$ is increased to a value at which essentially the core size is the same as when it is not collapsed.

\subsection{Truncation radius}
\label{sec: trunc_radius}

Also, we assumed that subhaloes follow a truncation radius that matches the same total enclosed mass for an initial NFW profile truncated at $r_{200}$. As briefly mentioned in Sec.~\ref{sec: discuss}, tidal effects in the cluster are expected to not only speed up the core-collapse process, but also strip down the subhaloes to become smaller in radius. Since the stripping will return some of the subhalo mass back to the ambient environment, and the number of subhaloes is conserved upon tidal stripping \footnote{We have neglected the complete disruption of lowest mass subhaloes, partly because the fraction is expected to be rather low given our formation ($z = 3$) and observed epoch ($z= 0.375$), and partly because those would contribute least towards the magnification distribution}, this would naturally lower the fraction of cluster mass that is taken up by subhaloes. To inspect what happens if the subhaloes are tidally stripped to smaller truncation radii, we repeat our calculation following the same suite of sampled NFW subhaloes truncated at $r_{200}$, then strip off their mass at $0.2\,r_{200}$ and redistribute back to the cluster halo. The SIDM subhaloes are then matched to the total mass of each of the corresponding subhaloes enclosed within $0.2\,r_{200}$, the same way as described in Sec.~\ref{sec: mag_map}. This way, the setup is identical to our original simulation but having subhaloes with small truncation radius. The choice of $0.2\,r_{200}$ is motivated by extrapolating the inference in \citet{Niemiec_2019, Sirks_2022,Chiang_2026}, which return roughly $\sim 60\%$ of the $r_{200}$ mass back to the cluster. Nevertheless, these works study the mass loss and stripping of more massive haloes ($\gtrsim 10^{9-10}\,M_{\odot}$) in simulated clusters, which are at least an order of magnitude more massive than our considered subhaloes. Our less massive, more compact \citep[given the mass-concentration relation,][]{Diemer_2015} subhaloes could have lost less mass \citep{Niemiec_2019} and sink more slowly than the massive subhaloes \citep{Sirks_2022}. We hereby stress that this is no more than a qualitative assessment of how the tidal stripping is expected to affect our result.

\begin{figure}
    \centering
    \includegraphics[width=\linewidth]{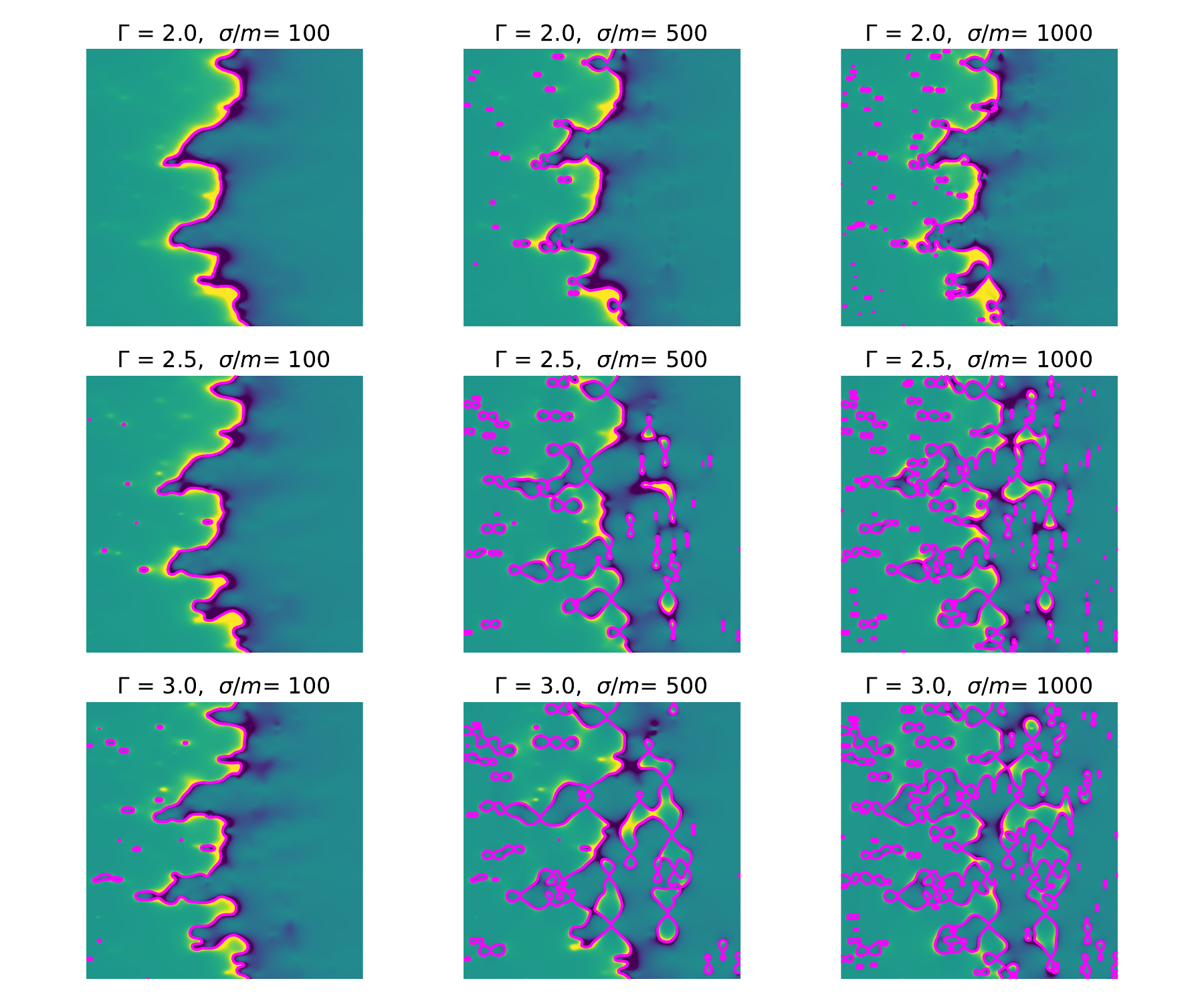}
    \caption{Same as Fig.~\ref{fig: mag_maps}, except we adopt a smaller truncation radius for the matched NFW subhaloes, $r_{\rm trun} = 0.2\,r_{200}$ instead of $r_{200}$, to mimic a scenario where our subhaloes are significantly stripped through tidal effects. }
    \label{fig: magmaps_smaller_trunc}
\end{figure}

We show all the revised magnification maps with a smaller truncation radius of $0.2\,r_{200}$ in Fig.~\ref{fig: magmaps_smaller_trunc}. Although the number density of the subhaloes is identical, the smaller truncation radii made them more compact, and formation of milli-CC is more vigorous even at a smaller $\Gamma \approx 2.5$ in the negative parity. The magnification distribution behaves no differently than the original simulation when $\sigma/m$ is small, and no subhaloes have undergone core-collapse; however, the steeper profile thus yields more abundant milli-CC in the negative parity, creating a negatively skewed and broader lensed star spatial distribution even at $\Gamma = 2.5$. The quality-of-fit is nearly unaffected for the case of $\Gamma = 3$, but improved significantly for the case of $\sigma = 500\,\rm{cm}^{2}/\rm{g}$ and $\sigma = 1000\,\rm{cm}^{2}/\rm{g}$ under $\Gamma = 2.5$ to $\chi^{2}_{\nu} = 1.06$ and $\chi^{2}_{\nu} = 0.87$, comparable (or even slightly better) than the original best-fitting SIDM model of $\Gamma = 3$ and $\sigma = 1000\,\rm{cm}^{2}/\rm{g}$ ($\chi^{2}_{\nu} = 1.15$). For the case of $\Gamma = 2$, despite the tidally truncated profiles being more compact, they failed to create sufficient milli-CCs on both sides of the CC that can redistribute the magnification and explain the observed width of the lensed star spatial distribution. The improvement in $\chi^{2}_{\nu}$ for $\Gamma = 2$ SIDM subhaloes under tidal truncation is within the seed noise. 

One extra contribution of tidal truncation we have neglected in our calculation is that it would simultaneously speed up the core collapse. \citet{Nishikawa_2020} demonstrated that NFW haloes truncated at $\sim 3\,r_{s}$ will core-collapse $\sim 2$ times faster than an untruncated halo. For our given range of subhalo mass, $3\,r_{s}$ would roughly correspond to $\sim 0.2\,r_{200}$, the same truncation radius we tested in this section. In other words, the preferred interaction cross-section could be read as up to $\sim 2$ times smaller when we consider tidal truncation at $0.2\,r_{200}$, i.e., preferring $\sigma/m$ to be $\gtrsim 250\,\rm{cm}^{2}/\rm{g}$ instead of $\gtrsim 500\,\rm{cm}^{2}/\rm{g}$.

This demonstrates that the truncation radius does affect our simulation and will affect the preferred interaction cross-section, $\sigma/m$, at least for the considered alternative truncation radius. Since this tested truncation radius is not a stringent result from tailored simulation and calculation, the tidal truncation of SIDM subhaloes in cluster environments has to be thoroughly explored (as much as how $\Gamma$ is coupled with $\sigma/m$) before we can refine our simulation to better understand the constraining power of the interaction cross-section of $10^{6}-10^{8}\,M_{\odot}$ subhaloes in clusters.

\subsection{Slope of luminosity function}
\label{sec: beta}

We adopted $\beta = 2$ for the slope of our adopted luminosity function (Eqn.~\ref{eqn: sLF}) -- as the slope of the luminosity function affects the degree of magnification bias \citep{Broadhurst_2025} and therefore the lensed star distribution, we follow the uncertainty inferred in \citet{Palencia_2026} and inspect how a different $\beta$ changes our results by repeating our calculations with $\beta = 1.5$ \citep[lower $2\sigma$ from][]{Palencia_2026} and $\beta = 2.5$ \citep[value deduced in][]{Diego_2024_3M, Broadhurst_2025}. The predicted spatial distributions are shown in Fig.~\ref{fig: b1p5} and Fig.~\ref{fig: b2p5}, respectively. 

The main difference with different $\beta$s is that a steeper luminosity function creates more events in the central region and makes all the fits worse. This is because of the magnification bias -- the increase in the number of fainter stars outperforms the reduction in source plane area at higher macroscopic magnification. 
While $\Gamma = 3$, $\sigma/m = 1000\,\rm{cm}^{2}/\rm{g}$ remains the best fit for all the $\beta$s considered, the statistical preference among models increases (decreases) for a steeper (shallower) luminosity function. As a steeper luminosity function makes all fits worse, the best-fitting SIDM model is now $\sim 6\sigma$ preferred over the smooth model, and $\sim 4\sigma$ preferred over the CDM model. One would just need $\sim 30$ extra events to distinguish between CDM and the best-fitting SIDM to $5\sigma$ confidence. With $\beta = 2.5$, the preference of $\Gamma \geq 2.5$ over $\Gamma = 2$ grows further to a Bayes factor of $\sim 10^{4}$, and the $\Gamma = 3$ model is definitively preferred over $\Gamma = 2.5$ with Bayes factor of $\sim 10^{2}$. Vice versa, swapping to $\beta = 1.5$ weakens the preference of the best-fitting SIDM model over the smooth model to $\sim 4\sigma$, and the CDM model is no longer significantly distinguished from the SIDM model ($\sim 2\sigma$). One would need a total of $\sim 300$ events to distinguish between CDM and SIDM, and $\sim 100$ events between the smooth and SIDM model, to $5\sigma$ confidence. While $\Gamma \geq 2.5$ is still strongly preferred over $\Gamma = 2$ with a Bayes factor of $\sim 20$, $\Gamma = 3$ and $\Gamma = 2.5$ are no longer distinguished with a Bayes factor of only $\sim 3$. Although the results of our work remain unchanged qualitatively, $\beta$ should be treated as a nuisance parameter for future work.

As a side note, the effect of $\beta$ is weak for the case of waveDM simulations in \citet{Broadhurst_2025}, like our case of SIDM model with $\Gamma = 3$ and $\sigma/m = 1000\,\rm{cm}^{2}/\rm{g}$. Swapping from $\beta = 1.5$ to $\beta = 2.5$ barely moves the reduced $\chi^{2}$ -- the reduced $\chi^{2}$s remain as $\sim 0.8$, $\sim 1$, and $\sim 2.4$ for the three de Broglie wavelengths of $15\,$pc, $20\,$pc and $40\,$pc, respectively. This is very likely arising from the fact that the magnification has been drastically redistributed across the CC in these cases (see, e.g., the orange curve in Fig.~\ref{fig: mag_vs_d}), further departing from the smooth CC case. The slope of the luminosity function controls the level of magnification bias \citep{Broadhurst_2025}, which determines the sensitivity of the lensed star detection rate towards higher magnification. A steeper slope, as we have shown earlier, gives rise to more events in the central region where the magnification is high. Once the high magnification is redistributed across the CC, the sensitivity becomes less relevant to the lensed star spatial distribution, as there is high magnification everywhere away from the original peak.

\subsection{Abundance of microlenses}

While the abundance of stellar microlenses of $\sim 10\,M_{\odot}/\rm{pc}^{2}$ is well motivated by intracluster light (ICL) analysis, portions of the Dragon arc are submerged by the stellar populations in cluster members. Although our subsample of lensed stars is far from such regions and the estimated ICL density should be robust, we briefly explore what would happen to our result if the abundance of microlenses increases by a factor of two, to $\sim 20\,M_{\odot}/\rm{pc}^{2}$, by rerunning our simulations with this revised abundance. This would also correspond to a possible case in which dark compact objects have taken up more fractions of DM than what was previously constrained \citep{Green_2021}. 

The difference between the reduced $\chi^{2}$ of the two most distinctive models, $\Gamma = 2$, $\sigma/m = 100\,\rm{cm}^{2}/\rm{g}$ and $\Gamma = 3$, $\sigma/m = 1000\,\rm{cm}^{2}/\rm{g}$, dropped from 2.89 and 1.15 (difference of 1.74) to 1.64 and 1.23 (difference of 0.4). Such a difference would then fall within the sampling uncertainty as shown earlier in Sec.~\ref{sec: discuss}. While the qualitative preference persists, the statistical significance in distinguishing the models dropped significantly. This is unsurprising -- an abundant amount of stellar microlenses will act as extra substructures and wash out the difference driven dominantly by DM substructures. Alternatively, we can bring down the abundance of stellar microlenses to better understand the situation. Half the original abundance of stellar microlenses, at $5\,M_{\odot}/\rm{pc}^{2}$, would lead to a more significant difference between the two aforementioned models, with $\chi^{2}_{\nu} = 4.93$ and $1.39$ respectively. The lensed star spatial distribution would peak more strongly at the center; thus, those models with little redistribution of macroscopic magnification (i.e., small $\Gamma$, and low core-collapse fraction, as shown in Fig.~\ref{fig: mag_maps}) will be more disfavoured by the data. This is similar to the case where we have a steeper $\beta$ for the background luminosity function, as discussed previously in Sec.~\ref{sec: beta}.

Nevertheless, we stress that this is purely a sensitivity test, as the uncertainty in the abundance of stellar microlenses should be much lower than the other unknowns given that it is purely photometric. Stacking repeated JWST observations would help increase the signal-to-noise in the ICL light, allowing for an even more precise determination of the abundance of stellar microlenses. Also, studying light curves of individual lensed stars could help one to distinguish the underlying low-mass initial mass function of the lensing population \citep[e.g.,][]{Kelly_2018}, further reducing the uncertainty in the abundance of stellar microlenses. Based on our results, lensed galaxies with a lower abundance of surface mass density would be beneficial in testing the nature of DM, given their relatively higher sensitivity to the underlying macroscopic magnification perturbed by DM substructures. Notice that background stars would be able to attain even higher maximum magnification given the lower abundance of stellar microlenses \citep{Li_2026_PCCE}, beneficial in increasing the number of lensed star detections, which is critical in distinguishing DM models.

\subsection{Globular clusters as extra perturbers}

We have neglected the existence of globular clusters (GCs) in our simulation. GCs are known to be well described by the very compact \citet{King_1962} profile, the same as $\Gamma = 3$ SIDM core-collapsed subhaloes. Although GCs typically have masses in a range lower ($10^{4}-10^{6}\,M_{\odot}$) than the subhaloes considered in our simulations, they can also perturb the magnification distribution locally \citep[e.g.,][]{Diego_2024_3M, Palencia_2025_microlensing} and lead to local over-density of lensed stars. The number density of GCs traces the ICL distribution closely, with both contributing, to the first order, the same surface mass density in galaxy clusters \citep{Diego_2023_GC}.

The latest deep observation of cluster AS1063 at $\sim 30.9\,$mag with GLIMPSE \citep{atek2025jwstsglimpseoverviewdeepest} revealed $\sim 30,000$ GCs, yielding a GC surface density scaling as $\sim (\kappa/0.4)^{1.4}$ with $\kappa$ the lensing convergence \citep{Diego_2026_GC}. Our subsample of lensed stars spans an area of $\sim 200\,\rm{kpc}^{2}$, with $0.5 < \kappa < 0.7$ according to the \citet{Diego_2025} lens model. If we assume the relation holds also at A370, this yields $\sim 15$ GCs that can act as potential perturbers. The GC mass function follows a log-normal distribution, centered at $\sim 2\times 10^{5}\,M_{\odot}$ with a width of $\sim 0.5\,\rm{dex}$ \citep[e.g.,][]{McLaughlin_2008, Baumgardt_2018, morenohilario2026reconstructingglobularclusterinitial}, meaning that we would expect only $\sim 1$ GC with $\sim 10^{6}\,M_{\odot}$ within our region of interest. The majority of the GCs would have mass significantly lower than our considered mass range of subhaloes, meaning that they would have limited perturbation towards the macroscopic magnification, and are rather unlikely to lead to extra bias in our result. Nevertheless, one strategy would be to avoid lensed stars discovered near massive GCs revealed by deep observations to minimize the effect of GCs on the inferred lensed star spatial distribution.

\section{Conclusion}

In this paper, we study for the first time the role of SIDM core-collapsed subhaloes in the spatial distribution of lensed stars now regularly found in deep, repeated HST and JWST cluster observations.  Adopting core-collapsed power-law profiles from \citet{Gilman_2023}, we explore how the steepness, core size of core-collapsed subhaloes, as well as the SIDM interaction cross-section affect the magnification distribution across the macroscopic CC and therefore the lensed star spatial distribution.

We found that the power-law slope, $\Gamma$, is determinant of the skewness of the lensed star spatial distribution -- only when $\Gamma$ is large enough ($\gtrsim 2.5$) can milliCCs be formed in the negative parity of the macro CC when the subhaloes are core-collapsed, therefore leading to negative skewness of lensed stars as observed in clusters. A larger interaction cross-section that leads to a larger fraction of subhaloes that have undergone core-collapse is a prerequisite for milliCCs to form in the negative parity with density slopes that are steep enough to explain the negative skewness. At face value, the observed lensed star spatial distribution qualitatively favours $\Gamma = 3$ and $500\lesssim \sigma/m \lesssim 1000\,\text{cm}^2/\text{g}$, 
which fits the observed lensed star spatial distribution in the Dragon galaxy \citep[$z = 0.7251$,][]{Kelly_2022_Flashlights, Fudamoto_2025, Palencia_2026} compared with the \citet{Diego_2025} lens model down to a reduced $\chi^{2}$ of $\sim 1.2$ (p-value of 0.45 from KS test). Moreover, models marginalizing over $\sigma/m$ indicate a definitive preference for $\Gamma \gtrsim 2.5$ over $\Gamma = 2$ with a Bayes factor of $\sim 10^{2}$, and moderately prefer $\Gamma = 3$ over $\Gamma = 2.5$ with a Bayes factor of $\sim 10$.

The Poissonian statistics, combined with the fact that the effect of tidal disruption upon the subhalo core collapse probability is unclear in a cluster environment, as well as the unknown correlation between $\Gamma$ and $\sigma/m$, made it impossible to draw solid constraints on the SIDM interaction cross-section in this work. Nevertheless, our methodology demonstrated that the lensed star spatial distribution can be used to constrain the SIDM interaction cross-section when sufficient data has been accumulated, and when the tidal effects are better understood. We also show that while both the best-fit claimed in \citet{Broadhurst_2025} -- wave DM with a de Broglie wavelength of $\sim 15\,$pc -- and our best-fitting SIDM can reproduce the current observations down to $\chi^{2}_{\nu} \approx 1$, the two models can be distinguished at $3\sigma$ level with a total of $\sim 200$ (extra $\sim 130$) cluster lensed star events, assuming that one model is the ground truth. To distinguish the best-fitting SIDM model and the standard CDM model, we would require a total of $\sim 160$ (extra $\sim 100$) events to reach $5\sigma$ confidence level.

We further explored the parameter space, including the subhalo mass fraction, the core size and truncation radius of the core-collapsed SIDM subhaloes, as well as the luminosity function of the background galaxy and the abundance of stellar microlenses. We found that the choice of these parameters does not affect our main result qualitatively, but affects the statistical significance of our results. We also deduce that the GC population does not carry a significant effect on our inference, but a careful treatment of excluding lensed stars near bright GCs (if any) would be beneficial. The ongoing monitoring campaign on the Dragon, combined with scheduled observations at many different clusters, is expected to return a suite of events and narrow down the uncertainty of some of the parameters (e.g., the slope of the luminosity function). This could eventually allow for distinguishing DM models based on lensed star observations, upon a careful treatment of lens model systematics and marginalization over parameter space. 

\begin{acknowledgements}
S.K.L. acknowledges the generous support from the Croucher Foundation through the Croucher Postdoctoral Fellowship. S.K.L., J.L. and A.A. acknowledge support from the Research Grants Council (RGC) of Hong Kong through the General Research Fund (GRF) 17312122 and 17302326. Y.H. and Y.Z. are supported by the GRF Grants No. 11302824 and 11310925, and PROCORE F-CityU106/25 from RGC, and the Grants No. 9610645 and No. 7020130 from the City University of Hong Kong.
J.M.D. acknowledges the support of projects PID2022-138896NB-C51 (MCIU/AEI/MINECO/FEDER, UE) Ministerio de Ciencia, Investigaci\'on y Universidades and SA101P24. 
A.A. acknowledges the support from the RGC Junior Research Fellow Scheme.
T.B. and P.M. are supported by the Spanish Grant PID2023-149016NB-I00 (MINECO/AEI/FEDER, UE). PM acknowledges financial support from fellowship PIF22/177 (UPV/EHU). T.B. is supported by the Basque Government Grant No. IT1977-26. S.C. acknowledges support from the EACOA Fellowship awarded by the East Asia Core Observatories Association. L.W.H.F. and R.M. acknowledge funding from the UK Science and Technology Facilities Council via grant ST/X001075/1.

The Instituto de Física de Cantabria (IFCA) has been accredited as Unidad de Excelencia María de Maeztu, a distinction awarded by the Spanish State Research Agency for the period 01/06/2026 - 31/05/2030. This publication is part of the grant «Excelencia María de Maeztu» CEX2025-001620-M, funded by MICIU/AEI/10.13039/501100011033.

We acknowledge the use of the following programs: Astropy \citep{astropy:2013, astropy:2018, astropy:2022}, Numpy \citep{numpy}, Matplotlib \citep{matplotlib}, and Scipy \citep{Scipy}.
Data will be shared upon reasonable request.

\end{acknowledgements}

%
\bibliographystyle{aa} 
\bibliography{bib} 

@article{vandenBosch:2026iyn,
    author = "van den Bosch, Frank C. and Tokayer, Yarone M.",
    title = "{Not so Fast: Revised Calibration of SIDM Core Collapse}",
    eprint = "2610.28890",
    archivePrefix = "arXiv",
    primaryClass = "astro-ph.CO",
    month = "10",
    year = "2026"
}

@article{Nadler:2020ulu,
    author = "Nadler, Ethan O. and Banerjee, Arka and Adhikari, Susmita and Mao, Yao-Yuan and Wechsler, Risa H.",
    title = "{Signatures of Velocity-Dependent Dark Matter Self-Interactions in Milky Way-mass Halos}",
    eprint = "2001.08754",
    archivePrefix = "arXiv",
    primaryClass = "astro-ph.CO",
    doi = "10.3847/1538-4357/ab94b0",
    journal = "Astrophys. J.",
    volume = "896",
    number = "2",
    pages = "112",
    year = "2020"
}

@article{Zeng:2023fnj,
    author = "Zeng, Zhichao Carton and Peter, Annika H. G. and Du, Xiaolong and Yang, Shengqi and Benson, Andrew and Cyr-Racine, Francis-Yan and Jiang, Fangzhou and Mace, Charlie and Metcalf, R. Benton",
    title = "{Evolution and properties of self-interacting dark matter subhalos until core collapse}",
    eprint = "2310.09910",
    archivePrefix = "arXiv",
    primaryClass = "astro-ph.GA",
    doi = "10.1103/PhysRevD.111.063001",
    journal = "Phys. Rev. D",
    volume = "111",
    number = "6",
    pages = "063001",
    year = "2025"
}

@article{Nishikawa:2019lsc,
    author = "Nishikawa, Hiroya and Boddy, Kimberly K. and Kaplinghat, Manoj",
    title = "{Accelerated core collapse in tidally stripped self-interacting dark matter halos}",
    eprint = "1901.00499",
    archivePrefix = "arXiv",
    primaryClass = "astro-ph.GA",
    doi = "10.1103/PhysRevD.101.063009",
    journal = "Phys. Rev. D",
    volume = "101",
    number = "6",
    pages = "063009",
    year = "2020"
}

@article{Pollack:2014rja,
    author = "Pollack, Jason and Spergel, David N. and Steinhardt, Paul J.",
    title = "{Supermassive Black Holes from Ultra-Strongly Self-Interacting Dark Matter}",
    eprint = "1501.00017",
    archivePrefix = "arXiv",
    primaryClass = "astro-ph.CO",
    reportNumber = "CALT-TH-2014-144",
    doi = "10.1088/0004-637X/804/2/131",
    journal = "Astrophys. J.",
    volume = "804",
    number = "2",
    pages = "131",
    year = "2015"
}

@article{Koda:2011yb,
    author = "Koda, Jun and Shapiro, Paul R.",
    title = "{Gravothermal collapse of isolated self-interacting dark matter haloes: N-body simulation versus the fluid model}",
    eprint = "1101.3097",
    archivePrefix = "arXiv",
    primaryClass = "astro-ph.CO",
    reportNumber = "TCC-001-11",
    doi = "10.1111/j.1365-2966.2011.18684.x",
    journal = "Mon. Not. Roy. Astron. Soc.",
    volume = "415",
    pages = "1125",
    year = "2011"
}

@article{Huo:2019yhk,
    author = "Huo, Ran and Yu, Hai-Bo and Zhong, Yi-Ming",
    title = "{The Structure of Dissipative Dark Matter Halos}",
    eprint = "1912.06757",
    archivePrefix = "arXiv",
    primaryClass = "astro-ph.CO",
    doi = "10.1088/1475-7516/2020/06/051",
    journal = "JCAP",
    volume = "06",
    pages = "051",
    year = "2020"
}

@article{Sanchez_Almeida_2025,
   title={Constraints on dark matter models from the stellar cores observed in ultra-faint dwarf galaxies: Self-interacting dark matter},
   volume={704},
   ISSN={1432-0746},
   url={http://dx.doi.org/10.1051/0004-6361/202557040},
   DOI={10.1051/0004-6361/202557040},
   journal={Astronomy \& Astrophysics},
   publisher={EDP Sciences},
   author={Sánchez Almeida, Jorge},
   year={2025},
   month=Dec, pages={A210} }

@ARTICLE{Moore_1996,
       author = {{Moore}, Ben and {Katz}, Neal and {Lake}, George and {Dressler}, Alan and {Oemler}, Augustus},
        title = "{Galaxy harassment and the evolution of clusters of galaxies}",
      journal = {\nat},
         year = 1996,
        month = feb,
       volume = {379},
       number = {6566},
        pages = {613-616},
          doi = {10.1038/379613a0},
archivePrefix = {arXiv},
       eprint = {astro-ph/9510034},
 primaryClass = {astro-ph},
       adsurl = {https://ui.adsabs.harvard.edu/abs/1996Natur.379..613M}
}

@article{Eid_2025,
   title={A New Framework for Understanding Systematic Errors in Cluster Lens Modeling. IV. Constraints from the Giant Arc in Abell 370},
   volume={990},
   ISSN={1538-4357},
   url={http://dx.doi.org/10.3847/1538-4357/adf4cb},
   DOI={10.3847/1538-4357/adf4cb},
   number={2},
   journal={The Astrophysical Journal},
   publisher={American Astronomical Society},
   author={Eid, Lana and Keeton, Charles R.},
   year={2025},
   month=Sept, pages={196} }

@article{McLaughlin_2008,
   title={Shaping the Globular Cluster Mass Function by Stellar‐Dynamical Evaporation},
   volume={679},
   ISSN={1538-4357},
   url={http://dx.doi.org/10.1086/533485},
   DOI={10.1086/533485},
   number={2},
   journal={The Astrophysical Journal},
   publisher={American Astronomical Society},
   author={McLaughlin, Dean E. and Fall, S. Michael},
   year={2008},
   month=June, pages={1272–1287} }

@misc{morenohilario2026reconstructingglobularclusterinitial,
      title={Reconstructing the Globular Cluster Initial Mass Function from Present-Day Globular Cluster Systems}, 
      author={Elizabeth Moreno-Hilario and Hui Li and Luis A. Martinez-Medina},
      year={2026},
      eprint={2605.27509},
      archivePrefix={arXiv},
      primaryClass={astro-ph.GA},
      url={https://arxiv.org/abs/2605.27509}, 
}

@misc{atek2025jwstsglimpseoverviewdeepest,
      title={JWST's GLIMPSE: an overview of the deepest probe of early galaxy formation and cosmic reionization}, 
      author={Hakim Atek and John Chisholm and Vasily Kokorev and Ryan Endsley and Richard Pan and Lukas Furtak and Iryna Chemerynska and Johan Richard and Adélaïde Claeyssens and Pascal Oesch and Seiji Fujimoto and Rohan Naidu and Damien Korber and Daniel Schaerer and Jeremy Blaizot and Joki Rosdahl and Angela Adamo and Yoshihisa Asada and Arghyadeep Basu and Benjamin Beauchesne and Danielle Berg and Rachel Bezanson and Rychard Bouwens and Gabriel Brammer and Miroslava Dessauges-Zavadsky and Amaël Ellien and Meriam Ezziati and Qinyue Fei and Ilias Goovaerts and Sylvain Heurtier and Tiger Yu-Yang Hsiao and Michelle Jecmen and Gourav Khullar and Jean-Paul Kneib and Ivo Labbé and Floriane Leclercq and Rui Marques-Chaves and Charlotte Mason and Kristen B. W. McQuinn and Julian B. Muñoz and Priyamvada Natarajan and Alberto Saldana-Lopez and Mabel G. Stephenson and Maxime Trebitsch and Marta Volonteri and Andrea Weibel and Adi Zitrin},
      year={2025},
      eprint={2511.07542},
      archivePrefix={arXiv},
      primaryClass={astro-ph.GA},
      url={https://arxiv.org/abs/2511.07542}, 
}

@ARTICLE{Baumgardt_2018,
       author = {{Baumgardt}, H. and {Hilker}, M.},
        title = "{A catalogue of masses, structural parameters, and velocity dispersion profiles of 112 Milky Way globular clusters}",
      journal = {\mnras},
         year = 2018,
        month = aug,
       volume = {478},
       number = {2},
        pages = {1520-1557},
          doi = {10.1093/mnras/sty1057},
archivePrefix = {arXiv},
       eprint = {1804.08359},
 primaryClass = {astro-ph.GA},
       adsurl = {https://ui.adsabs.harvard.edu/abs/2018MNRAS.478.1520B}
}

@article{Cha_2026,
   title={MrMARTIAN: A Multiresolution Mass Reconstruction Algorithm Combining Free-form and Analytic Components},
   volume={997},
   ISSN={1538-4357},
   url={http://dx.doi.org/10.3847/1538-4357/ae2258},
   DOI={10.3847/1538-4357/ae2258},
   number={1},
   journal={The Astrophysical Journal},
   publisher={American Astronomical Society},
   author={Cha, Sangjun and Jee, M. James},
   year={2026},
   month=Jan, pages={18} }

@article{Shirasaki_2022,
   title={Modelling self-interacting dark matter substructures – I.  Calibration with <i>N</i>-body simulations of a Milky-Way-sized halo and its satellite},
   volume={516},
   ISSN={1365-2966},
   url={http://dx.doi.org/10.1093/mnras/stac2539},
   DOI={10.1093/mnras/stac2539},
   number={3},
   journal={Monthly Notices of the Royal Astronomical Society},
   publisher={Oxford University Press (OUP)},
   author={Shirasaki, Masato and Okamoto, Takashi and Ando, Shin’ichiro},
   year={2022},
   month=Sept, pages={4594–4611} }

@article{Diego_2023_GC,
	author = {{Diego, J. M.} and {Pascale, M.} and {Frye, B.} and {Zitrin, A.} and {Broadhurst, T.} and {Mahler, G.} and {Caminha, G. B.} and {Jauzac, M.} and {Lee, M. G.} and {Bae, J. H.} and {Jang, I. S.} and {Montes, M.}},
	title = {Exploring the correlation between dark matter, intracluster light, and globular cluster distribution in SMACS0723},
	DOI= "10.1051/0004-6361/202345868",
	url= "https://doi.org/10.1051/0004-6361/202345868",
	journal = {A\&A},
	year = 2023,
	volume = 679,
	pages = "A159",
}

@article{Diego_2026_GC,
   title={A tight relation between the distribution of globular clusters and dark matter in AS1063},
   volume={709},
   ISSN={1432-0746},
   url={http://dx.doi.org/10.1051/0004-6361/202659579},
   DOI={10.1051/0004-6361/202659579},
   journal={Astronomy \& Astrophysics},
   publisher={EDP Sciences},
   author={Diego, J. M. and Goolsby, C. and Conselice, C. J. and Palencia, J. M.},
   year={2026},
   month=May, pages={A248} }

@article{Sirks_2022,
   title={The effects of self-interacting dark matter on the stripping of galaxies that fall into clusters},
   volume={511},
   ISSN={1365-2966},
   url={http://dx.doi.org/10.1093/mnras/stac406},
   DOI={10.1093/mnras/stac406},
   number={4},
   journal={Monthly Notices of the Royal Astronomical Society},
   publisher={Oxford University Press (OUP)},
   author={Sirks, Ellen L and Oman, Kyle A and Robertson, Andrew and Massey, Richard and Frenk, Carlos},
   year={2022},
   month=Feb, pages={5927–5935} }

@article{Perera_2026b,
   title={},
   journal={},
   author={Perera, Derek},
   year={in prep.}}

@ARTICLE{Mandelbaum_2008,
       author = {{Mandelbaum}, Rachel and {Seljak}, Uro{\v{s}} and {Hirata}, Christopher M.},
        title = "{A halo mass{\textemdash}concentration relation from weak lensing}",
      journal = {\jcap},
         year = 2008,
        month = aug,
       volume = {2008},
       number = {8},
          eid = {006},
        pages = {006},
          doi = {10.1088/1475-7516/2008/08/006},
archivePrefix = {arXiv},
       eprint = {0805.2552},
 primaryClass = {astro-ph},
       adsurl = {https://ui.adsabs.harvard.edu/abs/2008JCAP...08..006M}
}

@article{Cha_2025,
   title={A High-Caliber View of the Bullet Cluster through JWST Strong and Weak Lensing Analyses},
   volume={987},
   ISSN={2041-8213},
   url={http://dx.doi.org/10.3847/2041-8213/add2f0},
   DOI={10.3847/2041-8213/add2f0},
   number={1},
   journal={The Astrophysical Journal Letters},
   publisher={American Astronomical Society},
   author={Cha, Sangjun and Cho, Boseong Young and Joo, Hyungjin and Lee, Wonki and HyeongHan, Kim and Scofield, Zachary P. and Finner, Kyle and Jee, M. James},
   year={2025},
   month=June, pages={L15} }

@misc{gilman2026hintsselfinteractingdarkmatter,
      title={Hints of self-interacting dark matter from strong gravitational lensing}, 
      author={Daniel Gilman and Kalliopi Petraki and Thomas Pignard and James Gurian and Alex Kusenko and Anna Nierenberg},
      year={2026},
      eprint={2610.40766},
      archivePrefix={arXiv},
      primaryClass={astro-ph.CO},
      url={https://arxiv.org/abs/2610.40766}, 
}

@misc{gilman2026jwstlensedquasardark,
      title={JWST lensed quasar dark matter survey V: Hints of self-interacting dark matter from 29 quadruply imaged quasars}, 
      author={D. Gilman and A. M. Nierenberg and J. Gurian and H. Paugnat and C. Gannon and M. N. Martinez and T. Treu and K. N. Abazajian and T. Anguita and V. N. Bennert and A. J. Benson and S. Birrer and S. G. Djorgovski and S. F. Hoenig and R. E. Keeley and A. Kusenko and M. Millon and T. Morishita and L. A. Moustakas and P. Mozumdar and D. Paris and W. Sheu and D. Sluse and K. C. Wong},
      year={2026},
      eprint={2609.35974},
      archivePrefix={arXiv},
      primaryClass={astro-ph.CO},
      url={https://arxiv.org/abs/2609.35974}, 
}

@ARTICLE{Merten_2015,
       author = {{Merten}, J. and {Meneghetti}, M. and {Postman}, M. and {Umetsu}, K. and {Zitrin}, A. and {Medezinski}, E. and {Nonino}, M. and {Koekemoer}, A. and {Melchior}, P. and {Gruen}, D. and {Moustakas}, L.~A. and {Bartelmann}, M. and {Host}, O. and {Donahue}, M. and {Coe}, D. and {Molino}, A. and {Jouvel}, S. and {Monna}, A. and {Seitz}, S. and {Czakon}, N. and {Lemze}, D. and {Sayers}, J. and {Balestra}, I. and {Rosati}, P. and {Ben{\'\i}tez}, N. and {Biviano}, A. and {Bouwens}, R. and {Bradley}, L. and {Broadhurst}, T. and {Carrasco}, M. and {Ford}, H. and {Grillo}, C. and {Infante}, L. and {Kelson}, D. and {Lahav}, O. and {Massey}, R. and {Moustakas}, J. and {Rasia}, E. and {Rhodes}, J. and {Vega}, J. and {Zheng}, W.},
        title = "{CLASH: The Concentration-Mass Relation of Galaxy Clusters}",
      journal = {\apj},
         year = 2015,
        month = jun,
       volume = {806},
       number = {1},
          eid = {4},
        pages = {4},
          doi = {10.1088/0004-637X/806/1/4},
archivePrefix = {arXiv},
       eprint = {1404.1376},
 primaryClass = {astro-ph.CO},
       adsurl = {https://ui.adsabs.harvard.edu/abs/2015ApJ...806....4M}
}

@article{Nishikawa_2020,
   title={Accelerated core collapse in tidally stripped self-interacting dark matter halos},
   volume={101},
   ISSN={2470-0029},
   url={http://dx.doi.org/10.1103/PhysRevD.101.063009},
   DOI={10.1103/physrevd.101.063009},
   number={6},
   journal={Physical Review D},
   publisher={American Physical Society (APS)},
   author={Nishikawa, Hiroya and Boddy, Kimberly K. and Kaplinghat, Manoj},
   year={2020},
   month=Mar }

@ARTICLE{Li_2026_PCCE,
       author = {{Li}, Sung Kei and {Diego}, Jose M. and {Palencia}, Jose M.},
        title = "{Observing population III stars in the pristine region}",
      journal = {arXiv e-prints},
         year = 2026,
        month = sep,
          eid = {arXiv:2609.32347},
        pages = {arXiv:2609.32347},
          doi = {10.48550/arXiv.2609.32347},
archivePrefix = {arXiv},
       eprint = {2609.32347},
 primaryClass = {astro-ph.CO},
       adsurl = {https://ui.adsabs.harvard.edu/abs/2026arXiv260932347L}
}

@article{Niemiec_2019,
   title={Dark matter stripping in galaxy clusters: a look at the stellar-to-halo mass relation in the Illustris simulation},
   volume={487},
   ISSN={1365-2966},
   url={http://dx.doi.org/10.1093/mnras/stz1318},
   DOI={10.1093/mnras/stz1318},
   number={1},
   journal={Monthly Notices of the Royal Astronomical Society},
   publisher={Oxford University Press (OUP)},
   author={Niemiec, Anna and Jullo, Eric and Giocoli, Carlo and Limousin, Marceau and Jauzac, Mathilde},
   year={2019},
   month=May, pages={653–666} }

@ARTICLE{Diego_2024_3M,
       author = {{Diego}, Jose M. and {Li}, Sung Kei and {Amruth}, Alfred and {Meena}, Ashish K. and {Broadhurst}, Tom J. and {Kelly}, Patrick L. and {Filippenko}, Alexei V. and {Williams}, Liliya L.~R. and {Zitrin}, Adi and {Harris}, William E. and {Reina-Campos}, Marta and {Giocoli}, Carlo and {Dai}, Liang and {Struble}, Mitchell F. and {Treu}, Tommaso and {Fudamoto}, Yoshinobu and {Gilman}, Daniel and {Koekemoer}, Anton M. and {Lim}, Jeremy and {Palencia}, Jose Mar{\'\i}a and {Sun}, Fengwu and {Windhorst}, Rogier A.},
        title = "{Imaging dark matter at the smallest scales with z ≍ 1 lensed stars}",
      journal = {\aap},
         year = 2024,
        month = sep,
       volume = {689},
          eid = {A167},
        pages = {A167},
          doi = {10.1051/0004-6361/202450474},
archivePrefix = {arXiv},
       eprint = {2404.08033},
 primaryClass = {astro-ph.CO},
       adsurl = {https://ui.adsabs.harvard.edu/abs/2024A&A...689A.167D}
}

@article{Tajalli_2025,
   title={SHARP–IX. The dense, low-mass perturbers in B1938+666 and J0946+1006: implications for cold and self-interacting dark matter},
   volume={543},
   ISSN={1365-2966},
   url={http://dx.doi.org/10.1093/mnras/staf1357},
   DOI={10.1093/mnras/staf1357},
   number={1},
   journal={Monthly Notices of the Royal Astronomical Society},
   publisher={Oxford University Press (OUP)},
   author={Tajalli, M and Vegetti, S and O’Riordan, C M and White, S D M and Fassnacht, C D and Powell, D M and McKean, J P and Despali, G},
   year={2025},
   month=Aug, pages={540–564} }

@article{Chiang_2026,
   title={Constraining the Nature of Dark Matter from Tidal Radii of Cluster Galaxy Subhalos},
   volume={997},
   ISSN={1538-4357},
   url={http://dx.doi.org/10.3847/1538-4357/ae23c1},
   DOI={10.3847/1538-4357/ae23c1},
   number={1},
   journal={The Astrophysical Journal},
   publisher={American Astronomical Society},
   author={Chiang, Barry T. and Dutra, Isaque and Natarajan, Priyamvada},
   year={2026},
   month=Jan, pages={106} }

@article{Morishita_2017,
doi = {10.3847/1538-4357/aa8403},
url = {https://doi.org/10.3847/1538-4357/aa8403},
year = {2017},
month = {sep},
publisher = {The American Astronomical Society},
volume = {846},
number = {2},
pages = {139},
author = {Morishita, Takahiro and Abramson, Louis E. and Treu, Tommaso and Schmidt, Kasper B. and Vulcani, Benedetta and Wang, Xin},
title = {Characterizing Intracluster Light in the Hubble Frontier Fields},
journal = {The Astrophysical Journal}
}

@article{Vegetti_2014,
   title={Inference of the cold dark matter substructure mass function at z = 0.2 using strong gravitational lenses},
   volume={442},
   ISSN={0035-8711},
   url={http://dx.doi.org/10.1093/mnras/stu943},
   DOI={10.1093/mnras/stu943},
   number={3},
   journal={Monthly Notices of the Royal Astronomical Society},
   publisher={Oxford University Press (OUP)},
   author={Vegetti, S. and Koopmans, L. V. E. and Auger, M. W. and Treu, T. and Bolton, A. S.},
   year={2014},
   month=June, pages={2017–2035} }

@misc{cha2026testingoffsetsclusterscalehalos,
      title={Testing Offsets Between Cluster-Scale Halos and BCGs in Strong Lensing Models Using the Jackknife Method}, 
      author={Sangjun Cha and Marceau Limousin and M. James Jee},
      year={2026},
      eprint={2609.38310},
      archivePrefix={arXiv},
      primaryClass={astro-ph.CO},
      url={https://arxiv.org/abs/2609.38310}, 
}

@article{Natarajan_2026,
   title={New Cold Dark Matter Crisis Revealed by Multiscale Cluster Lensing},
   volume={1001},
   ISSN={2041-8213},
   url={http://dx.doi.org/10.3847/2041-8213/ae53ea},
   DOI={10.3847/2041-8213/ae53ea},
   number={1},
   journal={The Astrophysical Journal Letters},
   publisher={American Astronomical Society},
   author={Natarajan, Priyamvada and Chiang, Barry T. and Dutra, Isaque},
   year={2026},
   month=Apr, pages={L12} }

@article{Outmezguine_2023,
   title={Universal gravothermal evolution of isolated self-interacting dark matter halos for velocity-dependent cross-sections},
   volume={523},
   ISSN={1365-2966},
   url={http://dx.doi.org/10.1093/mnras/stad1705},
   DOI={10.1093/mnras/stad1705},
   number={3},
   journal={Monthly Notices of the Royal Astronomical Society},
   publisher={Oxford University Press (OUP)},
   author={Outmezguine, Nadav Joseph and Boddy, Kimberly K and Gad-Nasr, Sophia and Kaplinghat, Manoj and Sagunski, Laura},
   year={2023},
   month=June, pages={4786–4800} }

@article{Powell_2025,
   title={A million-solar-mass object detected at a cosmological distance using gravitational imaging},
   volume={9},
   ISSN={2397-3366},
   url={http://dx.doi.org/10.1038/s41550-025-02651-2},
   DOI={10.1038/s41550-025-02651-2},
   number={11},
   journal={Nature Astronomy},
   publisher={Springer Science and Business Media LLC},
   author={Powell, D. M. and McKean, J. P. and Vegetti, S. and Spingola, C. and White, S. D. M. and Fassnacht, C. D.},
   year={2025},
   month=Oct, pages={1714–1722} }

@article{Umetsu_2022,
doi = {10.3847/1538-4357/ac7a9e},
url = {https://doi.org/10.3847/1538-4357/ac7a9e},
year = {2022},
month = {aug},
publisher = {The American Astronomical Society},
volume = {934},
number = {2},
pages = {169},
author = {Umetsu, Keiichi and Ueda, Shutaro and Hsieh, Bau-Ching and Nonino, Mario and Chiu, I-Non and Oguri, Masamune and Molnar, Sandor M. and Koekemoer, Anton M. and Tam, Sut-Ieng},
title = {Line-of-sight Elongation and Hydrostatic Mass Bias of the Frontier Fields Galaxy Cluster Abell 370},
journal = {The Astrophysical Journal}
}

@article{Green_2021,
   title={Primordial black holes as a dark matter candidate},
   volume={48},
   ISSN={1361-6471},
   url={http://dx.doi.org/10.1088/1361-6471/abc534},
   DOI={10.1088/1361-6471/abc534},
   number={4},
   journal={Journal of Physics G: Nuclear and Particle Physics},
   publisher={IOP Publishing},
   author={Green, Anne M and Kavanagh, Bradley J},
   year={2021},
   month=Feb, pages={043001} }

@article{Dutra_2024,
   title={Self-interacting Dark Matter, Core Collapse, and the Galaxy–Galaxy Strong-lensing Discrepancy},
   volume={978},
   ISSN={1538-4357},
   url={http://dx.doi.org/10.3847/1538-4357/ad9b09},
   DOI={10.3847/1538-4357/ad9b09},
   number={1},
   journal={The Astrophysical Journal},
   publisher={American Astronomical Society},
   author={Dutra, Isaque and Natarajan, Priyamvada and Gilman, Daniel},
   year={2024},
   month=Dec, pages={38} }

@article{Perera_2026,
    author = {Perera, Derek and Gilman, Daniel and Williams, Liliya L R and Dai, Liang and Du, Xiaolong and Rihtaršič, Gregor and Becerra-Espinoza, Joaquin and Keen, Allison},
    title = {Dents in the mirror: a novel probe of dark matter substructure in galaxy clusters from the astrometric asymmetry of lensed arcs},
    journal = {Monthly Notices of the Royal Astronomical Society},
    volume = {547},
    number = {4},
    pages = {stag486},
    year = {2026},
    month = {04},
    issn = {0035-8711},
    doi = {10.1093/mnras/stag486},
    url = {https://doi.org/10.1093/mnras/stag486},
    eprint = {https://academic.oup.com/mnras/article-pdf/547/4/stag486/67300797/stag486.pdf},
}

@article{Hsueh_2020,
    author = {Hsueh, J-W and Enzi, W and Vegetti, S and Auger, M W and Fassnacht, C D and Despali, G and Koopmans, L V E and McKean, J P},
    title = {SHARP – VII. New constraints on the dark matter free-streaming properties and substructure abundance from gravitationally lensed quasars},
    journal = {Monthly Notices of the Royal Astronomical Society},
    volume = {492},
    number = {2},
    pages = {3047-3059},
    year = {2020},
    month = {02},
    issn = {0035-8711},
    doi = {10.1093/mnras/stz3177},
    url = {https://doi.org/10.1093/mnras/stz3177},
    eprint = {https://academic.oup.com/mnras/article-pdf/492/2/3047/43927118/stz3177.pdf},
}

@ARTICLE{Gnedin_1999,
       author = {{Gnedin}, Oleg Y. and {Hernquist}, Lars and {Ostriker}, Jeremiah P.},
        title = "{Tidal Shocking by Extended Mass Distributions}",
      journal = {\apj},
         year = 1999,
        month = mar,
       volume = {514},
       number = {1},
        pages = {109-118},
          doi = {10.1086/306910},
       adsurl = {https://ui.adsabs.harvard.edu/abs/1999ApJ...514..109G}
}

@article{Cowan_2011,
   title={Asymptotic formulae for likelihood-based tests of new physics},
   volume={71},
   ISSN={1434-6052},
   url={http://dx.doi.org/10.1140/epjc/s10052-011-1554-0},
   DOI={10.1140/epjc/s10052-011-1554-0},
   number={2},
   journal={The European Physical Journal C},
   publisher={Springer Science and Business Media LLC},
   author={Cowan, Glen and Cranmer, Kyle and Gross, Eilam and Vitells, Ofer},
   year={2011},
   month=Feb }

@article{Boddy:2016bbu,
    author = "Boddy, Kimberly K. and Kaplinghat, Manoj and Kwa, Anna and Peter, Annika H. G.",
    title = "{Hidden Sector Hydrogen as Dark Matter: Small-scale Structure Formation Predictions and the Importance of Hyperfine Interactions}",
    eprint = "1609.03592",
    archivePrefix = "arXiv",
    primaryClass = "hep-ph",
    doi = "10.1103/PhysRevD.94.123017",
    journal = "Phys. Rev. D",
    volume = "94",
    number = "12",
    pages = "123017",
    year = "2016"
}

@article{Buckley:2009in,
    author = "Buckley, Matthew R. and Fox, Patrick J.",
    title = "{Dark Matter Self-Interactions and Light Force Carriers}",
    eprint = "0911.3898",
    archivePrefix = "arXiv",
    primaryClass = "hep-ph",
    reportNumber = "FERMILAB-PUB-09-560-T",
    doi = "10.1103/PhysRevD.81.083522",
    journal = "Phys. Rev. D",
    volume = "81",
    pages = "083522",
    year = "2010"
}

@article{Feng:2009mn,
    author = "Feng, Jonathan L. and Kaplinghat, Manoj and Tu, Huitzu and Yu, Hai-Bo",
    title = "{Hidden Charged Dark Matter}",
    eprint = "0905.3039",
    archivePrefix = "arXiv",
    primaryClass = "hep-ph",
    reportNumber = "UCI-TR-2009-06",
    doi = "10.1088/1475-7516/2009/07/004",
    journal = "JCAP",
    volume = "07",
    pages = "004",
    year = "2009"
}

@article{Tulin:2017ara,
    author = "Tulin, Sean and Yu, Hai-Bo",
    title = "{Dark Matter Self-interactions and Small Scale Structure}",
    eprint = "1705.02358",
    archivePrefix = "arXiv",
    primaryClass = "hep-ph",
    doi = "10.1016/j.physrep.2017.11.004",
    journal = "Phys. Rept.",
    volume = "730",
    pages = "1--57",
    year = "2018"
}

@article{Spergel:1999mh,
    author = "Spergel, David N. and Steinhardt, Paul J.",
    title = "{Observational evidence for selfinteracting cold dark matter}",
    eprint = "astro-ph/9909386",
    archivePrefix = "arXiv",
    doi = "10.1103/PhysRevLett.84.3760",
    journal = "Phys. Rev. Lett.",
    volume = "84",
    pages = "3760--3763",
    year = "2000"
}

@article{Jiang:2022aqw,
    author = "Jiang, Fangzhou and others",
    title = "{A semi-analytic study of self-interacting dark-matter haloes with baryons}",
    eprint = "2206.12425",
    archivePrefix = "arXiv",
    primaryClass = "astro-ph.CO",
    doi = "10.1093/mnras/stad705",
    journal = "Mon. Not. Roy. Astron. Soc.",
    volume = "521",
    number = "3",
    pages = "4630--4644",
    year = "2023"
}

@article{Zhong:2023yzk,
    author = "Zhong, Yi-Ming and Yang, Daneng and Yu, Hai-Bo",
    title = "{The impact of baryonic potentials on the gravothermal evolution of self-interacting dark matter haloes}",
    eprint = "2306.08028",
    archivePrefix = "arXiv",
    primaryClass = "astro-ph.CO",
    doi = "10.1093/mnras/stad2765",
    journal = "Mon. Not. Roy. Astron. Soc.",
    volume = "526",
    number = "1",
    pages = "758--770",
    year = "2023"
}

@article{Turner:2020vlf,
    author = "Turner, Hannah C. and Lovell, Mark R. and Zavala, Jes{\'u}s and Vogelsberger, Mark",
    title = "{The onset of gravothermal core collapse in velocity-dependent self-interacting dark matter subhaloes}",
    eprint = "2010.02924",
    archivePrefix = "arXiv",
    primaryClass = "astro-ph.GA",
    doi = "10.1093/mnras/stab1725",
    journal = "Mon. Not. Roy. Astron. Soc.",
    volume = "505",
    number = "4",
    pages = "5327--5339",
    year = "2021"
}

@article{Diemer_2015,
   title={A UNIVERSAL MODEL FOR HALO CONCENTRATIONS},
   volume={799},
   ISSN={1538-4357},
   url={http://dx.doi.org/10.1088/0004-637X/799/1/108},
   DOI={10.1088/0004-637x/799/1/108},
   number={1},
   journal={The Astrophysical Journal},
   publisher={American Astronomical Society},
   author={Diemer, Benedikt and Kravtsov, Andrey V.},
   year={2015},
   month=Jan, pages={108} }

@article{Fischer:2025rky,
    author = "Fischer, Moritz S. and Yu, Hai-Bo and Dolag, Klaus",
    title = "{Accurately simulating core-collapse self-interacting dark matter halos}",
    eprint = "2506.06269",
    archivePrefix = "arXiv",
    primaryClass = "astro-ph.CO",
    doi = "10.1051/0004-6361/202556189",
    journal = "Astron. Astrophys.",
    volume = "703",
    pages = "A234",
    year = "2025"
}

@ARTICLE{Oke_1983,
       author = {{Oke}, J.~B. and {Gunn}, J.~E.},
        title = "{Secondary standard stars for absolute spectrophotometry.}",
      journal = {\apj},
         year = 1983,
        month = mar,
       volume = {266},
        pages = {713-717},
          doi = {10.1086/160817},
       adsurl = {https://ui.adsabs.harvard.edu/abs/1983ApJ...266..713O}
}

@BOOK{Schneider_1992,
       author = {{Schneider}, Peter and {Ehlers}, J{\"u}rgen and {Falco}, Emilio E.},
        title = "{Gravitational Lenses}",
         year = 1992,
          doi = {10.1007/978-3-662-03758-4},
       adsurl = {https://ui.adsabs.harvard.edu/abs/1992grle.book.....S}
}

@article{Diego_2019,
   title={The Universe at extreme magnification},
   volume={625},
   ISSN={1432-0746},
   url={http://dx.doi.org/10.1051/0004-6361/201833670},
   DOI={10.1051/0004-6361/201833670},
   journal={Astronomy \& Astrophysics},
   publisher={EDP Sciences},
   author={Diego, J. M.},
   year={2019},
   month=May, pages={A84} }

@ARTICLE{King_1962,
       author = {{King}, Ivan},
        title = "{The structure of star clusters. I. an empirical density law}",
      journal = {\aj},
         year = 1962,
        month = oct,
       volume = {67},
        pages = {471},
          doi = {10.1086/108756},
       adsurl = {https://ui.adsabs.harvard.edu/abs/1962AJ.....67..471K}
}

@article{Lovell_2014,
   title={The properties of warm dark matter haloes},
   volume={439},
   ISSN={0035-8711},
   url={http://dx.doi.org/10.1093/mnras/stt2431},
   DOI={10.1093/mnras/stt2431},
   number={1},
   journal={Monthly Notices of the Royal Astronomical Society},
   publisher={Oxford University Press (OUP)},
   author={Lovell, Mark R. and Frenk, Carlos S. and Eke, Vincent R. and Jenkins, Adrian and Gao, Liang and Theuns, Tom},
   year={2014},
   month=Feb, pages={300–317} }

@ARTICLE{Amruth_2023,
       author = {{Amruth}, Alfred and {Broadhurst}, Tom and {Lim}, Jeremy and {Oguri}, Masamune and {Smoot}, George F. and {Diego}, Jose M. and {Leung}, Enoch and {Emami}, Razieh and {Li}, Juno and {Chiueh}, Tzihong and {Schive}, Hsi-Yu and {Yeung}, Michael C.~H. and {Li}, Sung Kei},
        title = "{Einstein rings modulated by wavelike dark matter from anomalies in gravitationally lensed images}",
      journal = {Nature Astronomy},
         year = 2023,
        month = jun,
       volume = {7},
        pages = {736-747},
          doi = {10.1038/s41550-023-01943-9},
archivePrefix = {arXiv},
       eprint = {2304.09895},
 primaryClass = {astro-ph.CO},
       adsurl = {https://ui.adsabs.harvard.edu/abs/2023NatAs...7..736A}
}

@misc{enzi2024overconcentrateddarkhalostrong,
      title={The overconcentrated dark halo in the strong lens SDSS J0946+1006 is a subhalo: evidence for self interacting dark matter?}, 
      author={Wolfgang J. R. Enzi and Coleman M. Krawczyk and Daniel J. Ballard and Thomas E. Collett},
      year={2024},
      eprint={2411.08565},
      archivePrefix={arXiv},
      primaryClass={astro-ph.CO},
      url={https://arxiv.org/abs/2411.08565}, 
}

@article{Gilman_2023,
   title={Constraining resonant dark matter self-interactions with strong gravitational lenses},
   volume={107},
   ISSN={2470-0029},
   url={http://dx.doi.org/10.1103/PhysRevD.107.103008},
   DOI={10.1103/physrevd.107.103008},
   number={10},
   journal={Physical Review D},
   publisher={American Physical Society (APS)},
   author={Gilman, Daniel and Zhong, Yi-Ming and Bovy, Jo},
   year={2023},
   month=May }

@article{Robertson_2016,
   title={What does the Bullet Cluster tell us about self-interacting dark matter?},
   volume={465},
   ISSN={1365-2966},
   url={http://dx.doi.org/10.1093/mnras/stw2670},
   DOI={10.1093/mnras/stw2670},
   number={1},
   journal={Monthly Notices of the Royal Astronomical Society},
   publisher={Oxford University Press (OUP)},
   author={Robertson, Andrew and Massey, Richard and Eke, Vincent},
   year={2016},
   month=Oct, pages={569–587} }

@ARTICLE{Dai_2020,
       author = {{Dai}, Liang and {Kaurov}, Alexander A. and {Sharon}, Keren and {Florian}, Michael and {Miralda-Escud{\'e}}, Jordi and {Venumadhav}, Tejaswi and {Frye}, Brenda and {Rigby}, Jane R. and {Bayliss}, Matthew},
        title = "{Asymmetric surface brightness structure of caustic crossing arc in SDSS J1226+2152: a case for dark matter substructure}",
      journal = {\mnras},
         year = 2020,
        month = jan,
       volume = {495},
       number = {3},
        pages = {3192-3208},
          doi = {10.1093/mnras/staa1355},
archivePrefix = {arXiv},
       eprint = {2001.00261},
 primaryClass = {astro-ph.GA},
       adsurl = {https://ui.adsabs.harvard.edu/abs/2020MNRAS.495.3192D}
}

@article{Dai_2018,
doi = {10.3847/1538-4357/aae478},
url = {https://doi.org/10.3847/1538-4357/aae478},
year = {2018},
month = {oct},
publisher = {The American Astronomical Society},
volume = {867},
number = {1},
pages = {24},
author = {Dai, Liang and Venumadhav, Tejaswi and Kaurov, Alexander A. and Miralda-Escud, Jordi},
title = {Probing Dark Matter Subhalos in Galaxy Clusters Using Highly Magnified Stars},
journal = {The Astrophysical Journal}
}

@ARTICLE{Fudamoto_2025,
       author = {{Fudamoto}, Yoshinobu and {Sun}, Fengwu and {Diego}, Jose M. and {Dai}, Liang and {Oguri}, Masamune and {Zitrin}, Adi and {Zackrisson}, Erik and {Jauzac}, Mathilde and {Lagattuta}, David J. and {Egami}, Eiichi and {Iani}, Edoardo and {Windhorst}, Rogier A. and {Abe}, Katsuya T. and {Bauer}, Franz Erik and {Bian}, Fuyan and {Bhatawdekar}, Rachana and {Broadhurst}, Thomas J. and {Cai}, Zheng and {Chen}, Chian-Chou and {Chen}, Wenlei and {Cohen}, Seth H. and {Conselice}, Christopher J. and {Espada}, Daniel and {Foo}, Nicholas and {Frye}, Brenda L. and {Fujimoto}, Seiji and {Furtak}, Lukas J. and {Golubchik}, Miriam and {Hsiao}, Tiger Yu-Yang and {Jolly}, Jean-Baptiste and {Kawai}, Hiroki and {Kelly}, Patrick L. and {Koekemoer}, Anton M. and {Kohno}, Kotaro and {Kokorev}, Vasily and {Li}, Mingyu and {Li}, Zihao and {Lin}, Xiaojing and {Magdis}, Georgios E. and {Meena}, Ashish K. and {Niemiec}, Anna and {Nabizadeh}, Armin and {Richard}, Johan and {Steinhardt}, Charles L. and {Wu}, Yunjing and {Zhu}, Yongda and {Zou}, Siwei},
        title = "{Identification of more than 40 gravitationally magnified stars in a galaxy at redshift 0.725}",
      journal = {Nature Astronomy},
         year = 2025,
        month = mar,
       volume = {9},
        pages = {428-437},
          doi = {10.1038/s41550-024-02432-3},
archivePrefix = {arXiv},
       eprint = {2404.08045},
 primaryClass = {astro-ph.GA},
       adsurl = {https://ui.adsabs.harvard.edu/abs/2025NatAs...9..428F}
}

@article{Xie_2015,
   title={Assembly history of subhalo populations in galactic and cluster sized dark haloes},
   volume={454},
   ISSN={1365-2966},
   url={http://dx.doi.org/10.1093/mnras/stv2077},
   DOI={10.1093/mnras/stv2077},
   number={2},
   journal={Monthly Notices of the Royal Astronomical Society},
   publisher={Oxford University Press (OUP)},
   author={Xie, Lizhi and Gao, Liang},
   year={2015},
   month=Oct, pages={1697–1703} }

@ARTICLE{Diego_2025,
       author = {{Diego}, Jose M. and {Sun}, Fengwu and {Palencia}, Jose M. and {Lin}, Xiaojing and {Limousin}, Marceau and {Gledhill}, Rachel and {Niemiec}, Anna and {Chen}, Wenlei and {Windhorst}, Rogier A. and {Struble}, Mitchell F. and {Broadhurst}, Tom},
        title = "{JWST lens model for A370: A very low dark matter fraction for a brightest cluster galaxy and lensing properties for the Dragon arc}",
      journal = {\aap},
         year = 2025,
        month = nov,
       volume = {703},
          eid = {A207},
        pages = {A207},
          doi = {10.1051/0004-6361/202556062},
archivePrefix = {arXiv},
       eprint = {2506.11207},
 primaryClass = {astro-ph.CO},
       adsurl = {https://ui.adsabs.harvard.edu/abs/2025A&A...703A.207D}
}

@ARTICLE{Randall_2008,
       author = {{Randall}, Scott W. and {Markevitch}, Maxim and {Clowe}, Douglas and {Gonzalez}, Anthony H. and {Brada{\v{c}}}, Marusa},
        title = "{Constraints on the Self-Interaction Cross Section of Dark Matter from Numerical Simulations of the Merging Galaxy Cluster 1E 0657-56}",
      journal = {\apj},
         year = 2008,
        month = jun,
       volume = {679},
       number = {2},
        pages = {1173-1180},
          doi = {10.1086/587859},
archivePrefix = {arXiv},
       eprint = {0704.0261},
 primaryClass = {astro-ph},
       adsurl = {https://ui.adsabs.harvard.edu/abs/2008ApJ...679.1173R}
}

@article{Harvey_2014,
   title={On the cross-section of dark matter using substructure infall into galaxy clusters},
   volume={441},
   ISSN={0035-8711},
   url={http://dx.doi.org/10.1093/mnras/stu337},
   DOI={10.1093/mnras/stu337},
   number={1},
   journal={Monthly Notices of the Royal Astronomical Society},
   publisher={Oxford University Press (OUP)},
   author={Harvey, David and Tittley, Eric and Massey, Richard and Kitching, Thomas D. and Taylor, Andy and Pike, Simon R. and Kay, Scott T. and Lau, Erwin T. and Nagai, Daisuke},
   year={2014},
   month=Apr, pages={404–416} }

@ARTICLE{Meena_2025_IMF,
       author = {{Meena}, Ashish Kumar and {Li}, Sung Kei and {Zitrin}, Adi and {Kelly}, Patrick L. and {Broadhurst}, Tom and {Chen}, Wenlei and {Diego}, Jose M. and {Filippenko}, Alexei V. and {Furtak}, Lukas J. and {Williams}, Liliya L.~R.},
        title = "{Flashlights: Prospects for constraining the initial mass function around cosmic noon with caustic-crossing events}",
      journal = {\aap},
         year = 2025,
        month = jul,
       volume = {699},
          eid = {A299},
        pages = {A299},
          doi = {10.1051/0004-6361/202555023},
archivePrefix = {arXiv},
       eprint = {2503.21706},
 primaryClass = {astro-ph.GA},
       adsurl = {https://ui.adsabs.harvard.edu/abs/2025A&A...699A.299M}
}

@ARTICLE{Kelly_2022_Flashlights,
       author = {{Kelly}, Patrick L. and {Chen}, Wenlei and {Alfred}, Amruth and {Broadhurst}, Thomas J. and {Diego}, Jose M. and {Emami}, Najmeh and {Filippenko}, Alexei V. and {Keen}, Allison and {Li}, Sung Kei and {Lim}, Jeremy and {Meena}, Ashish K. and {Oguri}, Masamune and {Scarlata}, Claudia and {Treu}, Tommaso and {Williams}, Hayley and {Williams}, Liliya L.~R. and {Zhou}, Rui and {Zitrin}, Adi and {Foley}, Ryan J. and {Jha}, Saurabh W. and {Kaiser}, Nick and {Mehta}, Vihang and {Rieck}, Steven and {Salo}, Laura and {Smith}, Nathan and {Weisz}, Daniel R.},
        title = "{Flashlights: More than A Dozen High-Significance Microlensing Events of Extremely Magnified Stars in Galaxies at Redshifts z=0.7-1.5}",
      journal = {arXiv e-prints},
         year = 2022,
        month = nov,
          eid = {arXiv:2211.02670},
        pages = {arXiv:2211.02670},
          doi = {10.48550/arXiv.2211.02670},
archivePrefix = {arXiv},
       eprint = {2211.02670},
 primaryClass = {astro-ph.CO},
       adsurl = {https://ui.adsabs.harvard.edu/abs/2022arXiv221102670K}
}

@article{Jiang_2021,
   title={SatGen: a semi-analytical satellite galaxy generator – I. The model and its application to Local-Group satellite statistics},
   volume={502},
   ISSN={1365-2966},
   url={http://dx.doi.org/10.1093/mnras/staa4034},
   DOI={10.1093/mnras/staa4034},
   number={1},
   journal={Monthly Notices of the Royal Astronomical Society},
   publisher={Oxford University Press (OUP)},
   author={Jiang, Fangzhou and Dekel, Avishai and Freundlich, Jonathan and van den Bosch, Frank C and Green, Sheridan B and Hopkins, Philip F and Benson, Andrew and Du, Xiaolong},
   year={2021},
   month=Jan, pages={621–641} }

@ARTICLE{Kelly_2018,
       author = {{Kelly}, Patrick L. and {Diego}, Jose M. and {Rodney}, Steven and {Kaiser}, Nick and {Broadhurst}, Tom and {Zitrin}, Adi and {Treu}, Tommaso and {P{\'e}rez-Gonz{\'a}lez}, Pablo G. and {Morishita}, Takahiro and {Jauzac}, Mathilde and {Selsing}, Jonatan and {Oguri}, Masamune and {Pueyo}, Laurent and {Ross}, Timothy W. and {Filippenko}, Alexei V. and {Smith}, Nathan and {Hjorth}, Jens and {Cenko}, S. Bradley and {Wang}, Xin and {Howell}, D. Andrew and {Richard}, Johan and {Frye}, Brenda L. and {Jha}, Saurabh W. and {Foley}, Ryan J. and {Norman}, Colin and {Bradac}, Marusa and {Zheng}, Weikang and {Brammer}, Gabriel and {Benito}, Alberto Molino and {Cava}, Antonio and {Christensen}, Lise and {de Mink}, Selma E. and {Graur}, Or and {Grillo}, Claudio and {Kawamata}, Ryota and {Kneib}, Jean-Paul and {Matheson}, Thomas and {McCully}, Curtis and {Nonino}, Mario and {P{\'e}rez-Fournon}, Ismael and {Riess}, Adam G. and {Rosati}, Piero and {Schmidt}, Kasper Borello and {Sharon}, Keren and {Weiner}, Benjamin J.},
        title = "{Extreme magnification of an individual star at redshift 1.5 by a galaxy-cluster lens}",
      journal = {Nature Astronomy},
         year = 2018,
        month = apr,
       volume = {2},
        pages = {334-342},
          doi = {10.1038/s41550-018-0430-3},
archivePrefix = {arXiv},
       eprint = {1706.10279},
 primaryClass = {astro-ph.GA},
       adsurl = {https://ui.adsabs.harvard.edu/abs/2018NatAs...2..334K}
}

@ARTICLE{Yan_2023,
       author = {{Yan}, Haojing and {Ma}, Zhiyuan and {Sun}, Bangzheng and {Wang}, Lifan and {Kelly}, Patrick and {Diego}, Jos{\'e} M. and {Cohen}, Seth H. and {Windhorst}, Rogier A. and {Jansen}, Rolf A. and {Grogin}, Norman A. and {Beacom}, John F. and {Conselice}, Christopher J. and {Driver}, Simon P. and {Frye}, Brenda and {Coe}, Dan and {Marshall}, Madeline A. and {Koekemoer}, Anton and {Willmer}, Christopher N.~A. and {Robotham}, Aaron and {D'Silva}, Jordan C.~J. and {Summers}, Jake and {Nonino}, Mario and {Pirzkal}, Nor and {Ryan}, Russell E. and {Ortiz}, Rafael and {Tompkins}, Scott and {Bhatawdekar}, Rachana A. and {Cheng}, Cheng and {Zitrin}, Adi and {Willner}, S.~P.},
        title = "{JWST's PEARLS: Transients in the MACS J0416.1-2403 Field}",
      journal = {\apjs},
         year = 2023,
        month = dec,
       volume = {269},
       number = {2},
          eid = {43},
        pages = {43},
          doi = {10.3847/1538-4365/ad0298},
archivePrefix = {arXiv},
       eprint = {2307.07579},
 primaryClass = {astro-ph.GA},
       adsurl = {https://ui.adsabs.harvard.edu/abs/2023ApJS..269...43Y}
}

@article{Choi_2016,
   title={MESA ISOCHRONES AND STELLAR TRACKS (MIST). I. SOLAR-SCALED MODELS},
   volume={823},
   ISSN={1538-4357},
   url={http://dx.doi.org/10.3847/0004-637X/823/2/102},
   DOI={10.3847/0004-637x/823/2/102},
   number={2},
   journal={The Astrophysical Journal},
   publisher={American Astronomical Society},
   author={Choi, Jieun and Dotter, Aaron and Conroy, Charlie and Cantiello, Matteo and Paxton, Bill and Johnson, Benjamin D.},
   year={2016},
   month=May, pages={102} }

@ARTICLE{Diego_2024_buffaloflashlights,
       author = {{Diego}, Jose M. and {Li}, Sung Kei and {Meena}, Ashish K. and {Niemiec}, Anna and {Acebron}, Ana and {Jauzac}, Mathilde and {Struble}, Mitchell F. and {Amruth}, Alfred and {Broadhurst}, Tom J. and {Cerny}, Catherine and {Ebeling}, Harald and {Filippenko}, Alexei V. and {Jullo}, Eric and {Kelly}, Patrick and {Koekemoer}, Anton M. and {Lagattuta}, David and {Lim}, Jeremy and {Limousin}, Marceau and {Mahler}, Guillaume and {Patel}, Nency and {Remolina}, Juan and {Richard}, Johan and {Sharon}, Keren and {Steinhardt}, Charles and {Umetsu}, Keiichi and {Williams}, Liliya and {Zitrin}, Adi and {Palencia}, Jose Mar{\'\i}a and {Dai}, Liang and {Ji}, Lingyuan and {Pascale}, Massimo},
        title = "{BUFFALO/Flashlights: Constraints on the abundance of lensed supergiant stars in the Spock galaxy at redshift 1}",
      journal = {\aap},
         year = 2024,
        month = jan,
       volume = {681},
          eid = {A124},
        pages = {A124},
          doi = {10.1051/0004-6361/202346761},
archivePrefix = {arXiv},
       eprint = {2304.09222},
 primaryClass = {astro-ph.GA},
       adsurl = {https://ui.adsabs.harvard.edu/abs/2024A&A...681A.124D}
}

@ARTICLE{Li_2024,
       author = {{Li}, Sung Kei and {Diego}, Jose M. and {Kelly}, Patrick L. and {Lim}, Jeremy and {Chen}, WenLei and {Alfred}, Amruth and {Williams}, Liliya L.~R. and {Broadhurst}, Thomas J. and {Meena}, Ashish. K. and {Zitrin}, Adi and {Chow}, Alex},
        title = "{Flashlights: Transients among Gravitationally-Lensed Star Clusters in the Dragon Arc. I. Stellar Microlensing vs Stellar Outbursts}",
      journal = {arXiv e-prints},
         year = 2024,
        month = apr,
          eid = {arXiv:2404.08571},
        pages = {arXiv:2404.08571},
          doi = {10.48550/arXiv.2404.08571},
archivePrefix = {arXiv},
       eprint = {2404.08571},
 primaryClass = {astro-ph.CO},
       adsurl = {https://ui.adsabs.harvard.edu/abs/2024arXiv240408571L}
}

@article{Kim_2017,
    author = {Kim, Stacy Y. and Peter, Annika H. G. and Wittman, David},
    title = {In the wake of dark giants: new signatures of dark matter self-interactions in equal-mass mergers of galaxy clusters},
    journal = {Monthly Notices of the Royal Astronomical Society},
    volume = {469},
    number = {2},
    pages = {1414-1444},
    year = {2017},
    month = {08},
    issn = {0035-8711},
    doi = {10.1093/mnras/stx896},
    url = {https://doi.org/10.1093/mnras/stx896},
    eprint = {https://academic.oup.com/mnras/article-pdf/469/2/1414/17272506/stx896.pdf},
}

@ARTICLE{Gomez_2012,
       author = {{G{\'o}mez}, P.~L. and {Valkonen}, L.~E. and {Romer}, A.~K. and {Lloyd-Davies}, E. and {Verdugo}, T. and {Cantalupo}, C.~M. and {Daub}, M.~D. and {Goldstein}, J.~H. and {Kuo}, C.~L. and {Lange}, A.~E. and {Lueker}, M. and {Holzapfel}, W.~L. and {Peterson}, J.~B. and {Ruhl}, J. and {Runyan}, M.~C. and {Reichardt}, C.~L. and {Sabirli}, K.},
        title = "{Optical and X-Ray Observations of the Merging Cluster AS1063}",
      journal = {\aj},
         year = 2012,
        month = sep,
       volume = {144},
       number = {3},
          eid = {79},
        pages = {79},
          doi = {10.1088/0004-6256/144/3/79},
       adsurl = {https://ui.adsabs.harvard.edu/abs/2012AJ....144...79G}
}

@misc{diego2026largecoresdarkmatter,
      title={The large cores of dark matter and globular clusters in AS1063. Possible evidence of self-interacting dark matter. Or not}, 
      author={Jose M. Diego},
      year={2026},
      eprint={2602.15940},
      archivePrefix={arXiv},
      primaryClass={astro-ph.GA},
      url={https://arxiv.org/abs/2602.15940}, 
}

@misc{mace2026sensitivitysubstructurelensingsidm,
      title={The Sensitivity of Substructure Lensing to SIDM Core-collapse Model Variation}, 
      author={Charlie Mace and Birendra Dhanasingham and Zhichao Carton Zeng and Francis-Yan Cyr-Racine and Xiaolong Du and Annika H. G. Peter and Andrew Benson},
      year={2026},
      eprint={2605.24174},
      archivePrefix={arXiv},
      primaryClass={astro-ph.CO},
      url={https://arxiv.org/abs/2605.24174}, 
}

@ARTICLE{Williams_2024,
       author = {{Williams}, Liliya L.~R. and {Kelly}, Patrick L. and {Treu}, Tommaso and {Amruth}, Alfred and {Diego}, Jose M. and {Li}, Sung Kei and {Meena}, Ashish K. and {Zitrin}, Adi and {Broadhurst}, Thomas J. and {Filippenko}, Alexei V.},
        title = "{Flashlights: Properties of Highly Magnified Images Near Cluster Critical Curves in the Presence of Dark Matter Subhalos}",
      journal = {\apj},
         year = 2024,
        month = feb,
       volume = {961},
       number = {2},
          eid = {200},
        pages = {200},
          doi = {10.3847/1538-4357/ad1660},
archivePrefix = {arXiv},
       eprint = {2304.06064},
 primaryClass = {astro-ph.CO},
       adsurl = {https://ui.adsabs.harvard.edu/abs/2024ApJ...961..200W}
}

@ARTICLE{Zhou_2025,
       author = {{Zhou}, Ruwen and {Dai}, Liang and {Ji}, Lingyuan and {Pascale}, Massimo and {Diego}, Jose M. and {Sun}, Fengwu and {Fudamoto}, Yoshinobu},
        title = "{Kinematic Mapping of Giant Arcs: A New Method to Locate Lensing Critical Curves}",
      journal = {arXiv e-prints},
         year = 2025,
        month = nov,
          eid = {arXiv:2511.18479},
        pages = {arXiv:2511.18479},
          doi = {10.48550/arXiv.2511.18479},
archivePrefix = {arXiv},
       eprint = {2511.18479},
 primaryClass = {astro-ph.GA},
       adsurl = {https://ui.adsabs.harvard.edu/abs/2025arXiv251118479Z}
}

@article{Vegetti_2026,
   title={A possible challenge for cold and warm dark matter},
   volume={10},
   ISSN={2397-3366},
   url={http://dx.doi.org/10.1038/s41550-025-02746-w},
   DOI={10.1038/s41550-025-02746-w},
   number={3},
   journal={Nature Astronomy},
   publisher={Springer Science and Business Media LLC},
   author={Vegetti, Simona and White, Simon D. M. and McKean, John P. and Powell, Devon M. and Spingola, Cristiana and Massari, Davide and Despali, Giulia and Fassnacht, Christopher D.},
   year={2026},
   month=Jan, pages={440–447} }

@ARTICLE{Planck18,
       author = {{Planck Collaboration} and {Aghanim}, N. and {Akrami}, Y. and {Ashdown}, M. and {Aumont}, J. and {Baccigalupi}, C. and {Ballardini}, M. and {Banday}, A.~J. and {Barreiro}, R.~B. and {Bartolo}, N. and {Basak}, S. and {Battye}, R. and {Benabed}, K. and {Bernard}, J.-P. and {Bersanelli}, M. and {Bielewicz}, P. and {Bock}, J.~J. and {Bond}, J.~R. and {Borrill}, J. and {Bouchet}, F.~R. and {Boulanger}, F. and {Bucher}, M. and {Burigana}, C. and {Butler}, R.~C. and {Calabrese}, E. and {Cardoso}, J.-F. and {Carron}, J. and {Challinor}, A. and {Chiang}, H.~C. and {Chluba}, J. and {Colombo}, L.~P.~L. and {Combet}, C. and {Contreras}, D. and {Crill}, B.~P. and {Cuttaia}, F. and {de Bernardis}, P. and {de Zotti}, G. and {Delabrouille}, J. and {Delouis}, J.-M. and {Di Valentino}, E. and {Diego}, J.~M. and {Dor{\'e}}, O. and {Douspis}, M. and {Ducout}, A. and {Dupac}, X. and {Dusini}, S. and {Efstathiou}, G. and {Elsner}, F. and {En{\ss}lin}, T.~A. and {Eriksen}, H.~K. and {Fantaye}, Y. and {Farhang}, M. and {Fergusson}, J. and {Fernandez-Cobos}, R. and {Finelli}, F. and {Forastieri}, F. and {Frailis}, M. and {Fraisse}, A.~A. and {Franceschi}, E. and {Frolov}, A. and {Galeotta}, S. and {Galli}, S. and {Ganga}, K. and {G{\'e}nova-Santos}, R.~T. and {Gerbino}, M. and {Ghosh}, T. and {Gonz{\'a}lez-Nuevo}, J. and {G{\'o}rski}, K.~M. and {Gratton}, S. and {Gruppuso}, A. and {Gudmundsson}, J.~E. and {Hamann}, J. and {Handley}, W. and {Hansen}, F.~K. and {Herranz}, D. and {Hildebrandt}, S.~R. and {Hivon}, E. and {Huang}, Z. and {Jaffe}, A.~H. and {Jones}, W.~C. and {Karakci}, A. and {Keih{\"a}nen}, E. and {Keskitalo}, R. and {Kiiveri}, K. and {Kim}, J. and {Kisner}, T.~S. and {Knox}, L. and {Krachmalnicoff}, N. and {Kunz}, M. and {Kurki-Suonio}, H. and {Lagache}, G. and {Lamarre}, J.-M. and {Lasenby}, A. and {Lattanzi}, M. and {Lawrence}, C.~R. and {Le Jeune}, M. and {Lemos}, P. and {Lesgourgues}, J. and {Levrier}, F. and {Lewis}, A. and {Liguori}, M. and {Lilje}, P.~B. and {Lilley}, M. and {Lindholm}, V. and {L{\'o}pez-Caniego}, M. and {Lubin}, P.~M. and {Ma}, Y.-Z. and {Mac{\'\i}as-P{\'e}rez}, J.~F. and {Maggio}, G. and {Maino}, D. and {Mandolesi}, N. and {Mangilli}, A. and {Marcos-Caballero}, A. and {Maris}, M. and {Martin}, P.~G. and {Martinelli}, M. and {Mart{\'\i}nez-Gonz{\'a}lez}, E. and {Matarrese}, S. and {Mauri}, N. and {McEwen}, J.~D. and {Meinhold}, P.~R. and {Melchiorri}, A. and {Mennella}, A. and {Migliaccio}, M. and {Millea}, M. and {Mitra}, S. and {Miville-Desch{\^e}nes}, M.-A. and {Molinari}, D. and {Montier}, L. and {Morgante}, G. and {Moss}, A. and {Natoli}, P. and {N{\o}rgaard-Nielsen}, H.~U. and {Pagano}, L. and {Paoletti}, D. and {Partridge}, B. and {Patanchon}, G. and {Peiris}, H.~V. and {Perrotta}, F. and {Pettorino}, V. and {Piacentini}, F. and {Polastri}, L. and {Polenta}, G. and {Puget}, J.-L. and {Rachen}, J.~P. and {Reinecke}, M. and {Remazeilles}, M. and {Renzi}, A. and {Rocha}, G. and {Rosset}, C. and {Roudier}, G. and {Rubi{\~n}o-Mart{\'\i}n}, J.~A. and {Ruiz-Granados}, B. and {Salvati}, L. and {Sandri}, M. and {Savelainen}, M. and {Scott}, D. and {Shellard}, E.~P.~S. and {Sirignano}, C. and {Sirri}, G. and {Spencer}, L.~D. and {Sunyaev}, R. and {Suur-Uski}, A.-S. and {Tauber}, J.~A. and {Tavagnacco}, D. and {Tenti}, M. and {Toffolatti}, L. and {Tomasi}, M. and {Trombetti}, T. and {Valenziano}, L. and {Valiviita}, J. and {Van Tent}, B. and {Vibert}, L. and {Vielva}, P. and {Villa}, F. and {Vittorio}, N. and {Wandelt}, B.~D. and {Wehus}, I.~K. and {White}, M. and {White}, S.~D.~M. and {Zacchei}, A. and {Zonca}, A.},
        title = "{Planck 2018 results. VI. Cosmological parameters}",
      journal = {\aap},
         year = 2020,
        month = sep,
       volume = {641},
          eid = {A6},
        pages = {A6},
          doi = {10.1051/0004-6361/201833910},
archivePrefix = {arXiv},
       eprint = {1807.06209},
 primaryClass = {astro-ph.CO},
       adsurl = {https://ui.adsabs.harvard.edu/abs/2020A&A...641A...6P}
}

@article{astropy:2013,
Adsurl = {http://adsabs.harvard.edu/abs/2013A%26A...558A..33A},
Archiveprefix = {arXiv},
Author = {{Astropy Collaboration} and {Robitaille}, T.~P. and {Tollerud}, E.~J. and {Greenfield}, P. and {Droettboom}, M. and {Bray}, E. and {Aldcroft}, T. and {Davis}, M. and {Ginsburg}, A. and {Price-Whelan}, A.~M. and {Kerzendorf}, W.~E. and {Conley}, A. and {Crighton}, N. and {Barbary}, K. and {Muna}, D. and {Ferguson}, H. and {Grollier}, F. and {Parikh}, M.~M. and {Nair}, P.~H. and {Unther}, H.~M. and {Deil}, C. and {Woillez}, J. and {Conseil}, S. and {Kramer}, R. and {Turner}, J.~E.~H. and {Singer}, L. and {Fox}, R. and {Weaver}, B.~A. and {Zabalza}, V. and {Edwards}, Z.~I. and {Azalee Bostroem}, K. and {Burke}, D.~J. and {Casey}, A.~R. and {Crawford}, S.~M. and {Dencheva}, N. and {Ely}, J. and {Jenness}, T. and {Labrie}, K. and {Lim}, P.~L. and {Pierfederici}, F. and {Pontzen}, A. and {Ptak}, A. and {Refsdal}, B. and {Servillat}, M. and {Streicher}, O.},
Doi = {10.1051/0004-6361/201322068},
Eid = {A33},
Eprint = {1307.6212},
Journal = {\aap},
Month = oct,
Pages = {A33},
Primaryclass = {astro-ph.IM},
Title = {{Astropy: A community Python package for astronomy}},
Volume = 558,
Year = 2013}

@ARTICLE{astropy:2018,
       author = {{Astropy Collaboration} and {Price-Whelan}, A.~M. and
         {Sip{\H{o}}cz}, B.~M. and {G{\"u}nther}, H.~M. and {Lim}, P.~L. and
         {Crawford}, S.~M. and {Conseil}, S. and {Shupe}, D.~L. and
         {Craig}, M.~W. and {Dencheva}, N. and {Ginsburg}, A. and {Vand
        erPlas}, J.~T. and {Bradley}, L.~D. and {P{\'e}rez-Su{\'a}rez}, D. and
         {de Val-Borro}, M. and {Aldcroft}, T.~L. and {Cruz}, K.~L. and
         {Robitaille}, T.~P. and {Tollerud}, E.~J. and {Ardelean}, C. and
         {Babej}, T. and {Bach}, Y.~P. and {Bachetti}, M. and {Bakanov}, A.~V. and
         {Bamford}, S.~P. and {Barentsen}, G. and {Barmby}, P. and
         {Baumbach}, A. and {Berry}, K.~L. and {Biscani}, F. and {Boquien}, M. and
         {Bostroem}, K.~A. and {Bouma}, L.~G. and {Brammer}, G.~B. and
         {Bray}, E.~M. and {Breytenbach}, H. and {Buddelmeijer}, H. and
         {Burke}, D.~J. and {Calderone}, G. and {Cano Rodr{\'\i}guez}, J.~L. and
         {Cara}, M. and {Cardoso}, J.~V.~M. and {Cheedella}, S. and {Copin}, Y. and
         {Corrales}, L. and {Crichton}, D. and {D'Avella}, D. and {Deil}, C. and
         {Depagne}, {\'E}. and {Dietrich}, J.~P. and {Donath}, A. and
         {Droettboom}, M. and {Earl}, N. and {Erben}, T. and {Fabbro}, S. and
         {Ferreira}, L.~A. and {Finethy}, T. and {Fox}, R.~T. and
         {Garrison}, L.~H. and {Gibbons}, S.~L.~J. and {Goldstein}, D.~A. and
         {Gommers}, R. and {Greco}, J.~P. and {Greenfield}, P. and
         {Groener}, A.~M. and {Grollier}, F. and {Hagen}, A. and {Hirst}, P. and
         {Homeier}, D. and {Horton}, A.~J. and {Hosseinzadeh}, G. and {Hu}, L. and
         {Hunkeler}, J.~S. and {Ivezi{\'c}}, {\v{Z}}. and {Jain}, A. and
         {Jenness}, T. and {Kanarek}, G. and {Kendrew}, S. and {Kern}, N.~S. and
         {Kerzendorf}, W.~E. and {Khvalko}, A. and {King}, J. and {Kirkby}, D. and
         {Kulkarni}, A.~M. and {Kumar}, A. and {Lee}, A. and {Lenz}, D. and
         {Littlefair}, S.~P. and {Ma}, Z. and {Macleod}, D.~M. and
         {Mastropietro}, M. and {McCully}, C. and {Montagnac}, S. and
         {Morris}, B.~M. and {Mueller}, M. and {Mumford}, S.~J. and {Muna}, D. and
         {Murphy}, N.~A. and {Nelson}, S. and {Nguyen}, G.~H. and
         {Ninan}, J.~P. and {N{\"o}the}, M. and {Ogaz}, S. and {Oh}, S. and
         {Parejko}, J.~K. and {Parley}, N. and {Pascual}, S. and {Patil}, R. and
         {Patil}, A.~A. and {Plunkett}, A.~L. and {Prochaska}, J.~X. and
         {Rastogi}, T. and {Reddy Janga}, V. and {Sabater}, J. and
         {Sakurikar}, P. and {Seifert}, M. and {Sherbert}, L.~E. and
         {Sherwood-Taylor}, H. and {Shih}, A.~Y. and {Sick}, J. and
         {Silbiger}, M.~T. and {Singanamalla}, S. and {Singer}, L.~P. and
         {Sladen}, P.~H. and {Sooley}, K.~A. and {Sornarajah}, S. and
         {Streicher}, O. and {Teuben}, P. and {Thomas}, S.~W. and
         {Tremblay}, G.~R. and {Turner}, J.~E.~H. and {Terr{\'o}n}, V. and
         {van Kerkwijk}, M.~H. and {de la Vega}, A. and {Watkins}, L.~L. and
         {Weaver}, B.~A. and {Whitmore}, J.~B. and {Woillez}, J. and
         {Zabalza}, V. and {Astropy Contributors}},
        title = "{The Astropy Project: Building an Open-science Project and Status of the v2.0 Core Package}",
      journal = {\aj},
         year = 2018,
        month = sep,
       volume = {156},
       number = {3},
          eid = {123},
        pages = {123},
          doi = {10.3847/1538-3881/aabc4f},
archivePrefix = {arXiv},
       eprint = {1801.02634},
 primaryClass = {astro-ph.IM},
       adsurl = {https://ui.adsabs.harvard.edu/abs/2018AJ....156..123A}
}

@ARTICLE{astropy:2022,
       author = {{Astropy Collaboration} and {Price-Whelan}, Adrian M. and {Lim}, Pey Lian and {Earl}, Nicholas and {Starkman}, Nathaniel and {Bradley}, Larry and {Shupe}, David L. and {Patil}, Aarya A. and {Corrales}, Lia and {Brasseur}, C.~E. and {N{\"o}the}, Maximilian and {Donath}, Axel and {Tollerud}, Erik and {Morris}, Brett M. and {Ginsburg}, Adam and {Vaher}, Eero and {Weaver}, Benjamin A. and {Tocknell}, James and {Jamieson}, William and {van Kerkwijk}, Marten H. and {Robitaille}, Thomas P. and {Merry}, Bruce and {Bachetti}, Matteo and {G{\"u}nther}, H. Moritz and {Aldcroft}, Thomas L. and {Alvarado-Montes}, Jaime A. and {Archibald}, Anne M. and {B{\'o}di}, Attila and {Bapat}, Shreyas and {Barentsen}, Geert and {Baz{\'a}n}, Juanjo and {Biswas}, Manish and {Boquien}, M{\'e}d{\'e}ric and {Burke}, D.~J. and {Cara}, Daria and {Cara}, Mihai and {Conroy}, Kyle E. and {Conseil}, Simon and {Craig}, Matthew W. and {Cross}, Robert M. and {Cruz}, Kelle L. and {D'Eugenio}, Francesco and {Dencheva}, Nadia and {Devillepoix}, Hadrien A.~R. and {Dietrich}, J{\"o}rg P. and {Eigenbrot}, Arthur Davis and {Erben}, Thomas and {Ferreira}, Leonardo and {Foreman-Mackey}, Daniel and {Fox}, Ryan and {Freij}, Nabil and {Garg}, Suyog and {Geda}, Robel and {Glattly}, Lauren and {Gondhalekar}, Yash and {Gordon}, Karl D. and {Grant}, David and {Greenfield}, Perry and {Groener}, Austen M. and {Guest}, Steve and {Gurovich}, Sebastian and {Handberg}, Rasmus and {Hart}, Akeem and {Hatfield-Dodds}, Zac and {Homeier}, Derek and {Hosseinzadeh}, Griffin and {Jenness}, Tim and {Jones}, Craig K. and {Joseph}, Prajwel and {Kalmbach}, J. Bryce and {Karamehmetoglu}, Emir and {Ka{\l}uszy{\'n}ski}, Miko{\l}aj and {Kelley}, Michael S.~P. and {Kern}, Nicholas and {Kerzendorf}, Wolfgang E. and {Koch}, Eric W. and {Kulumani}, Shankar and {Lee}, Antony and {Ly}, Chun and {Ma}, Zhiyuan and {MacBride}, Conor and {Maljaars}, Jakob M. and {Muna}, Demitri and {Murphy}, N.~A. and {Norman}, Henrik and {O'Steen}, Richard and {Oman}, Kyle A. and {Pacifici}, Camilla and {Pascual}, Sergio and {Pascual-Granado}, J. and {Patil}, Rohit R. and {Perren}, Gabriel I. and {Pickering}, Timothy E. and {Rastogi}, Tanuj and {Roulston}, Benjamin R. and {Ryan}, Daniel F. and {Rykoff}, Eli S. and {Sabater}, Jose and {Sakurikar}, Parikshit and {Salgado}, Jes{\'u}s and {Sanghi}, Aniket and {Saunders}, Nicholas and {Savchenko}, Volodymyr and {Schwardt}, Ludwig and {Seifert-Eckert}, Michael and {Shih}, Albert Y. and {Jain}, Anany Shrey and {Shukla}, Gyanendra and {Sick}, Jonathan and {Simpson}, Chris and {Singanamalla}, Sudheesh and {Singer}, Leo P. and {Singhal}, Jaladh and {Sinha}, Manodeep and {Sip{\H{o}}cz}, Brigitta M. and {Spitler}, Lee R. and {Stansby}, David and {Streicher}, Ole and {{\v{S}}umak}, Jani and {Swinbank}, John D. and {Taranu}, Dan S. and {Tewary}, Nikita and {Tremblay}, Grant R. and {de Val-Borro}, Miguel and {Van Kooten}, Samuel J. and {Vasovi{\'c}}, Zlatan and {Verma}, Shresth and {de Miranda Cardoso}, Jos{\'e} Vin{\'\i}cius and {Williams}, Peter K.~G. and {Wilson}, Tom J. and {Winkel}, Benjamin and {Wood-Vasey}, W.~M. and {Xue}, Rui and {Yoachim}, Peter and {Zhang}, Chen and {Zonca}, Andrea and {Astropy Project Contributors}},
        title = "{The Astropy Project: Sustaining and Growing a Community-oriented Open-source Project and the Latest Major Release (v5.0) of the Core Package}",
      journal = {\apj},
         year = 2022,
        month = aug,
       volume = {935},
       number = {2},
          eid = {167},
        pages = {167},
          doi = {10.3847/1538-4357/ac7c74},
archivePrefix = {arXiv},
       eprint = {2206.14220},
 primaryClass = {astro-ph.IM},
       adsurl = {https://ui.adsabs.harvard.edu/abs/2022ApJ...935..167A}
}

@Article{matplotlib,
  Author    = {Hunter, J. D.},
  Title     = {Matplotlib: A 2D graphics environment},
  Journal   = {Computing in Science \& Engineering},
  Volume    = {9},
  Number    = {3},
  Pages     = {90--95},
  publisher = {IEEE COMPUTER SOC},
  doi       = {10.1109/MCSE.2007.55},
  year      = 2007
}

@Article{numpy,
 title         = {Array programming with {NumPy}},
 author        = {Charles R. Harris and K. Jarrod Millman and St{\'{e}}fan J.
                 van der Walt and Ralf Gommers and Pauli Virtanen and David
                 Cournapeau and Eric Wieser and Julian Taylor and Sebastian
                 Berg and Nathaniel J. Smith and Robert Kern and Matti Picus
                 and Stephan Hoyer and Marten H. van Kerkwijk and Matthew
                 Brett and Allan Haldane and Jaime Fern{\'{a}}ndez del
                 R{\'{i}}o and Mark Wiebe and Pearu Peterson and Pierre
                 G{\'{e}}rard-Marchant and Kevin Sheppard and Tyler Reddy and
                 Warren Weckesser and Hameer Abbasi and Christoph Gohlke and
                 Travis E. Oliphant},
 year          = {2020},
 month         = sep,
 journal       = {Nature},
 volume        = {585},
 number        = {7825},
 pages         = {357--362},
 doi           = {10.1038/s41586-020-2649-2},
 publisher     = {Springer Science and Business Media {LLC}},
 url           = {https://doi.org/10.1038/s41586-020-2649-2}
}

@ARTICLE{Scipy,
  author  = {Virtanen, Pauli and Gommers, Ralf and Oliphant, Travis E. and
            Haberland, Matt and Reddy, Tyler and Cournapeau, David and
            Burovski, Evgeni and Peterson, Pearu and Weckesser, Warren and
            Bright, Jonathan and {van der Walt}, St{\'e}fan J. and
            Brett, Matthew and Wilson, Joshua and Millman, K. Jarrod and
            Mayorov, Nikolay and Nelson, Andrew R. J. and Jones, Eric and
            Kern, Robert and Larson, Eric and Carey, C J and
            Polat, {\.I}lhan and Feng, Yu and Moore, Eric W. and
            {VanderPlas}, Jake and Laxalde, Denis and Perktold, Josef and
            Cimrman, Robert and Henriksen, Ian and Quintero, E. A. and
            Harris, Charles R. and Archibald, Anne M. and
            Ribeiro, Ant{\^o}nio H. and Pedregosa, Fabian and
            {van Mulbregt}, Paul and {SciPy 1.0 Contributors}},
  title   = {{{SciPy} 1.0: Fundamental Algorithms for Scientific
            Computing in Python}},
  journal = {Nature Methods},
  year    = {2020},
  volume  = {17},
  pages   = {261--272},
  adsurl  = {https://rdcu.be/b08Wh},
  doi     = {10.1038/s41592-019-0686-2},
}

@article{Yang_2024,
   title={A parametric model for self-interacting dark matter halos},
   volume={2024},
   ISSN={1475-7516},
   url={http://dx.doi.org/10.1088/1475-7516/2024/02/032},
   DOI={10.1088/1475-7516/2024/02/032},
   number={02},
   journal={Journal of Cosmology and Astroparticle Physics},
   publisher={IOP Publishing},
   author={Yang, Daneng and Nadler, Ethan O. and Yu, Hai-Bo and Zhong, Yi-Ming},
   year={2024},
   month=Feb, pages={032} }

@misc{he2026probingcollapseddarkmatter,
      title={Probing Collapsed Dark Matter Halos with Fast Radio Bursts}, 
      author={Yuxuan He and Weiyang Wang and Chen Zhang and Yi-Ming Zhong},
      year={2026},
      eprint={2604.12189},
      archivePrefix={arXiv},
      primaryClass={astro-ph.CO},
      url={https://arxiv.org/abs/2604.12189}, 
}

@article{Essig_2019,
  title = {Constraining Dissipative Dark Matter Self-Interactions},
  author = {Essig, Rouven and McDermott, Samuel D. and Yu, Hai-Bo and Zhong, Yi-Ming},
  journal = {Phys. Rev. Lett.},
  volume = {123},
  issue = {12},
  pages = {121102},
  numpages = {6},
  year = {2019},
  month = {Sep},
  publisher = {American Physical Society},
  doi = {10.1103/PhysRevLett.123.121102},
  url = {https://link.aps.org/doi/10.1103/PhysRevLett.123.121102}
}

@ARTICLE{Navarro_1997,
       author = {{Navarro}, Julio F. and {Frenk}, Carlos S. and {White}, Simon D.~M.},
        title = "{A Universal Density Profile from Hierarchical Clustering}",
      journal = {\apj},
         year = 1997,
        month = dec,
       volume = {490},
       number = {2},
        pages = {493-508},
          doi = {10.1086/304888},
archivePrefix = {arXiv},
       eprint = {astro-ph/9611107},
 primaryClass = {astro-ph},
       adsurl = {https://ui.adsabs.harvard.edu/abs/1997ApJ...490..493N}
}

@ARTICLE{Li_2025_Horseshoe,
       author = {{Li}, Sung Kei and {Weisenbach}, Luke and {Collett}, Thomas E. and {Diego}, Jose M. and {Lim}, Jeremy and {Broadhurst}, Thomas J. and {Chow}, Alex and {Enzi}, Wolfgang J.~R. and {Kelly}, Patrick L. and {Melo-Carneiro}, Carlos R. and {Palencia}, Jose M. and {Williams}, Liliya L.~R. and {Zhang}, Jiashuo},
        title = "{Lensed stars in galaxy─galaxy strong lensing ─ a JWST prediction for the Cosmic Horseshoe}",
      journal = {\mnras},
         year = 2025,
        month = dec,
       volume = {544},
       number = {4},
        pages = {4469-4481},
          doi = {10.1093/mnras/staf1959},
archivePrefix = {arXiv},
       eprint = {2509.16154},
 primaryClass = {astro-ph.CO},
       adsurl = {https://ui.adsabs.harvard.edu/abs/2025MNRAS.544.4469L}
}

@ARTICLE{Li_2025_IMF,
       author = {{Li}, Sung Kei and {Diego}, Jose M. and {Meena}, Ashish K. and {Lim}, Jeremy and {Fung}, Leo W.~H. and {Levitskiy}, Arsen and {Nianias}, James and {Palencia}, Jose M. and {Williams}, Hayley and {Zhang}, Jiashuo and {Amruth}, Alfred and {Broadhurst}, Thomas J. and {Chen}, WenLei and {Filippenko}, Alexei V. and {Kelly}, Patrick L. and {Koekemoer}, Anton M. and {Perera}, Derek and {Sun}, Bangzheng and {Williams}, Liliya L.~R. and {Windhorst}, Rogier A. and {Yan}, Haojin and {Zitrin}, Adi},
        title = "{Constraining the z {\ensuremath{\sim}} 1 Initial Mass Function with HST and JWST Lensed Stars in MACS J0416.1{\ensuremath{-}}2403}",
      journal = {\apj},
         year = 2025,
        month = aug,
       volume = {988},
       number = {2},
          eid = {178},
        pages = {178},
          doi = {10.3847/1538-4357/ade4bd},
archivePrefix = {arXiv},
       eprint = {2504.06992},
 primaryClass = {astro-ph.CO},
       adsurl = {https://ui.adsabs.harvard.edu/abs/2025ApJ...988..178L}
}

@ARTICLE{Palencia_2025_microlensing,
       author = {{Palencia}, J.~M. and {Diego}, J.~M. and {Dai}, L. and {Pascale}, M. and {Windhorst}, R. and {Koekemoer}, A.~M. and {Li}, Sung Kei and {Kavanagh}, B.~J. and {Sun}, Fengwu and {Alfred}, Amruth and {Meena}, Ashish K. and {Broadhurst}, Thomas J. and {Kelly}, Patrick L. and {Perera}, Derek and {Williams}, Hayley and {Zitrin}, Adi},
        title = "{Microlensing at cosmological distances: Event rate predictions in the Warhol arc of MACS 0416}",
      journal = {\aap},
         year = 2025,
        month = jul,
       volume = {699},
          eid = {A295},
        pages = {A295},
          doi = {10.1051/0004-6361/202555447},
archivePrefix = {arXiv},
       eprint = {2504.07039},
 primaryClass = {astro-ph.CO},
       adsurl = {https://ui.adsabs.harvard.edu/abs/2025A&A...699A.295P}
}

@ARTICLE{Broadhurst_2025,
       author = {{Broadhurst}, Tom and {Li}, Sung Kei and {Alfred}, Amruth and {Diego}, Jose M. and {Morilla}, Paloma and {Kelly}, Patrick L. and {Sun}, Fengwu and {Oguri}, Masamune and {Williams}, Hayley and {Windhorst}, Rogier and {Zitrin}, Adi and {Abe}, Katsuya T. and {Chen}, Wenlei and {Dai}, Liang and {Fudamoto}, Yoshinobu and {Kawai}, Hiroki and {Lim}, Jeremy and {Liu}, Tao and {Meena}, Ashish K. and {Palencia}, Jose M. and {Smoot}, George F. and {Williams}, Liliya L.~R.},
        title = "{Dark Matter Distinguished by Skewed Microlensing in the ``Dragon Arc''}",
      journal = {\apjl},
         year = 2025,
        month = jan,
       volume = {978},
       number = {1},
          eid = {L5},
        pages = {L5},
          doi = {10.3847/2041-8213/ad9aa8},
archivePrefix = {arXiv},
       eprint = {2405.19422},
 primaryClass = {astro-ph.CO},
       adsurl = {https://ui.adsabs.harvard.edu/abs/2025ApJ...978L...5B}
}

@ARTICLE{Palencia_2026,
       author = {{Palencia}, J.~M. and {Sun}, Fengwu and {Diego}, J.~M. and {Fudamoto}, Yoshinobu and {Koekemoer}, Anton M. and {Willmer}, Christopher N.~A. and {Iani}, Eduardo and {Lin}, Xiaojing and {Pierel}, Justin D.~R. and {Amruth}, Alfred and {Broadhurst}, Tom and {Chen}, W. and {Dai}, Liang and {Espada}, Daniel and {Filippenko}, Alexei V. and {Fujimoto}, Seiji and {Kelly}, Patrick L. and {Li}, Mingyu and {Li}, Sung Kei and {Meena}, Ashish Kumar and {Miralda-Escud{\'e}}, Jordi and {Morilla}, P. and {Struble}, Mitchell F. and {Williams}, Hayley and {Windhorst}, Rogier A. and {Zackrisson}, E. and {Zhou}, Ruwen and {Zitrin}, Adi},
        title = "{First Statistical Study of Over 100 Magnified Stellar Events at Redshift $z \approx 0.725$ with JWST}",
      journal = {arXiv e-prints},
         year = 2026,
        month = apr,
          eid = {arXiv:2604.22702},
        pages = {arXiv:2604.22702},
          doi = {10.48550/arXiv.2604.22702},
archivePrefix = {arXiv},
       eprint = {2604.22702},
 primaryClass = {astro-ph.GA},
       adsurl = {https://ui.adsabs.harvard.edu/abs/2026arXiv260422702P}
}

@ARTICLE{Palencia_2024,
       author = {{Palencia}, J.~M. and {Diego}, J.~M. and {Kavanagh}, B.~J. and {Mart{\'\i}nez-Arrizabalaga}, J.},
        title = "{Statistics of magnification for extremely lensed high redshift stars}",
      journal = {\aap},
         year = 2024,
        month = jul,
       volume = {687},
          eid = {A81},
        pages = {A81},
          doi = {10.1051/0004-6361/202347492},
archivePrefix = {arXiv},
       eprint = {2307.09505},
 primaryClass = {astro-ph.CO},
       adsurl = {https://ui.adsabs.harvard.edu/abs/2024A&A...687A..81P}
}

\appendix

\section{Analytical understanding of formation of milliCCs}
\label{sec: analytical}

Our numerical simulations have shown that milliCCs always form at positive parity, and only form at negative parity when $\Gamma$ is large enough and the core size is small after core collapse. Here, we perform an analytical derivation to better understand why this would happen.

The formation of (milli)CC happens when the sign of the magnification changes. In other words, there exists a criticality for which this happens. We can rewrite Eqn.~\ref{eqn: magnification} in terms of lensing convergence, $\kappa$, and shear, $\gamma$, such that for a tangential arc we are interested in this work, the criticality happens when the tangential eigenvalue, $1-\kappa-\gamma = 0$, given that the radial eigenvalue, $1-\kappa+\gamma > 0$. When a subhalo (subscript, $H$) is perturbing the macroscopic cluster environment (subscript, $m$), we can rewrite the convergence and shear as:

\begin{equation}
\kappa' = \kappa_m + \kappa_H, \qquad \gamma' = \gamma_m + \gamma_H .
\label{eqn: two_case}
\end{equation}

\noindent The Abel projection of Eqn.~\ref{eqn: cusp_SIDM_NFW}, assuming $r>>r_c$, gives:

\begin{equation}
    \kappa(r) \propto r^{1-\Gamma}.
\end{equation}

\noindent The shear produced by the subhalo as a function of distance from the center is:

\begin{equation}
    \gamma(r) = \bar\kappa(<r) - \kappa(r),
\end{equation}

\noindent with $\bar\kappa(<r)$ the mean enclosed convergence at $r$:

\begin{equation}
    \bar\kappa(<r) = \frac{2}{r^2}\int_0^r \kappa(r')\,r'\,dr'.
\end{equation}

\noindent Substituting and rearranging these gives:

\begin{equation}
    \frac{\gamma(r)}{\kappa(r)} =  \frac{\Gamma-1}{3-\Gamma}.
\end{equation}

\noindent One can see that when $\Gamma = 2$, the ratio is one. But then when $\Gamma$ increases, this ratio grows steeply and eventually diverges at $\Gamma = 3$.
This aligns with one's expectation, as a steeper profile has more mass ($\kappa$) towards the center. In other words, for the range of $2 \leq \Gamma \leq 3$ considered in this work, the shear component of subhaloes at any given radius will become strong and stronger than the convergence component when we increase $\Gamma$.

With these ingredients, we can ask when a subhalo perturbing the cluster field nucleates a new (milli)CC. The combination in Eqn.~\ref{eqn: two_case}, $\gamma_H$ adding directly onto $\gamma_m$, is only valid along one specific direction: the line joining the subhalo to the cluster center, where the subhalo's own shear (circular about its own center) happens to point the same way as the macro shear (circular about the cluster center). This is the direction that controls the tangential eigenvalue, $1-\kappa'-\gamma'$, and since $\kappa_H,\gamma_H\geq0$ can only push this quantity down, it can only be driven through zero from the positive-parity side, explaining why regardless of the subhalo concentration, CCs are always nucleating in the positive parity. 

Perpendicular to that line, the subhalo's shear instead opposes $\gamma_m$ rather than reinforcing it, and it is the radial eigenvalue, $1-\kappa'+\gamma'$, that gets pushed toward zero there. Because $1-\kappa_m+\gamma_m>0$ everywhere in the negative-parity region by definition, this channel is available throughout that region, not just at its boundary.

Substituting $\gamma' = \gamma_m + \gamma_H$ (Eqn.~\ref{eqn: two_case}) into each of these two conditions in turn gives two thresholds on $\Gamma$ where CC can form in the two respective parities:

\begin{equation}
\Gamma \geq 3-\frac{2\kappa_H}{1-\kappa_m-\gamma_m} \qquad \text{(positive parity)},
\label{eqn: pos_parity}
\end{equation}
\begin{equation}
\Gamma \geq 3-\frac{2\kappa_H}{1-\kappa_m+\gamma_m} \qquad \text{(negative parity)}.
\label{eqn: neg_parity}
\end{equation}

Since $1-\kappa_m-\gamma_m$ and $1-\kappa_m+\gamma_m$ are different numbers whenever $\gamma_m\neq0$ --- a subhalo at fixed $(\kappa_m,\gamma_m)$ can satisfy one comfortably while falling short of the other. Since $\gamma_m>0$ for any subhalo embedded in the macro-model shear field, $1-\kappa_m-\gamma_m < 1-\kappa_m+\gamma_m$, so the positive-parity threshold $3-2\kappa_H/(1-\kappa_m-\gamma_m)$ is always lower than its negative-parity counterpart $3-2\kappa_H/(1-\kappa_m+\gamma_m)$: the positive-parity condition is the easier of the two to satisfy, for any $(\kappa_m,\gamma_m,\kappa_H)$. As $\kappa_H\to0$, both thresholds approach $\Gamma\to3$, so only the steepest subhalos can nucleate a milliCC at all; as $\kappa_H$ grows relative to $1-\kappa_m\mp\gamma_m$, both thresholds drop well below 3, admitting progressively shallower profiles. Because the positive-parity threshold has the smaller denominator, it falls faster as $\kappa_H$ increases, so a subhalo of intermediate steepness can cross it while the negative-parity threshold remains out of reach. This is why steep, core-collapsed subhalos nucleate milliCCs on both sides of the macroscopic CC, while shallower ones are confined to the positive parity side alone. This also explains why the orientation of the hourglass-shaped milliCCs are orthogonal in the two parities, as shown in the literature \citep[e.g., ][]{Diego_2019, Williams_2024}, as well as in Fig.~\ref{fig: mag_maps}.

\clearpage 
\onecolumn
\section{Spatial distribution for different $\beta$s}

\begin{figure*}[!b]
    \centering
    \includegraphics[width=0.8\linewidth]{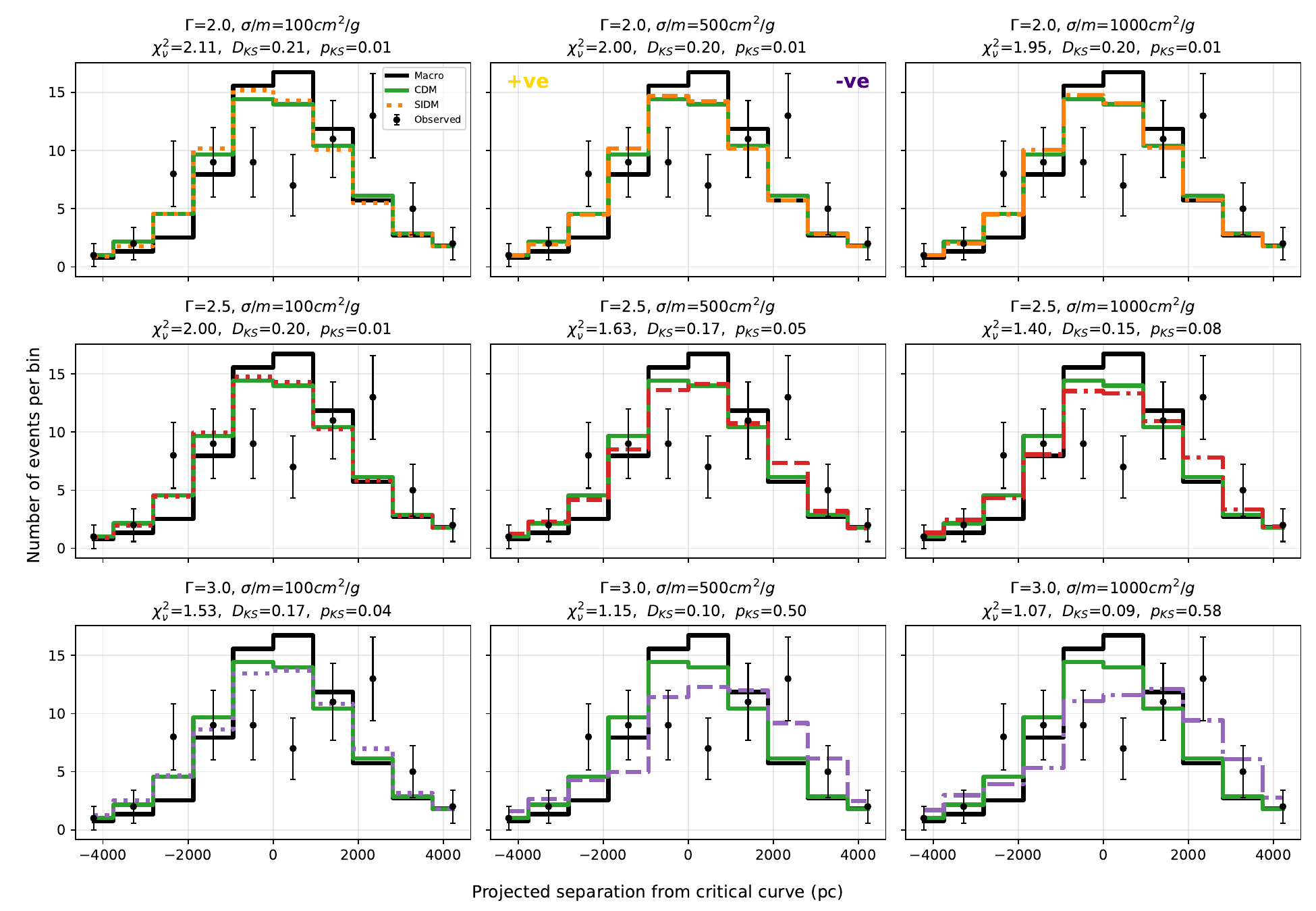}
    \caption{Same as Fig.~\ref{fig: spatial_distribution}, but with $\beta = 1.5$. A shallower slope washes down the central concentration and leads to broader distributions, diminishing the statistical significance among different models. }
    \label{fig: b1p5}
\end{figure*}

\begin{figure*}[!b]
    \centering
    \includegraphics[width=0.8\linewidth]{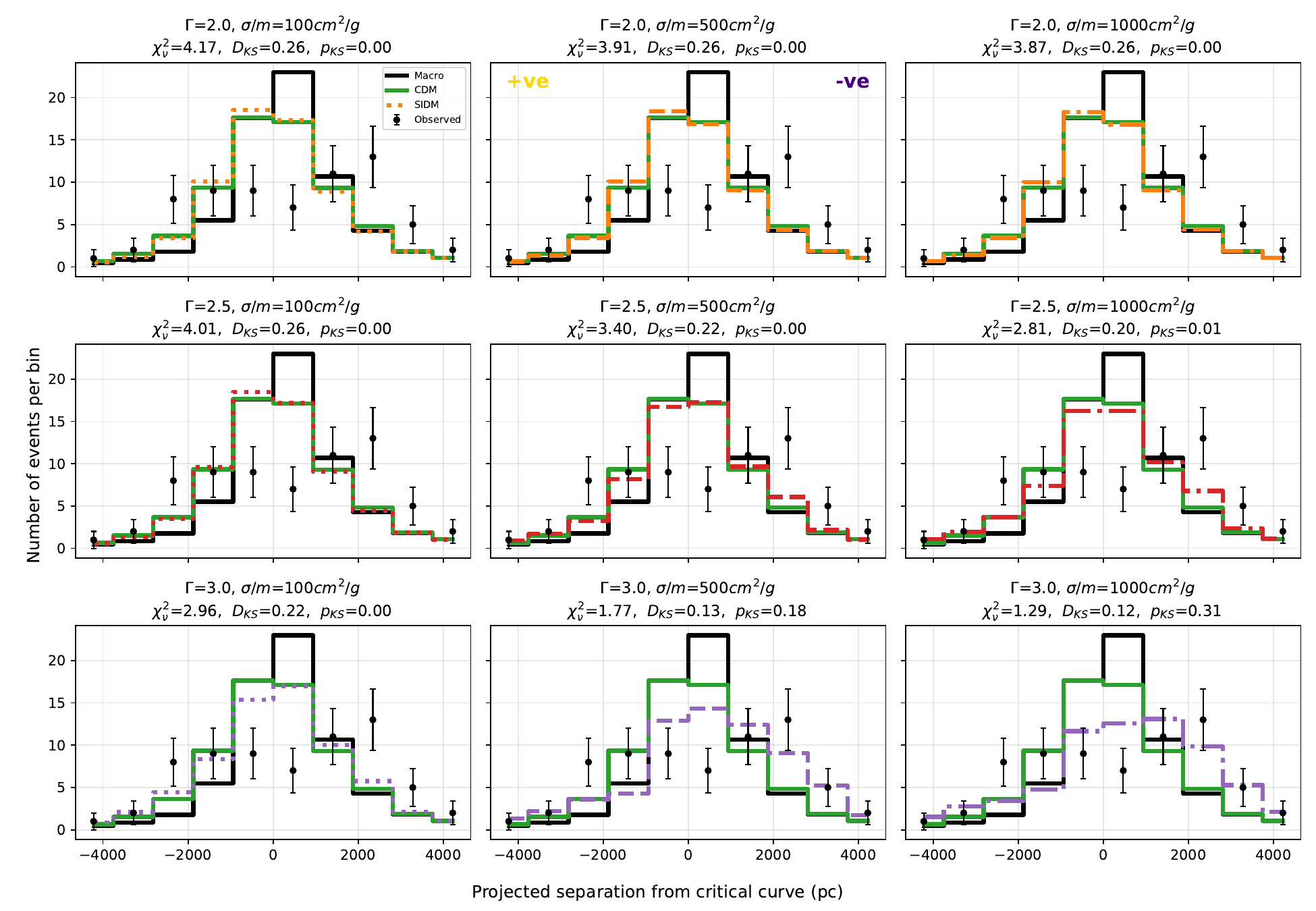}
    \caption{Same as Fig.~\ref{fig: spatial_distribution}, but with $\beta = 2.5$. One can see that most distributions are strongly peaked in the center, promoting the statistical significance betwen different models. }
    \label{fig: b2p5}
\end{figure*}







   
  




\end{document}